\documentclass[12pt]{article}
\usepackage[utf8]{inputenc}
\usepackage[top=50pt,bottom=50pt,left=68pt,right=66pt]{geometry}
\usepackage{amsmath}
\usepackage{bm}
\usepackage{booktabs}
\usepackage{amssymb}
\usepackage[titletoc,title]{appendix}
\usepackage{mathrsfs}
\usepackage{cite}
\usepackage{float}
\usepackage{xcolor}
\usepackage{multirow}
\usepackage{graphicx,caption,subcaption}
\usepackage[space]{grffile}
\usepackage{ upgreek }

\makeatletter
\def\@hangfrom#1{\setbox\@tempboxa\hbox{{#1}}%
      \hangindent 0pt
      \noindent\box\@tempboxa}
\makeatother

\newcommand\rring[1]{%
  {
   \mathop{\kern0pt #1}\limits^{
     \vbox to-1.85ex{
       \kern-2ex 
       \hbox to 0pt{\hss\normalfont\kern.1em \r{}\kern-.45em \r{}\hss}%
       \vss 
     }
   }
  }
}

\renewcommand{\arraystretch}{1.3}
\definecolor{darkblue}{rgb}{0.1,0.1,.7}
\usepackage[breaklinks=true,linktocpage=true,
colorlinks=true,urlcolor=darkblue,linkcolor=darkblue,
citecolor=darkblue,pdfpagelabels=true,hypertexnames=true,
plainpages=false,naturalnames=false,]{hyperref}

\usepackage{datetime}
\newdateformat{monthyeardate}{%
  \monthname[\THEMONTH], \THEYEAR}
\date{\monthyeardate\today}

\def\cg{{\cal G}}

\def\cl{{\cal L}}

\def\co{{\cal O}}

\begin{document}

\renewcommand{\arraystretch}{1.3}
\thispagestyle{empty}

{\hbox to\hsize{\vbox{\noindent\monthyeardate\today}}}

\noindent
\vskip2.0cm
\begin{center}

{\Large\bf Bouncing cosmologies from Born--Infeld-type gravity}

\vglue.3in

Yermek Aldabergenov,${}^{a,}$\footnote{ayermek@fudan.edu.cn (corresponding author)} Wei Lin,${}^{a,}$\footnote{wlin24@m.fudan.edu.cn} Rongjian Li,${}^{a,}$\footnote{24210190018@m.fudan.edu.cn} Ding Ding,${}^{b,}$\footnote{iasdding@ust.hk} Yidun Wan${}^{a,c,}$\footnote{ydwan@fudan.edu.cn (corresponding author)}
\vglue.1in

${}^a$~{\it State Key Laboratory of Surface Physics, Center for Astronomy and Astrophysics, Department of Physics, Center for Field Theory and Particle Physics, and Institute for Nanoelectronic devices and Quantum computing, Fudan University,
 2005 Songhu Road, Shanghai 200433, China}\\
${}^b$~{\it The HKUST Jockey Club Institute for Advanced Study,
The Hong Kong University of Science and Technology,
Clear Water Bay, Kowloon, Hong Kong, P.R. China}\\
${}^c$~{\it Shanghai Research Center for Quantum Sciences, 99 Xiupu Road, Shanghai 201315, China}\\
\vglue.1in

\end{center}

\vglue.3in

\begin{center}
{\Large\bf Abstract}
\vglue.2in
\end{center}

We construct a Born--Infeld-type $f(R,\cg)$ modification of gravity, where $\cg$ is the Gauss--Bonnet term, by embedding Born--Infeld electrodynamics in a five-dimensional pure modified gravity. This method leads to the correspondence between curvature scalars and electromagnetic field strength scalars -- $R\leftrightarrow F_{\mu\nu}F^{\mu\nu}$ and $\cg\leftrightarrow (\epsilon_{\mu\nu\rho\sigma}F^{\mu\nu}F^{\rho\sigma})^2$ -- allowing us to replicate the structure of Born--Infeld electrodynamics in the gravitational sector. The resulting Born--Infeld-type gravity is a ghost-free $f(R,\cg)$ theory which reduces to Einstein gravity in the low-energy limit. In this work, we focus on bouncing cosmological solutions of such a theory, which require positive spatial curvature.  By using both the Jordan and Einstein frame analyses, we find a vast space of bouncing solutions with different asymptotic behaviors, including solutions with multiple bounces grouped together. Observational consequences of such solutions will be investigated in the future.

\newpage

\tableofcontents

\setcounter{footnote}{0}

\section{Introduction}

Non-linear electrodynamics, first proposed by Born and Infeld in 1934 \cite{Born:1934gh}, has several remarkable properties discovered over the years. It curbs the divergent self-energy of a point particle, exhibits electromagnetic duality symmetry, and has connections to non-linear supersymmetry and string theory. Various attempts have been made to replicate the structure of the Born--Infeld (BI) theory in the gravitational sector in order to remove singularities plaguing general relativity (GR). In this work, we reconstruct a BI-type modified gravity based on the Kaluza--Klein (KK) approach of embedding electromagnetism in a higher-dimensional gravitational theory. We obtain the following two key results:
\begin{itemize}
    \item By embedding the BI electrodynamics in a five-dimensional (modified) gravity, we reconstruct a ghost-free BI-type $f(R,\cg)$ gravity in four dimensions, where $\cg$ is the Gauss--Bonnet (GB) term;
    \item We find and classify non-singular bouncing cosmological solutions of this theory, among which there are solutions with multiple (but finite number of) bounces.
\end{itemize}
Possible observational consequences of such multi-bounce solutions, as well as the generation of cosmological perturbations in this model will be investigated in the future.

Various approaches to construct BI-like modifications of gravity can be found in the literature, starting from Refs. \cite{Deser:1998rj,Vollick:2003qp,Vollick:2006qd,Nieto:2004qj} where the focus was on replicating the ``square-root of determinant" structure (see Eq. \eqref{L_BI_det}) in the gravity sector. Such theories in general suffer from Ostrogradski ghosts, which must be cancelled order-by-order in curvature, as discussed in \cite{Deser:1998rj}. A new problem then arises -- these ghost-free completions are no longer unique. A different approach can also be taken where the determinant structure is abandoned, and one considers instead a square-root of a certain combination of curvature scalars, at the expense of introducing new degrees of freedom (see, e.g., \cite{Feigenbaum:1997pf,Feigenbaum:1998wy,Comelli:2005tn,Garcia-Salcedo:2009lui,Kruglov:2012ja,Aldabergenov:2018qhs}, and for a more detailed review of BI-like gravities see \cite{BeltranJimenez:2017doy}). Theories constructed in this way belong to the $f(R,\cg,\ldots)$ class and have closed-form Lagrangians that are (under appropriate conditions on the parameters) ghost-free. Our proposed construction belongs to this class. Our approach is based on the idea of Kaluza \cite{Kaluza:1921tu} and Klein \cite{Klein:1926tv} that Maxwell electrodynamics (and 4D GR) can be embedded in a five-dimensional gravity. This implies the connection between the Einstein--Hilbert term and the Maxwell term $R\leftrightarrow F_{\mu\nu}F^{\mu\nu}$. Following this logic, we attempt to reconstruct an extension of GR from BI electrodynamics written (in 4D) in the form \eqref{L_BI} below, rather than in the determinant form. As the main goal of the BI-like gravity theories is often to tame spacetime singularities, in this paper we focus on non-singular bouncing cosmological solutions of our theory, and leave the study of black hole solutions for future work. Bouncing cosmologies in modified gravity is a large and active area of research on its own -- see for instance \cite{Barragan:2009sq,Saidov:2010wx,Bamba:2013fha,Bamba:2014zoa,Amani:2015upn,Odintsov:2015ynk,Nojiri:2017ncd,Terrucha:2019jpm,Barros:2019pvc,Elizalde:2020zcb,Odintsov:2020vjb,Odintsov:2021urx,Nojiri:2022xdo,Miranda:2022wvr,Singh:2022gln,Oikonomou:2022irx,Singh:2023gxd,Klinkhamer:2019dzc,Battista:2020lqv} -- and, hopefully, our methods for finding and studying bouncing solutions based on both the Einstein and Jordan frame analyses can be useful in other theories of modified gravity, beyond our particular models.

Our paper is organized as follows. Section \ref{sec_BI} recalls the main features of BI electrodynamics. Section \ref{sec_BIKK} uses the KK approach to embed 4D BI electrodynamics in a 5D modified gravity, and by performing dimensional reduction, derives 4D modified $f(R,\cg)$ gravity non-minimally coupled to BI electrodynamics. The setup is particularly simple, containing a single BI parameter (in addition to the Planck mass), and whose low-energy limit is the Einstein--Maxwell theory. Section \ref{Sec_Bounce} studies bouncing cosmological solutions in the gravitational sector by ignoring the electromagnetic field. The main features of our BI-like gravity are captured by a simplified $f(R)$ model without the Gauss--Bonnet term, which we study first. We will show that positive spatial curvature is a crucial ingredient for the bouncing solutions in this model. Section \ref{Sec_GB} introduces the GB term with an additional independent parameter for generality. Our analysis is done in both the Einstein frame and the Jordan frame because the models can be formulated as scalar-tensor theories with a single (non-ghost) scalaron field. This helps understand the bouncing dynamics in terms of the scalaron potential. Section \ref{Sec_Conclusion} gives some final comments and concluding remarks.

\section{Overview of the Born--Infeld theory}\label{sec_BI}

In this section, we briefly review the Born--Infeld non-linear electrodynamics. The BI Lagrangian has a supersymmetric version and various formulations (details can be found in \cite{Tseytlin:1999dj,Ketov:2001dq}). The BI Lagrangian has a usual determinant form~\footnote{We will work in the reduced Planck units, $M_P=1$, unless otherwise mentioned.}:
\begin{equation}
    {\cal L}_{\rm BI}=\frac{1}{b^2}\left(1-\sqrt{-{\rm det}(g_{\mu\nu}+bF_{\mu\nu})}\right)~,\quad 
    F_{\mu\nu}=\partial_\mu A_\nu-\partial_\nu A_\mu~,\label{L_BI_det}
\end{equation}
where $A_\mu$ is the (abelian) gauge field, and $b$ is the BI parameter. In four spacetime dimensions, \eqref{L_BI_det} can be written as
\begin{equation}
    {\cal L}_{\rm BI}=\frac{\sqrt{-g}}{b^2}\left(1-\sqrt{1+\frac{b^2}{2}F^2-\frac{b^4}{16}(F\tilde{F})^2}\right)~,\label{L_BI}
\end{equation}
where we define $\tilde{F}^{\mu\nu}\equiv\frac{1}{2}\epsilon^{\mu\nu\kappa\lambda}F_{\kappa\lambda}$,  
$F^2\equiv F_{\mu\nu}F^{\mu\nu}$ and $F\tilde{F}\equiv\frac{1}{2}\epsilon^{\mu\nu\kappa\lambda}F_{\mu\nu}F_{\kappa\lambda}$ with the Levi-Civita tensor $\epsilon$.

The equations of motion (EOM) derived from \eqref{L_BI} and the Bianchi identity for the gauge field can be written 
as
\begin{align}
    \nabla^{\mu}G_{\mu\nu}=0~,\label{dG=0}\\
    \nabla^{\mu}\tilde{F}_{\mu\nu}=0~,\label{F_Bianchi}
\end{align}
where
\begin{equation}
    G_{\mu\nu}\equiv -2\frac{\partial{\cal L}}{\partial F^{\mu\nu}}~.\label{G_def}
\end{equation}
The system comprised of \eqref{dG=0} and \eqref{F_Bianchi} is invariant under the $SO(2)$ electromagnetic duality rotations:
\begin{align}
    F_{\mu\nu}\rightarrow\cos\theta\, F_{\mu\nu}+\sin\theta\,\tilde{G}_{\mu\nu}~,\nonumber\\
    G_{\mu\nu}\rightarrow\cos\theta\, G_{\mu\nu}+\sin\theta\,\tilde{F}_{\mu\nu}~,\label{emduality}
\end{align}
where $\tilde{G}^{mn}\equiv\frac{1}{2}\epsilon^{mnkl}G_{kl}$.

The square root structure of the BI theory puts an upper bound on the field strength of $A_\mu$, and famously helps to tame the diverging self-energy of a point charge. Born--Infeld theory was later discovered as an effective action of open string theories \cite{Fradkin:1985qd,Abouelsaood:1986gd,Bergshoeff:1987at,Metsaev:1987qp} and D-branes \cite{Tseytlin:1996it,Bagger:1996wp,Rocek:1997hi}, where it is associated with partial supersymmetry breaking on a brane.

\section{Born--Infeld from Kaluza--Klein}\label{sec_BIKK}

Here, we will try to obtain (4D) Born--Infeld theory from a 5D gravitational theory, by identifying the abelian BI gauge field with the non-diagonal component of the 5D metric.

We take the following KK metric ansatz,
\begin{equation}\label{KK_metric}
\hat g_{MN}=
\begin{pmatrix}
g_{\mu\nu}+2\Phi^2A_\mu A_\nu & \sqrt{2}\Phi^2 A_\mu\\
\sqrt{2}\Phi^2 A_\nu & \Phi^2
\end{pmatrix}~,
\end{equation}
where $\Phi$ is a scalar field controlling the size of the compact dimension. In this work, we set $\Phi=1$, and its stabilization issue will be addressed elsewhere. We also assume that 4D fields do not depend on the compact dimension and only consider KK zero modes.

In order to derive the Born-Infeld non-linear electrodynamics from a 5D gravity, we note that the dimensional reduction of 5D scalar curvature $\hat R$ and Gauss--Bonnet term $\hat\cg$ yields
\begin{align}
    \hat R &=-\tfrac{1}{2}F_{\mu\nu}F^{\mu\nu}+\ldots~,\label{Rhat_F}\\
    \hat\cg &=\tfrac{3}{2}(F_{\mu\nu}\tilde F^{\mu\nu})^2-4\nabla_\lambda F^{\lambda\mu}\nabla^\nu F_{\nu\mu}+2\nabla_\lambda F_{\mu\nu}\nabla^\lambda F^{\mu\nu}+\ldots~,\label{Ghat_F}
\end{align}
where we take $\Phi=1$, and $\ldots$ stands for 4D curvature terms. In the first step, we ignore the 4D curvature terms and try to reconstruct the 5D gravity via a ``bottom-up" approach by demanding that in 4D and flat spacetime (Minkowski limit), the theory reduces to the BI electrodynamics, at least up to derivatives of the field strength.

Looking at the BI Lagrangian \eqref{L_BI}, one can notice that the field strength only enters it through a linear combination of $F^2$ and $(F\tilde F)^2$. Therefore, after taking into account the relations \eqref{Rhat_F} and \eqref{Ghat_F}, we choose as our starting point a 5D modified gravity defined by a function $f(\hat L)$, where $\hat L\equiv \hat R+\frac{b^2}{24}\hat\cg$, and $b$ is a real constant of inverse mass dimension (to be identified with the BI parameter later):
\begin{equation}\label{S_5d_f(L)}
    S_{5d}=\tfrac{1}{2}{\hat M}_P^3\int d^5x\sqrt{-\hat g}\,f(\hat L)~.
\end{equation}
Here, $\hat M_P$ is the 5D Planck mass. The action \eqref{S_5d_f(L)} can be written in the form
\begin{equation}\label{S_5d_Sigma}
    S_{5d}=\tfrac{1}{2}{\hat M}_P^3\int d^5x\sqrt{-\hat g}\big[f'(\Sigma)\hat L-\Sigma f'(\Sigma)+f(\Sigma)\big]~,
\end{equation}
with the help of an auxiliary scalar $\Sigma$. Varying \eqref{S_5d_Sigma} with respect to $\Sigma$ leads back to \eqref{S_5d_f(L)}, provided that $f''(\Sigma)\neq 0$.

We now perform dimensional reduction of \eqref{S_5d_Sigma} using the metric \eqref{KK_metric} (dimensional reduction of the relevant quantities can be found in the Appendix \ref{App_dimred}). Integrating over the compact coordinate $z$ with $\hat M^3_{\rm P}\int dz=M^2_{\rm P}=1$ (4D Planck units), we get
\begin{align}\label{S_4d_Sigma}
\begin{aligned}
    S_{4d}=\tfrac{1}{2}\int d^4x\sqrt{-g}\Big\{f'\Big[R-\tfrac{1}{2}F^2+\tfrac{b^2}{24}\Big(\cg &-RF^2-4R_{\mu\nu}F^{\mu\lambda}{F_\lambda}^\nu-R^{\mu\nu\rho\sigma}F_{\mu\nu}F_{\rho\sigma}+2\Box F^2\\&+4\nabla_\lambda\nabla^\mu(F_{\mu\nu}F^{\nu\lambda})+\tfrac{3}{2}(F\tilde F)^2\Big)\Big]-\Sigma f'+f\Big\}~,
\end{aligned}
\end{align}
where $f$ is a function of $\Sigma$. At this point, we introduce the function $f(\Sigma)$ leading to BI electrodynamics,
\begin{equation}\label{f_BI}
    f(\Sigma)=\frac{2}{b^2}(1-\sqrt{1-b^2\Sigma})~.
\end{equation}
It can be shown that eliminating $\Sigma$ and fixing the background metric to be Minkowski, $g_{\mu\nu}=\eta_{\mu\nu}$, leads to the Lagrangian
\begin{equation}\label{L_BI+derivatives}
    \cl=b^{-2}\bigg[1-\sqrt{1+\frac{b^2}{2}F^2-\frac{b^4}{16}(F\tilde F)^2-\frac{b^4}{12}\big(\Box F^2+2\partial_\lambda\partial^\mu(F_{\mu\nu}F^{\nu\lambda})\big)}~\bigg]~,
\end{equation}
which up to the field strength derivatives,~\footnote{Different sets of field strength derivative corrections can also be found in D-brane BI actions in superstring and bosonic string theories \cite{Wyllard:2001ye}. Such corrections appear to be different from what we obtained in \eqref{L_BI+derivatives}.} is equal to the BI theory \eqref{L_BI}. Notably, these derivative corrections do not lead to higher derivatives of the gauge potential $A_\mu$ in its equation of motion (in a full theory with gravitational degrees of freedom), and therefore, new degrees of freedom of the gauge field do not arise. This can be seen from the scalar-tensor formulation \eqref{S_5d_Sigma}, which is a non-minimally coupled gravitational theory with scalar--Gauss--Bonnet coupling, and has second order equations of motion. The resulting 4D scalar-vector-tensor theory (derived below) inherits this property.

Starting from Eq. \eqref{S_4d_Sigma}, if we set $A_\mu=0$ and focus on the gravitational action, after eliminating $\Sigma$ (and using the $f(\Sigma)$ function \eqref{f_BI}) we get a BI-like $f(R,\cg)$ gravity:
\begin{equation}\label{L_BIG}
    \cl=b^{-2}\sqrt{-g}\Big(1-\sqrt{1-b^2R-\tfrac{b^4}{24}\cg}\,\Big)~,
\end{equation}
which reduces to GR in the $b\rightarrow 0$ limit.

Going back to the action \eqref{S_4d_Sigma}, we can perform the Weyl rescaling $g_{\mu\nu}\rightarrow g_{\mu\nu}/f'$ to bring the action to the Einstein frame. In terms of the canonical scalaron $\varphi$, defined as $f'=e^{\sqrt{2/3}\varphi}$, the resulting Lagrangian reads (up to total derivatives)
\begin{align}
\begin{aligned}\label{L_BI_SVT}
    \sqrt{-g}^{\,-1}\cl &=\tfrac{1}{2}R-\tfrac{1}{2}\partial\varphi\partial\varphi-\tfrac{1}{4}e^{\sqrt{\frac{2}{3}}\varphi}F^2-V(\varphi)+\tfrac{b^2}{48}e^{\sqrt{\frac{2}{3}}\varphi}\Big[\cg+\tfrac{8}{3}G^{\mu\nu}\partial_\mu\varphi\partial_\nu\varphi\\
    &-\sqrt{\tfrac{8}{3}}\partial\varphi\partial\varphi\Box\varphi-e^{\sqrt{\frac{2}{3}}\varphi}(RF^2+4R_{\mu\nu}F^{\mu\lambda}{F_\lambda}^\nu+R^{\mu\nu\rho\sigma}F_{\mu\nu}F_{\rho\sigma})\\
    &+e^{\sqrt{\frac{2}{3}}\varphi}\Big(\sqrt{\tfrac{2}{3}}\Box\varphi+\tfrac{4}{3}\partial\varphi\partial\varphi\Big)F^2+2e^{\sqrt{\frac{2}{3}}\varphi}\Big(\sqrt{\tfrac{2}{3}}\nabla_\mu\nabla_\nu\varphi+\tfrac{7}{3}\partial_\mu\varphi\partial_\nu\varphi\Big)F^{\mu\lambda}{F_\lambda}^\nu\\
    &+\tfrac{3}{2}e^{2\sqrt{\frac{2}{3}}\varphi}(F\tilde F)^2\Big]~,
\end{aligned}
\end{align}
where the scalar potential is
\begin{equation}
    V(\varphi)=\frac{1}{2b^2}e^{-\sqrt{\frac{2}{3}}\varphi}\big(1-e^{-\sqrt{\frac{2}{3}}\varphi}\big)^2~.
\end{equation}
The Lagrangian \eqref{L_BI_SVT} belongs to a scalar-vector-tensor generalization of the Horndeski theory discussed in \cite{Heisenberg:2018acv,Mironov:2024umy,Mironov:2025dzz} and leads to second-order equations of motion for the scalaron $\varphi$, metric $g_{\mu\nu}$, and the gauge field $A_\mu$. In the limit $b\rightarrow 0$, the theory reduces to the Einstein--Maxwell, provided that the scalaron is stabilized at $\varphi=0$, which yields a stable Minkowski vacuum. A runaway vacuum at large $\varphi$ should also be noted. The potential and its effects on the bouncing solutions will be discussed in the following sections.

\section{Bouncing solutions in the BI-type gravity}\label{Sec_Bounce}

In the study of cosmological solutions, we use the Friedmann--Lema\^{\i}tre--Robertson--Walker (FLRW) metric with spatial curvature parameter $K$,
\begin{equation}\label{FLRW_K}
    ds^2 = -dt^2 + a(t)^2 \Big[ \frac{dr^2}{1-Kr^2} + r^2(d\theta^2 + \sin^2\theta d\phi^2) \Big]~,
\end{equation}
where $K=0,+1,-1$ describes a flat, closed, and open universe, respectively.~\footnote{Although current cosmological observations are compatible with a flat universe, there seems to be a slight preference (though inconclusive) for small positive spatial curvature \cite{Planck:2018vyg}.} When using the metric \eqref{FLRW_K}, the scalar curvature and the GB term become
\begin{align}
    R &=6\Big(\dot H+2H^2+\frac{K}{a^2}\Big)~,\label{R_FLRW_K}\\
    \cg &=24\Big(H^4+\dot HH^2+H^2\frac{K}{a^2}+\dot H\frac{K}{a^2}\Big)~.\label{G_FLRW_K}
\end{align}

We will consider cosmological solutions in pure gravitational theory \eqref{L_BIG} where the gauge field is set to zero.

\subsection{Simplified model: Born--Infeld-type $f(R)$ gravity}\label{FR_type}

First, consider a simplified model without the GB term,
\begin{equation}\label{L_sBI_gravity}
    \cl=b^{-2}\sqrt{-g}\Big(1-\sqrt{1-b^2R}\,\Big)~,
\end{equation}
which can be obtained from the 5D action \eqref{S_5d_f(L)} by discarding the GB term. Upon dimensional reduction, the vector part of the simplified theory is the minimal BI theory, $\cl\propto 1-\sqrt{1+\tfrac{1}{2}b^2F^2}$, which coincides with the electrostatic limit of the full BI theory and is often considered in the context of black hole solutions \cite{Cai:2004eh,Yang:2020jno,He:2022opa}.

The Lagrangian \eqref{L_sBI_gravity} is of $f(R)$ gravity type, and leads to the Einstein equations
\begin{equation}\label{fR_EFE}
    f_RR_{\mu\nu}-\tfrac{1}{2}fg_{\mu\nu}-\nabla_\mu\nabla_\nu f_R+g_{\mu\nu}\Box f_R=0~,
\end{equation}
where $f_R\equiv df/dR$, and in our case
\begin{equation}\label{f(R)_BI}
f=\frac{2}{b^2}\Big(1-\sqrt{1-b^2R}\Big)~.
\end{equation}

By using the FLRW metric \eqref{FLRW_K}, from the $00$ Einstein equation we get
\begin{equation}\label{EFE_00_fR}
H\dot f_R - (\dot H + H^2)f_R + \frac{1}{3b^2}(1-f_R^{-1}) = 0~,
\end{equation}
where, by using \eqref{f(R)_BI}, we wrote $f$ in terms of $f_R$ as $f=2b^{-2}(1-f_R^{-1})$. In an FLRW space, $f_R$ is
\begin{equation}
    f_R=(1-b^2R)^{-1/2}=\big[1-6b^2(\dot H+2H^2+K/a^2)\big]^{-1/2}~.
\end{equation}

Before solving Eq. \eqref{EFE_00_fR}, it is convenient to nondimensionalize it by introducing the rescaled scale factor and time,
\begin{equation}\label{nondim}
    \bar a\equiv a/b~,~~~\bar t\equiv t/b~,
\end{equation}
assuming $b\neq 0$. The corresponding rescaled Hubble function is $\bar H\equiv bH$, and the time derivative with respect to $\bar t$ is denoted $\mathring A=b\dot A$. In terms of $\bar a$ and $\bar t$, Eq. \eqref{EFE_00_fR} can be rewritten as
\begin{equation}\label{EFE_00_fR_no_b}
    \bar H\mathring f_R - (\mathring{\bar H} + \bar H^2)f_R + \tfrac{1}{3}(1-f_R^{-1}) = 0~,
\end{equation}
with
\begin{equation}
    f_R=\big[1-6(\mathring{\bar H}+2\bar H^2+K/\bar a^2)\big]^{-1/2}~,
\end{equation}
where explicit dependence on $b$ is removed.

Equation \eqref{EFE_00_fR_no_b} is a third-order equation for the scale factor $\bar a(\bar t)$, requiring three initial conditions for numerical integration. In particular, we are interested in bouncing solutions where $\mathring{\bar a}=0$ and $\bar a,\rring{\bar a}>0$ at the bounce point; since \eqref{EFE_00_fR} is an autonomous equation, we can fix the bounce point as $\bar t=0$ without loss of generality. Equation \eqref{EFE_00_fR_no_b} itself severely restricts the space of consistent initial conditions for bouncing solutions because the first term of Eq. \eqref{EFE_00_fR_no_b}, which includes $\rring{\bar H}$ (i.e., the third derivative of $\bar a$), multiplies $\bar H$. Since at the bounce point $\bar H(0)=0$, from \eqref{EFE_00_fR_no_b} we get an equation that $\bar a(0)$, $\mathring{\bar a}(0)$, and $\rring{\bar a}(0)$ must satisfy for a given curvature $K$, namely
\begin{equation}\label{H_eq_0}
    1-3\mathring{\bar H}(0)-6\frac{K}{\bar a^2(0)}=\sqrt{1-6\Big(\mathring{\bar H}(0)+\frac{K}{\bar a^2(0)}\Big)}~.
\end{equation}
This equation can be solved for $\mathring{\bar H}(0)$ as
\begin{equation}\label{H_root_0}
    \mathring{\bar H}(0)=-2\frac{K}{\bar a^2(0)}\pm 2\sqrt{\frac{K}{6\bar a^2(0)}}~,
\end{equation}
where the ``$+$" branch is restricted to $\bar a(0)>\sqrt{6}$, or $\mathring{\bar H}>0$ (otherwise it is incompatible with \eqref{H_eq_0}). One can immediately see from \eqref{H_root_0} that a flat universe ($K=0$) sets $\mathring{\bar H}(0)=0$, contradicting the bouncing requirement $\rring{\bar a}>0$~\footnote{An exception from this requirement is a bouncing scenario such as $\bar a\sim \bar t^4$, so that $\rring{\bar a}(0)=0$. Our numerical tests of Eq. \eqref{EFE_00_fR_no_b}, however, suggest that with $\rring{\bar a}(0)=0$ it is difficult to find numerically well-behaved bouncing solutions.}. An open ($K=-1$) bouncing universe is also excluded, as it leads to complex $\mathring{\bar H}(0)$. This leaves us with a closed universe, $K=+1$,~\footnote{See also Refs. \cite{Haro:2015eza,Matsui:2019ygj,Gungor:2020fce,Daniel:2022ppp} for other models utilizing positive spatial curvature to achieve successful bounce.} which is compatible with the bounce condition $\mathring{\bar H}(0)>0$ if $\bar a(0)>\sqrt{6}$. In summary, bouncing solutions (if any) to Eq. \eqref{EFE_00_fR_no_b} require $K=+1$, and consistent initial conditions for it must satisfy
\begin{equation}\label{a_IC_0}
    \bar a(0)>\sqrt{6}~,~~~\mathring{\bar a}(0)=0~,~~~\rring{\bar a}(0)=\frac{2}{\sqrt{6}}-\frac{2}{\bar a(0)}~.
\end{equation}

Numerically, it can be problematic to start from the initial condition $\mathring{\bar a}(0)=0$ because the coefficient of the highest-order term of \eqref{EFE_00_fR_no_b} vanishes at this point, leading to an apparent singularity. Therefore, we shift the initial time to $\bar t=\bar t_i$ with $\bar t_i\ll 1$ and expand $\bar a(\bar t_i)$ and its derivatives around $\bar a(0)$ at the linear order,
\begin{equation}
    \bar a(\bar t_i) \simeq \bar a(0)+\mathring{\bar a}(0)\bar t_i~,~~~\mathring{\bar a}(\bar t_i) \simeq \mathring{\bar a}(0)+\rring{\bar a}(0)\bar t_i~,~~~\rring{\bar a}(\bar t_i) \simeq \rring{\bar a}(0)+{\bar a}^{(3)}(0)\bar t_i~,
\end{equation}
where $\bar a^{(3)}\equiv d^3\bar a/d\bar t^3$. Assuming that $\bar a(\bar t)$ is an even function (at least around $\bar t=0$), so that $\bar a^{(3)}(0)=0$, we can completely fix the initial conditions at some small $\bar t_i$ from the initial conditions \eqref{a_IC_0} at $\bar t=0$. In particular, we get arbitrarily small but non-zero $\mathring{\bar a}(\bar t_i)$ to avoid the apparent initial singularity for numerical integration.

\subsubsection*{Runaway and oscillatory bouncing solutions}

\begin{figure}
\includegraphics[width=1\textwidth]{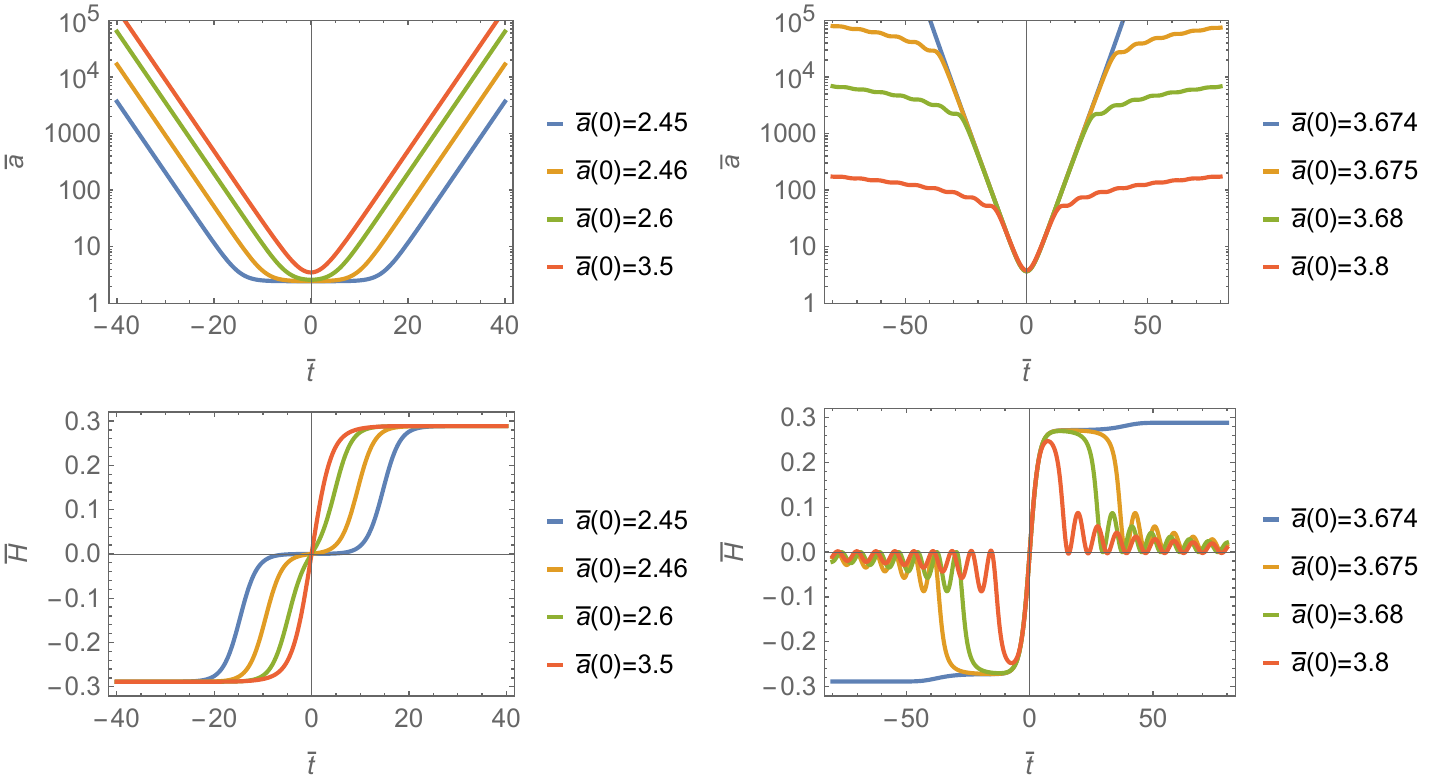}
\centering
\caption{Bouncing closed universe ($K=+1$) solutions in the minimal BI-like model \eqref{L_sBI_gravity} for different values of $\bar a(0)$.}
\label{Fig_R_bounce}
\end{figure}

Taking into account the above initial conditions, we perform numerical integration of the third-order equation \eqref{EFE_00_fR_no_b} for $\bar a(\bar t)$ with $K=+1$. Since there are no (explicit) free parameters, and $\mathring{\bar a}(0)$ and $\rring{\bar a}(0)$ are fixed, the bounce solution depends only on the initial value of the scale factor $\bar a(0)$ (which is bounded from below by $\sqrt{6}\approx 2.449$). Figure \ref{Fig_R_bounce} shows the bouncing solutions found numerically for different choices of $\bar a(0)$. For $2.45\lesssim \bar a(0)\lesssim 3.674$, the bouncing solutions have their Hubble function approaching a constant value at distant past and future (an analytical approximation is available in these regimes, as will be shown below) -- we will call this class of solutions ``runaway" solutions. For $3.675\lesssim \bar a(0)\lesssim 3.8$, the Hubble function, instead of plateauing, falls off and starts oscillating around $\bar H=0$ -- we will call these solutions ``oscillatory" solutions. During these oscillations, $\bar H$ is always positive in our examples, except for the choice $\bar a(0)=3.8$, where the minima of the oscillations reach small negative values, $\bar H\sim -0.002$. The two classes of bouncing solutions are separated by a critical value of the initial scale factor, $\bar a_{\rm cr}(0)$, somewhere between $3.674$ and $3.675$, as can be seen from the bottom-right plot of Fig. \ref{Fig_R_bounce}. In the next subsection, we will show that the exact critical value is $\bar a_{\rm cr}(0)=\tfrac{3}{2}\sqrt{6}\approx 3.6742$, and it is related to the local maximum of the scalaron potential.

\begin{figure}
 \includegraphics[width=1\textwidth]{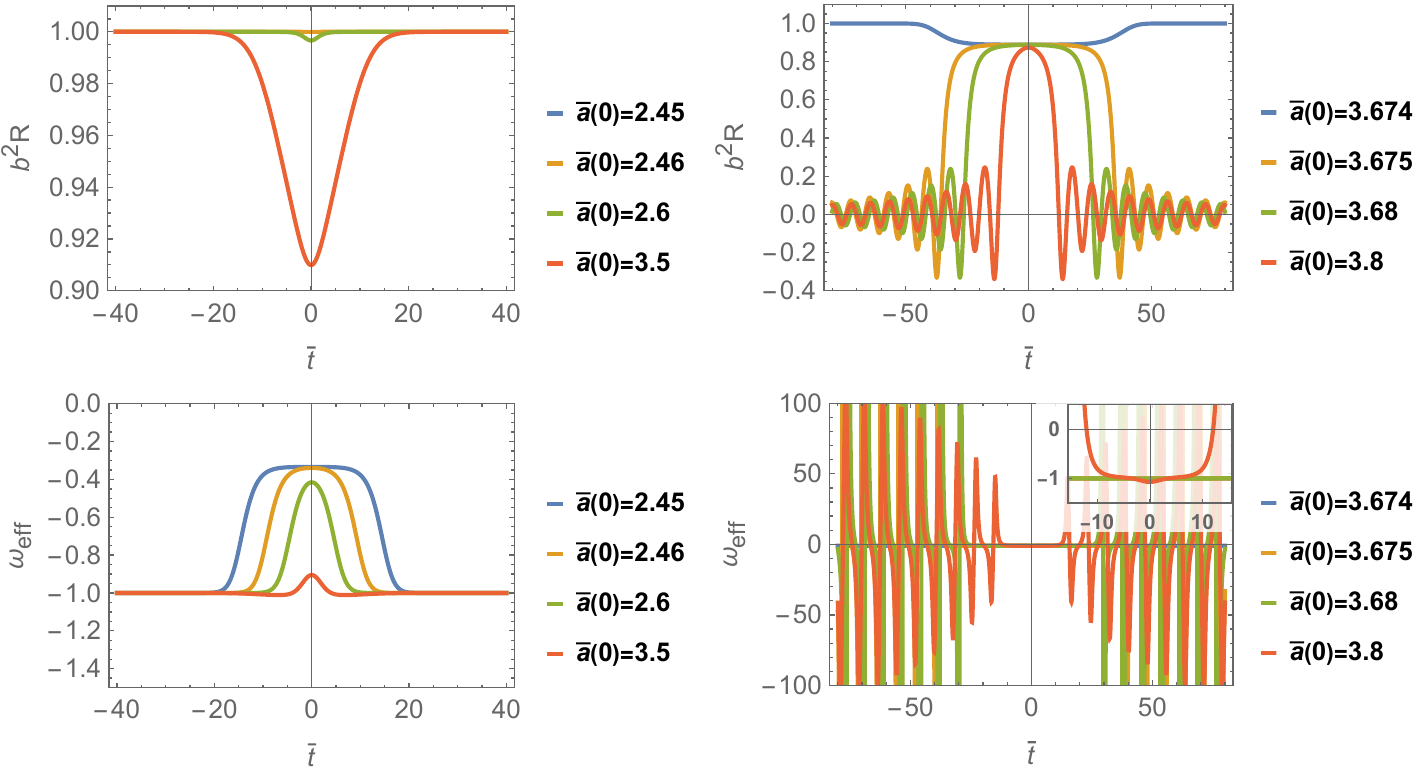}
 \centering
\caption{Evolution of the scalar curvature and effective EOS parameter for the bouncing solutions of Fig. \ref{Fig_R_bounce}.}
\label{Fig_R_bounce_curvature}
\end{figure}

It is also useful to look at the evolution of the effective equation of state (EOS) $\omega_{\rm eff}\equiv p_{\rm eff}/\rho_{\rm eff}$, where effective pressure and density are given by
\begin{align}
    b^2p_{\rm eff} &=\rring{f}_R+2\bar H\mathring{f}_R-\Big(\mathring{\bar H}+3\bar H^2+2\frac{K}{\bar a^2}\Big)f_R-2\mathring{\bar H}-3\bar H^3-\frac{K}{\bar a^2}+1-\frac{1}{f_R}~,\label{p_eff_f(R)}\\
    b^2\rho_{\rm eff} &=3\bar H\mathring{f}_R-3(\mathring{\bar H}+\bar H^2)f_R-3\Big(\bar H^2+\frac{K}{\bar{a}^2}\Big)+1-\frac{1}{f_R}~.\label{rho_eff_f(R)}
\end{align}
The Friedmann equations can then be written as
\begin{align}
    \bar H^2+\frac{K}{\bar a^2} &=\tfrac{1}{3}b^2\rho_{\rm eff}~,\label{Fried1_EOS}\\
    \mathring{\bar H}+2\bar H^2+\frac{K}{\bar a^2} &=\tfrac{1}{6}(1-3\omega_{\rm eff})b^2\rho_{\rm eff}~.\label{Fried2_EOS}
\end{align}
By using \eqref{Fried1_EOS} and \eqref{Fried2_EOS} one can write the effective EOS as
\begin{equation}
    \omega_{\rm eff}=-\frac{2\mathring{\bar H}+3\bar H^2+K/\bar a^2}{3(\bar H^2+K/\bar a^2)}~.
\end{equation}
The evolution of the scalar curvature and effective EOS is shown in Fig. \ref{Fig_R_bounce_curvature}, for the bouncing solutions of Fig. \ref{Fig_R_bounce}. 

In the case of runaway bouncing solutions (the top-left plot of Fig. \ref{Fig_R_bounce_curvature} and the blue curve of the top-right plot), scalar curvature is always positive and approaches its upper bound ($R\rightarrow 1/b^2$) in distant past and future, while reaching its lowest value at the bounce point. By contrast, for the oscillatory bouncing solutions (top-right plot) scalar curvature reaches its maximum at the bounce, and undergoes damped oscillations around $R=0$ in asymptotic past and future. The bottom row of Fig. \ref{Fig_R_bounce_curvature} shows the effective EOS parameter. According to the bottom-left plot, runaway solutions mimic dark energy/cosmological constant, $\omega_{\rm eff}\simeq -1$, everywhere except near the bounce point, where $-1<\omega_{\rm eff}<0$ (the exception is the case with $\bar a(0)=3.674$, where $\omega_{\rm eff}\simeq -1$ at all times). For the oscillatory solutions, $\omega_{\rm eff}\simeq -1$ around the bounce point, while it enters an oscillatory regime away from the bounce.

Let us go back to the asymptotic behavior (at $\bar t\rightarrow \pm\infty$) of the runaway solutions. In the limits $\bar t\rightarrow \pm\infty$, their scalar curvature approaches a constant $R\rightarrow 1/b^2$. Consequently, since $f_R=(1-b^2R)^{-1/2}$, we get $f_R\rightarrow\infty$, which in turn implies that Eq. \eqref{EFE_00_fR_no_b} can be approximated by $\mathring f_R\simeq 0$. This is consistent with the constant $R$ limit. Therefore, it suffices to solve $b^2R=6(\mathring{\bar H}+2{\bar H}^2+1/{\bar a}^2)\simeq 1$. This equation can be further simplified as the term $1/{\bar a}^2$ can be ignored (we have $\bar a\gg 1$ at $\bar t\rightarrow\pm\infty$), and the end result is a simple equation $\mathring{\bar H}+2{\bar H}^2\simeq 1/6$, which can be solved as
\begin{equation}\label{H_tanh}
    \bar H\simeq\frac{1}{\sqrt{12}}\tanh\Big({\frac{\bar t}{\sqrt{3}}\Big)}~,
\end{equation}
by setting the integration constant to zero. Asymptotically, $\bar H$ approaches $\pm 1/\sqrt{12}\approx\pm 0.29$, in agreement with the bottom left plot of Fig. \ref{Fig_R_bounce}.

\subsubsection*{Odd-bounce solutions}

\begin{figure}
 \includegraphics[width=1\textwidth]{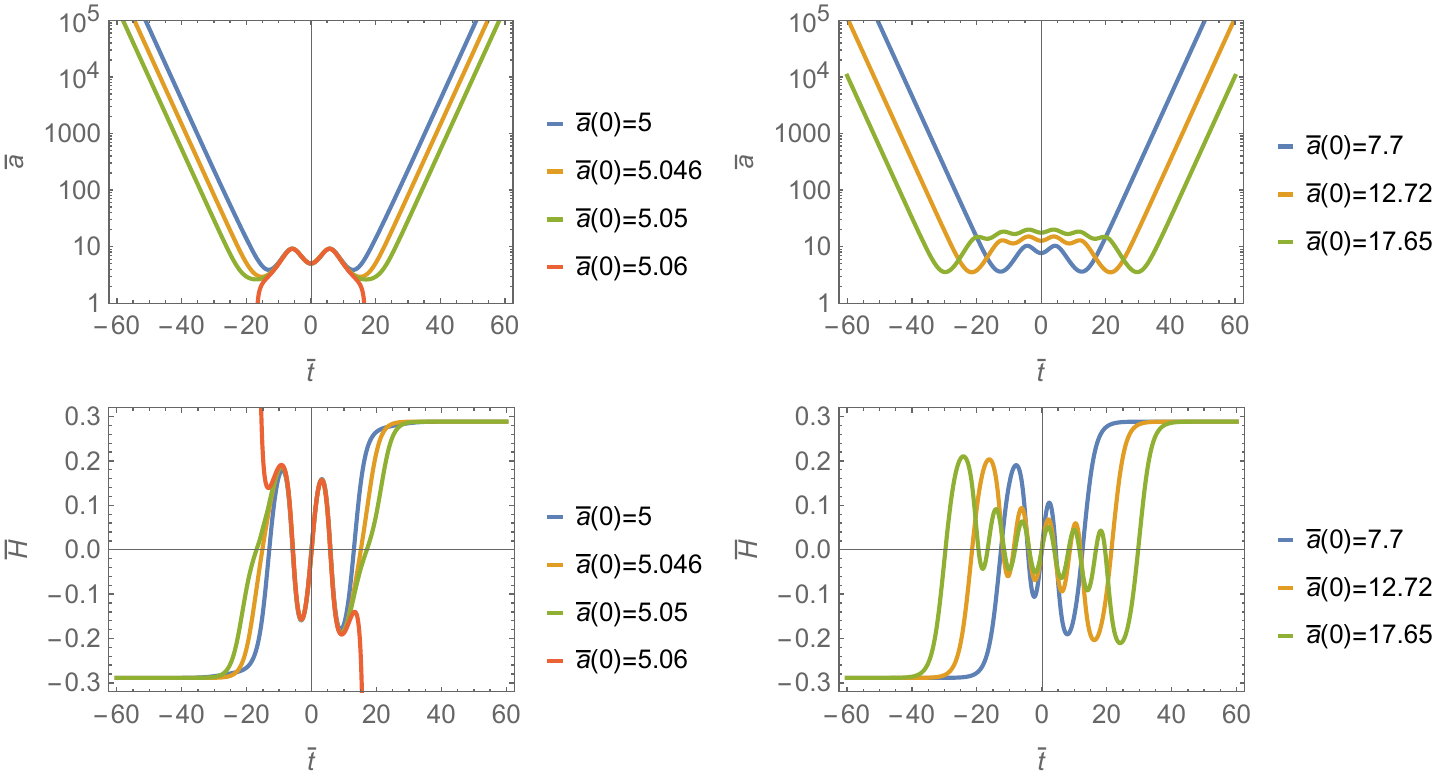}
 \centering
\caption{Runaway solutions with odd number of bounces. Left column shows triple-bounce solutions with $\bar a(0)$ close to $5$, as well as an example (red curve) of a singular solution. Right column shows examples with three, five, and seven bounces.}
\label{Fig_R_odd_bounce}
\end{figure}

In addition to the solutions described above, we find a generalization of runaway solutions (asymptotically described by Eq. \eqref{H_tanh}) which undergo an odd number of successive bounces centered around $\bar t=0$. Examples of such solutions are shown in Fig. \ref{Fig_R_odd_bounce}. Our numerical results show that the odd-bounce solutions with runaway behavior are confined to several ``islands", such as $5\lesssim\bar a(0)\lesssim 5.05$ and $7.63\lesssim\bar a(0)\lesssim 7.75$ for triple-bounce solutions, $12.7\lesssim\bar a(0)\lesssim 12.74$ for five-bounce solutions, $17.64\lesssim\bar a(0)\lesssim 17.66$ for seven-bounce solutions, etc. Between these islands, the solutions develop singularities away from $\bar t=0$. For example, seen from the top-left and bottom-left plots of Fig. \ref{Fig_R_odd_bounce}, for $\bar a(0)=5.06$, singularities (i.e., a vanishing scale factor) develop around $\bar t=\pm 20$.

The appearance of an odd number of bounces can partly be explained by the presence of time reflection symmetry of the system, as well as our choice of the initial conditions that implies the presence of at least one bounce at $\bar t=0$. Any additional bounces will be reflected in time (as long as the scale factor is even at the origin, so that odd derivatives vanish at this point), which leads to an odd total number of possible bounces.

\subsubsection*{Even-bounce solutions}

\begin{figure}
 \includegraphics[width=1\textwidth]{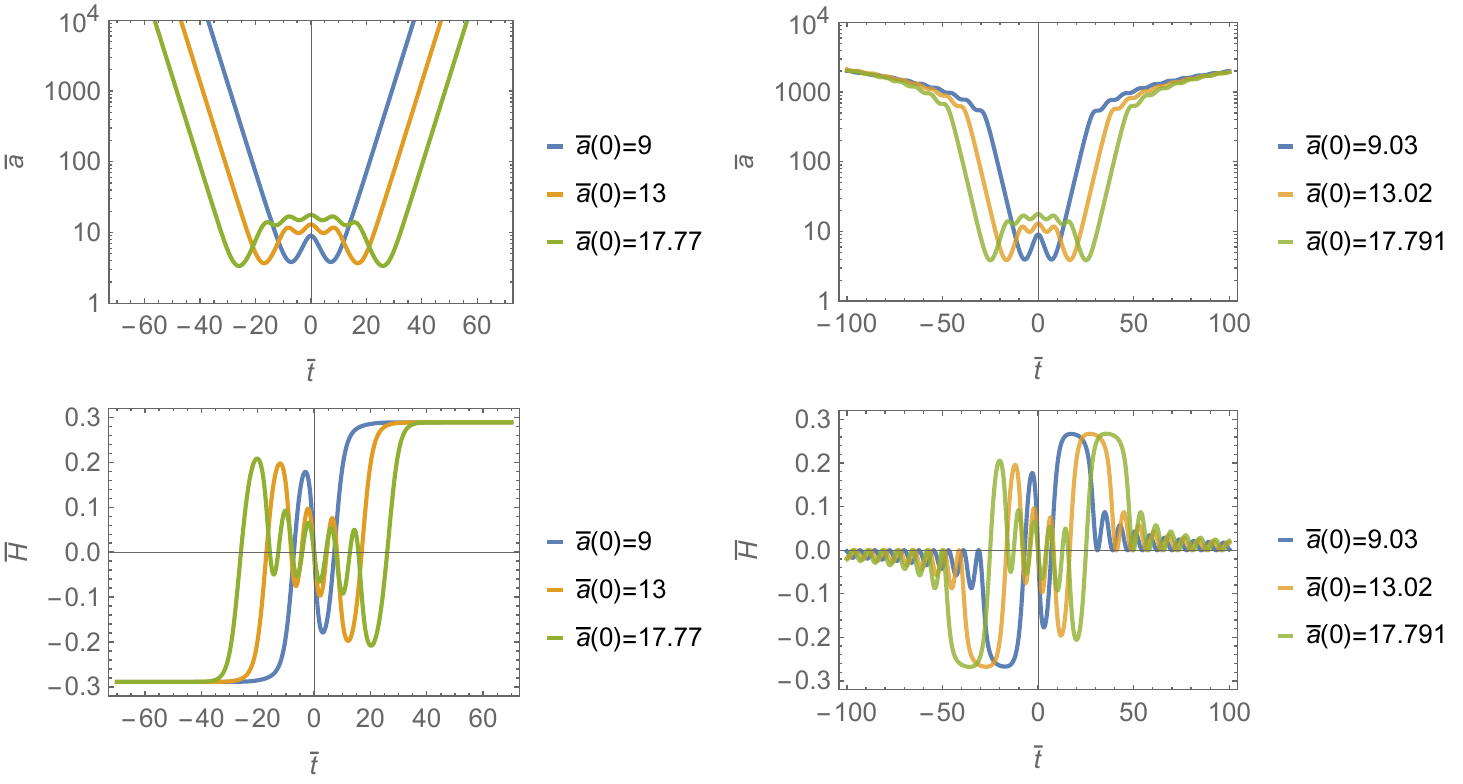}
 \centering
\caption{Solutions with even number of bounces. Left column shows runaway even-bounce solutions, right column shows oscillatory even-bounce solutions.}
\label{Fig_R_even_bounce}
\end{figure}

Having shown the existence of multi-bounce solutions with an odd number of bounces, it is natural to ask if similar solutions with an even number of bounces exist as well. The answer is yes, and to search for such solutions we need to modify the initial conditions that we used, as our previous choice of the initial acceleration, $\rring{\bar a}(0)>0$, implies at least one bounce at the origin. For an even number of bounces, and assuming time-reflection symmetry $t\rightarrow -t$, we should instead impose $\rring{\bar a}(0)<0$, such that $\bar t=0$ corresponds to an ``anti-bounce" point, i.e. an expansion-to-contraction bounce, where the scale factor is at its local maximum.~\footnote{Solutions with a single anti-bounce are not physically relevant, because they describe an expansion-to-contraction bounce. Therefore we are only interested in solutions where the anti-bounce at $t=0$ is sandwiched between an even number of (contraction-to-expansion) bounces. Such solutions lead to an expansion in asymptotic future, and contraction in asymptotic past.} Such a condition does not guarantee the existence of bounces, thus we simply scan the space of possible initial values of $\bar a$. As before, the initial acceleration $\rring{\bar a}(0)$ can be found from the $00$ Einstein equation. For an anti-bounce at the origin, we simply choose the $``-"$ branch in Eq. \eqref{H_root_0}, such that
\begin{equation}\label{anti_boun_IC}
    \rring{\bar a}(0)=-\frac{2}{\sqrt{6}}-\frac{2}{\bar a(0)}~.    
\end{equation}
By scanning the parameter space of $\bar a(0)$, we find several even-bounce runaway solutions: For $8.87\lesssim\bar a(0)\lesssim 9.02$, we find two-bounce solutions, for $\bar a\approx 9.38$ and $12.96\lesssim\bar a(0)\lesssim 13.01$, we find four-bounce solutions, for $17.76\lesssim\bar a(0)\lesssim 17.79$, six-bounce solutions. Oscillatory even-bounce solutions also exist if $\bar a(0)$ is chosen close to, but outside the aforementioned numerical ranges. We present the examples of even-bounce solution in Fig. \ref{Fig_R_even_bounce}.

According to our numerical results, oscillatory solutions (both single-bounce and multi-bounce) can in general suffer from future/past singularities in Minkowski vacuum. Below we discuss such singularities and their resolution in the presence of a cosmological constant.

\subsubsection*{Singularities of the oscillatory solutions (and how to resolve them)}

\begin{figure}
 \includegraphics[width=.9\textwidth]{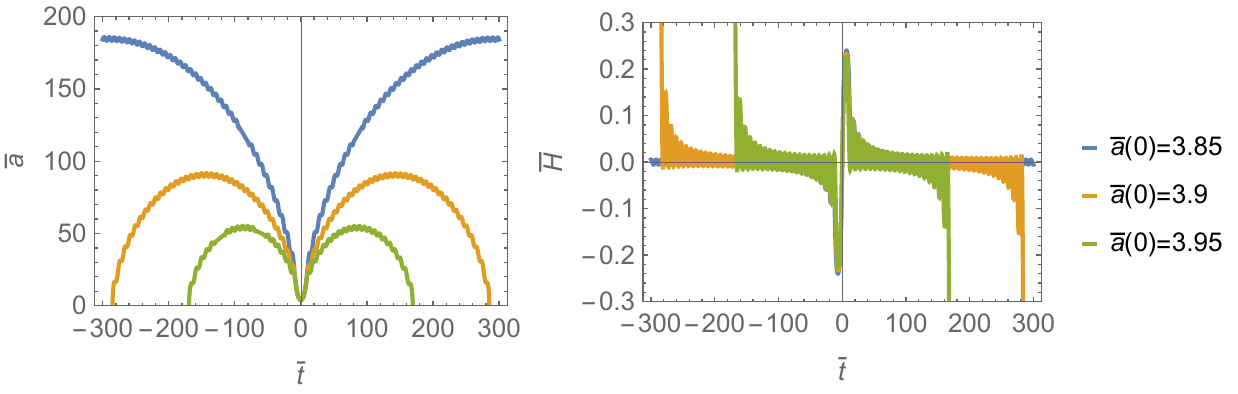}
 \centering
\caption{Oscillatory bouncing solutions with future/past singularities.}
\label{Fig_R_bounce_singular}
\end{figure}

In the oscillatory bouncing solutions, singularities can develop away from the bounce point, as demonstrated in Fig. \ref{Fig_R_bounce_singular}, where the initial value of $\bar a$ is close to $4$. The left-side plot of Fig. \ref{Fig_R_bounce_singular} shows the scale factor evolution, where $\bar a$ vanishes at large negative $\bar t$ (i.e., the Big Bang singularity appears before the bounce) and large positive $\bar t$ (the universe eventually collapses some time after the bounce). By gradually decreasing $\bar a(0)$ from $\bar a(0)=4$, one can move the singularities arbitrarily farther away in time (from the bounce point), but their presence seems to be a generic feature of the oscillatory solutions in Minkowski vacuum. Once a large enough positive cosmological constant is added, however, it can remove the singularities by preventing the contraction phase after the bounce (and the expansion phase before the bounce). This is analogous to what happens in the standard $\Lambda$CDM cosmology with positive spatial curvature: without the cosmological constant, the Universe would eventually enter a contracting phase culminating in the collapse (Big Crunch). A positive cosmological constant counteracts the contraction. In the next subsection (where we analyze the solutions in the scalar-tensor formulation of the theory), it will be clearer why the collapse happens in the oscillatory solutions but not in the runaway solutions.

Let us show how a positive cosmological constant affects the oscillatory solutions. Although we consider a true cosmological constant, it can also represent some slowly-changing scalar potential, such as an inflationary one. The presence of a cosmological constant requires a modification of the expression \eqref{a_IC_0} for the initial acceleration $\rring{\bar a}(0)$ as follows. A cosmological constant $\Lambda$ can be included in the $f(R)$ function as
\begin{equation}\label{f(R)_BI_Lambda}
f=\frac{2}{b^2}\Big(1-\Lambda-\sqrt{1-b^2R}\Big)~.
\end{equation}
After nondimensionalization via Eq. \eqref{nondim},  $00$ Einstein equation yields
\begin{equation}\label{EFE_00_fR_no_b_Lambda}
    \bar H\mathring f_R - (\mathring{\bar H} + \bar H^2)f_R + \tfrac{1}{3}(1-\bar\Lambda-f_R^{-1}) = 0~,
\end{equation}
where $\bar\Lambda\equiv b^2\Lambda$. Repeating the calculation that leads to Eq. \eqref{H_root_0}, we get its $\Lambda$-extension (choosing the $``+"$ branch),
\begin{equation}\label{Hdot_IC_Lambda}
    \mathring{\bar H}(0)=\lambda-\frac{2}{\bar a^2(0)}+(1-\bar\Lambda)\sqrt{\frac{2}{3\bar a^2(0)}-\frac{\lambda}{3}}~,
\end{equation}
where $\lambda\equiv\frac{1}{3}\bar\Lambda(2-\bar\Lambda)$, and we assumed $0\leq\bar\Lambda<1$ and $K=+1$. Equation \eqref{Hdot_IC_Lambda} fixes the $\rring{\bar a}$ initial condition in terms of $\bar a(0)$ and $\bar\Lambda$. For $\bar\Lambda=0$, Eq. \eqref{H_root_0} is recovered.

\begin{figure}
 \includegraphics[width=.9\textwidth]{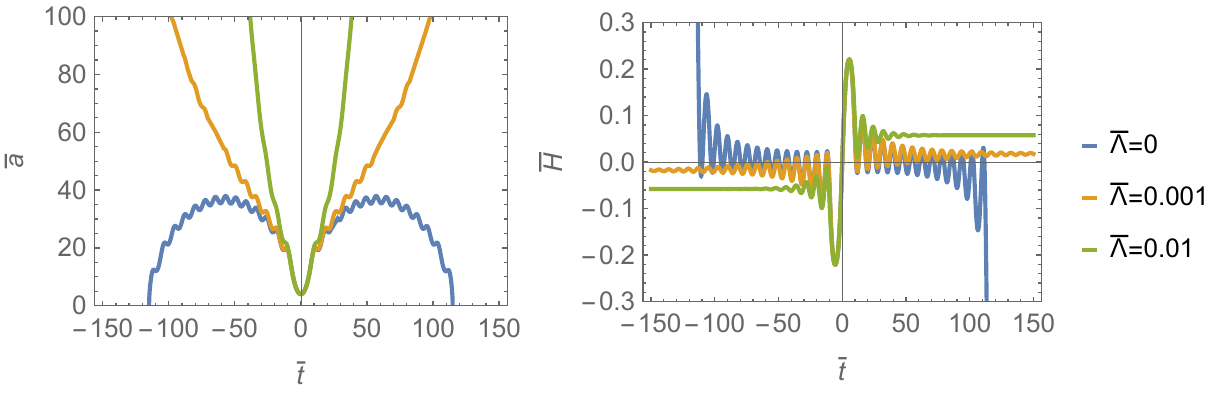}
 \centering
\caption{Effects of a (positive) cosmological constant on an oscillatory solution (with $\bar a(0)=4$) with singularities. The singularities are removed for large enough positive $\bar\Lambda$.}
\label{Fig_R_bounce_CC}
\end{figure}

Figure \ref{Fig_R_bounce_CC} shows the effects of a positive cosmological constant on a singular oscillatory solution: A large enough cosmological constant prevents singularities from forming by introducing a constant contraction phase before the bounce, and a transition to a de Sitter expansion phase after the bounce. Generalization to multi-bounce oscillatory solutions is straightforward.

\subsubsection*{Asymmetric solutions}

So far we considered only symmetric (w.r.t. the time reflection $t\rightarrow -t$) solutions, by providing symmetric initial conditions with vanishing odd derivatives of the scale factor at $\bar t=0$. Here, we will show how asymmetric bouncing solutions can be obtained. The idea is to deform known symmetric bouncing solutions by including asymmetric perturbations to the initial conditions. First, we take one of the symmetric bouncing solutions that were found earlier (these are characterized solely by a choice of $\bar a(0)$), then we add a small $\mathring{\bar a}(0)$. By gradually increasing $|\mathring{\bar a}(0)|$, one can smoothly deform the symmetric solution in a controllable way, in order to avoid the appearance of singularities. This will also somewhat shift the bounce point away from the origin. Each symmetric bouncing solution can be deformed in this way.

\begin{figure}
 \includegraphics[width=.9\textwidth]{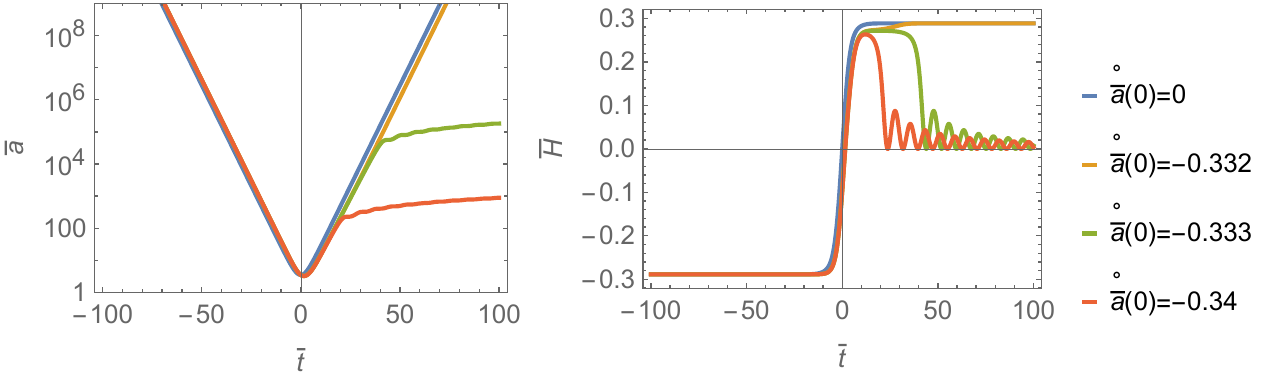}
 \centering
\caption{Bouncing solutions with $\bar a(0)=3.5$ and different choices of $\mathring{\bar a}(0)$. Certain values of $\mathring{\bar a}(0)$ admit simultaneous presence of runaway and oscillatory behavior -- one pre-bounce, and the other post-bounce or vice-versa.}
\label{Fig_R_bounce_asymmetric}
\end{figure}

Among the asymmetric solutions, there exist solutions that interpolate between the runaway evolution at negative (positive) $\bar t$ and the oscillatory evolution at positive (negative) $\bar t$. Examples of such solutions are shown in Fig. \ref{Fig_R_bounce_asymmetric}, where for a given $\bar a(0)$ we vary $\mathring{\bar a}(0)$. At $\mathring{\bar a}(0)=0$, the solution is runaway, but for a large enough negative $\mathring{\bar a}(0)$, its post-bounce evolution changes to an oscillatory one. By changing the sign of $\mathring{\bar a}(0)$, one can flip the solutions in time.

\subsection{Bouncing solutions in the presence of matter}

\begin{figure}
 \includegraphics[width=1\textwidth]{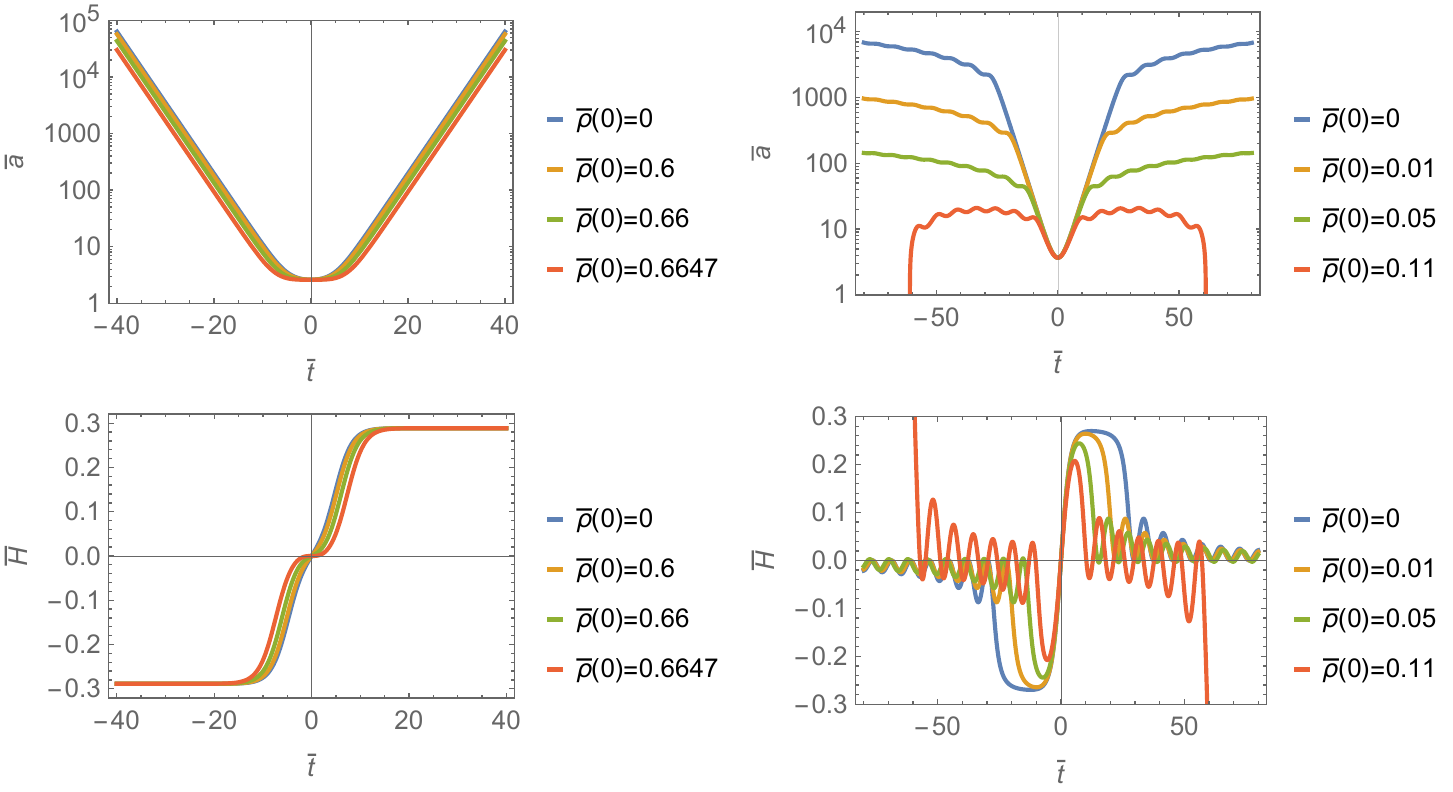}
 \centering
\caption{Simplest bouncing solutions in the presence of matter with $\omega=0$ (dust). Left column: runaway solution with $\bar a(0)=2.6$. Right column: oscillatory solution with $\bar a(0)=3.68$.}
\label{Fig_R_bounce_matter}
\end{figure}

Here, we study the impact of (minimally coupled) perfect fluid matter on the bouncing solutions. In the presence of matter, Eq. \eqref{EFE_00_fR_no_b} becomes
\begin{equation}\label{EFE_00_fR_matter}
    \bar H\mathring f_R - (\mathring{\bar H} + \bar H^2)f_R + \tfrac{1}{3}(1-f_R^{-1}) = \tfrac{1}{3}\bar\rho~,
\end{equation}
where the dimensionless matter energy density $\bar\rho\equiv b^2\rho$ evolves as
\begin{equation}
    \bar\rho(\bar t)=\bar\rho(0)\Big(\frac{\bar a(\bar t)}{\bar a(0)}\Big)^{-3(1+\omega)}~,
\end{equation}
where $\bar\rho(0)$ is the initial density at the bounce, and $\omega$ is the matter equation of state. Since the parameter $b$ has inverse mass dimension, we can write $b\equiv 1/M_{\rm BI}$, with the BI mass parameter $M_{\rm BI}$. It is reasonable to assume that the initial matter density is smaller than the BI scale (for example $M_{\rm BI}$ can be close to inflationary scale or higher), $|\rho(0)|\ll M^2_{\rm BI}$, or after restoring the Planck mass, $|\rho(0)|\ll M^2_{\rm BI}M_{\rm P}^2$. This implies that $|\bar\rho(0)|\ll 1$. This assumption also ensures that the presence of such matter does not invalidate the bouncing solutions. Nontheless, for the bounce initial conditions $\bar a(0)$ and $\rring{\bar a}(0)$ (here we consider the simplest symmetric bouncing solutions), we must take into account the presence of a new number $\bar\rho(0)$. From \eqref{EFE_00_fR_matter} at $\bar t=0$, we get
\begin{equation}
    \rring{\bar a}(0)=\frac{\bar a(0)}{3}\big[-S+\sqrt{S}\big(1-\bar\rho(0)\big)\big]~,~~~S\equiv \frac{6K}{\bar a^2(0)}-2\bar\rho(0)+\bar\rho^2(0)~.
\end{equation}
The initial scale factor is bounded as $\bar a(0)>\sqrt{6}$, as before. We can now solve \eqref{EFE_00_fR_matter} after fixing the EOS parameter. In Fig. \ref{Fig_R_bounce_matter}, we show the runaway and oscillatory bouncing solutions, and how they are affected by the presence of dust matter ($\omega=0$) as an example. It can be seen that for the runaway solution, the dust matter has a slight flattening effect on the scale factor and Hubble function at the bounce, but a noticeable change in the solution requires a relatively large initial density $\bar\rho(0)\sim 0.664$. For the oscillatory solution, the presence of dust matter shortens the post-bounce expansion phase and introduces singularities (or, rather, brings them closer to $\bar t=0$). This can be counteracted by a large enough cosmological constant. We do not include the plots of negative $\bar\rho(0)$, because in that case the effects of matter are negligible, even for $|\bar\rho(0)|=\co(1)$. For radiation, $\omega=1/3$, the picture is qualitatively the same as what is shown in Fig. \ref{Fig_R_bounce_matter}, while the cosmological constant case ($\omega=-1$) was already considered in Sec. \ref{FR_type}.

In summary, a reasonable initial distribution of matter (at the bounce) does not significantly, or qualitatively, affect the bouncing solutions. Here, ``a reasonable initial distribution" means that the initial energy density $\rho(0)$ is much smaller than the BI density $M^2_{\rm BI}M^2_{\rm P}$.

\subsection{$f(R)$ frame versus Einstein frame}\label{subsec_frames}

It is useful to analyze the bouncing solutions of our $f(R)$ theory \eqref{L_sBI_gravity} in the equivalent scalar-tensor formulation in order to better understand the properties of these solutions. When going to the Einstein frame, the modified gravity effects are encoded solely in the scalar potential of the scalaron field, which allows us to interpret these effects in terms of the potential's shape. It should be noted that although the $f(R)$ frame and the Einstein frame are equivalent, the observables, such as the scale factor and Hubble function, do not coincide -- they are related by a Weyl/conformal transformation.

The Einstein frame Lagrangian corresponding to the $f(R)$ model \eqref{L_sBI_gravity} is
\begin{align}
    \sqrt{-g}^{\,-1}\cl &=\tfrac{1}{2}R-\tfrac{1}{2}\partial\varphi\partial\varphi-V~,\label{varphi_L}\\
    V &=\frac{1}{2b^2}e^{-\sqrt{\frac{2}{3}}\varphi}\big(1-e^{-\sqrt{\frac{2}{3}}\varphi}\big)^2~.\label{varphi_V}
\end{align}
The potential \eqref{varphi_V} has a stable Minkowski minimum at $\varphi=0$ and a runaway minimum at $\varphi\rightarrow\infty$, separated by a local maximum at $e^{-\sqrt{2/3}\varphi}=1/3$. It should also be noted that, as $\varphi$ is a canonical scalar with the correct sign of the kinetic term, null energy condition ($p+\rho\geq 0$) is not violated.

As before, it is convenient to remove the $b$-dependence of the FLRW equations by the nondimensionalization \eqref{nondim}. This leads to the EOM
\begin{align}
    \rring{\varphi}+3\bar H\mathring{\varphi}+\bar V_{,\varphi} &=0~,\label{varphi_FLRW_KG}\\
    3\Big(\bar H^2+\frac{K}{\bar a^2}\Big)-\tfrac{1}{2}\mathring{\varphi}^2-\bar V &=0~,\label{varphi_FLRW_Fried1}\\
    \mathring{\bar H}-\frac{K}{\bar a^2}+\tfrac{1}{2}\mathring{\varphi}^2 &=0~,\label{varphi_FLRW_Fried2}
\end{align}
where $\bar V\equiv b^2V$ and $K=+1$.

In order to reproduce the bouncing solutions that we found in the $f(R)$ formulation, we need to find the exact transformation of the scale factor between the two frames. In this subsection, we will denote the $f(R)$ frame quantities by subscript ``$R$", for example, the $f(R)$ frame scale factor (nondimensionalized) is $\bar a_R$. The transformation to the Einstein frame involves the scalaron field as
\begin{equation}
    (g_R)_{\mu\nu}=g_{\mu\nu}y~,~~~y\equiv e^{-\sqrt{\frac{2}{3}}\varphi}~,\label{Weyl_varphi_frame}
\end{equation}
where $g_R$ is the $f(R)$ frame metric and the shorthand $y$ is introduced for convenience. Since the time components $(g_R)_{00}$ and $g_{00}$ are both set to $-1$ in the respective FLRW backgrounds, the $00$ part of the transformation \eqref{Weyl_varphi_frame} can instead be understood as the transformation between the time coordinates $t_R$ and $t$:
\begin{equation}\label{t_R_t}
    \frac{dt_R(t)}{dt}=\sqrt{y(t)}~.
\end{equation}
The scale factor transformation can be obtained from \eqref{Weyl_varphi_frame} as
\begin{equation}\label{a_R_a}
    a_R=a\sqrt{y}~.
\end{equation}
Equations \eqref{t_R_t} and \eqref{a_R_a} rescale the (FLRW) line element, $ds^2_R=yds^2$, so they are equivalent to \eqref{Weyl_varphi_frame}. Knowing the transformation rules, we can relate the $f(R)$ frame Hubble function $H_R=a_R^{-1}da_R/dt_R$ to the Einstein frame Hubble function $H=a^{-1}da/dt$ as
\begin{equation}\label{H_R_H}
    H_R=\frac{1}{\sqrt{y}}\big(H-\frac{1}{\sqrt{6}}\dot{\varphi}\big)~,
\end{equation}
where $\dot\varphi\equiv d\varphi/dt$. After nondimensionalization, $\{a,a_R,t,t_R,H,H_R\}$ are simply replaced by their ``barred" versions, while $\varphi$ is unaffected. Below we will work with the nondimensionalized quantities.

In the $f(R)$ frame, the third-order equation for the scale factor requires three initial conditions $\{\bar a(0),\mathring{\bar a}(0),\rring{\bar a}(0)\}$ (subscript $R$ implied). In the Einstein frame, the system of \eqref{varphi_FLRW_KG}, \eqref{varphi_FLRW_Fried1}, and \eqref{varphi_FLRW_Fried2} requires at least three initial conditions $\{\bar a(0),\varphi(0),\mathring{\varphi}(0)\}$. One of the bounce conditions in the $f(R)$ frame is $d{\bar a}_R/dt_R=0$ or $\bar H_R=0$. Equation \eqref{H_R_H} then implies that either $\bar H=\mathring\varphi=0$ or $\bar H=\mathring{\varphi}/\sqrt{6}\neq 0$ (hereafter, $\mathring{\varphi}\equiv d\varphi/d\bar t$). The latter implies that the bounce point in the $f(R)$ frame does not translate into a bounce point in the Einstein frame. The initial velocity $\mathring{\varphi}(0)$ can be shown to be proportional to $\bar a_R^{(3)}(0)$ in the $f(R)$ frame (see Appendix \ref{App_B}), so assuming $\bar a_R^{(3)}(0)=0$, we can set $\bar H(0)=\mathring{\varphi}(0)=0$~\footnote{When classical velocity of the scalaron vanishes at the bounce, quantum diffusion can become important \cite{starobinsky1986stochastic,Salopek:1990nonlinear,Vennin:2015correlation,Aldabergenov:2024fws,Aldabergenov:2025ulq}. This deserves a separate investigation once the generation of realistic perturbations in this model is addressed.}. From Eq. \eqref{varphi_FLRW_Fried1} we then obtain
\begin{equation}\label{a0_y0}
    \bar a^2(0)=\frac{3}{\bar V(\varphi(0))}=\frac{6}{y(0)[1-y(0)]^2}~,
\end{equation}
which relates $\bar a$ and $y\equiv e^{-\sqrt{2/3}\varphi}$ initial conditions. Another requirement for the $f(R)$ frame bounce is
\begin{equation}
    \frac{d\bar H_R(0)}{d\bar t_R}=\frac{1}{y(0)}\Big[\mathring{\bar H}(0)-\frac{1}{\sqrt{6}}\rring\varphi(0)\Big]>0~,
\end{equation}
where we used $\bar H(0)=\mathring{\varphi}(0)=0$. Since $y$ is by definition positive, we have (assuming $\mathring{\bar H}(0)$ is positive)
\begin{equation}\label{dH_ddvarphi}
    \mathring{\bar H}(0)>\frac{1}{\sqrt{6}}\rring\varphi(0)~.
\end{equation}
By combining \eqref{dH_ddvarphi} and the EOM \eqref{varphi_FLRW_KG}, \eqref{varphi_FLRW_Fried1}, \eqref{varphi_FLRW_Fried2} at the bounce point, we can obtain the following constraint on $y(0)$,
\begin{equation}\label{y0_range}
    y\frac{V_{,y}}{V}\Big|_{t=0}=\frac{1-3y(0)}{1-y(0)}<1~~\Rightarrow~~0<y(0)<1~.
\end{equation}

Going back to Eq. \eqref{a0_y0}, one notices that the condition \eqref{y0_range} implies a lower bound on $\bar a(0)$, which can be obtained by minimizing \eqref{a0_y0} with respect to $y(0)$. The critical value of $y(0)$ that minimizes the scale factor is $y_{\rm cr}=1/3$, and the minimized scale factor is $\bar a_{\rm cr}=9/\sqrt{2}$. Notably, the value $y_{\rm cr}=1/3$ is also the local maximum of the potential \eqref{varphi_V}. Next, by using $\bar a_R=\bar a\sqrt{y}$, one can find that the corresponding critical value of the $f(R)$-frame scale factor is $\bar a_{R,{\rm cr}}=\frac{3}{2}\sqrt{6}\approx 3.6742$. This is precisely the value that separates the oscillatory and the runaway bouncing solutions described in the previous subsection.

\begin{figure}
 \includegraphics[width=1\textwidth]{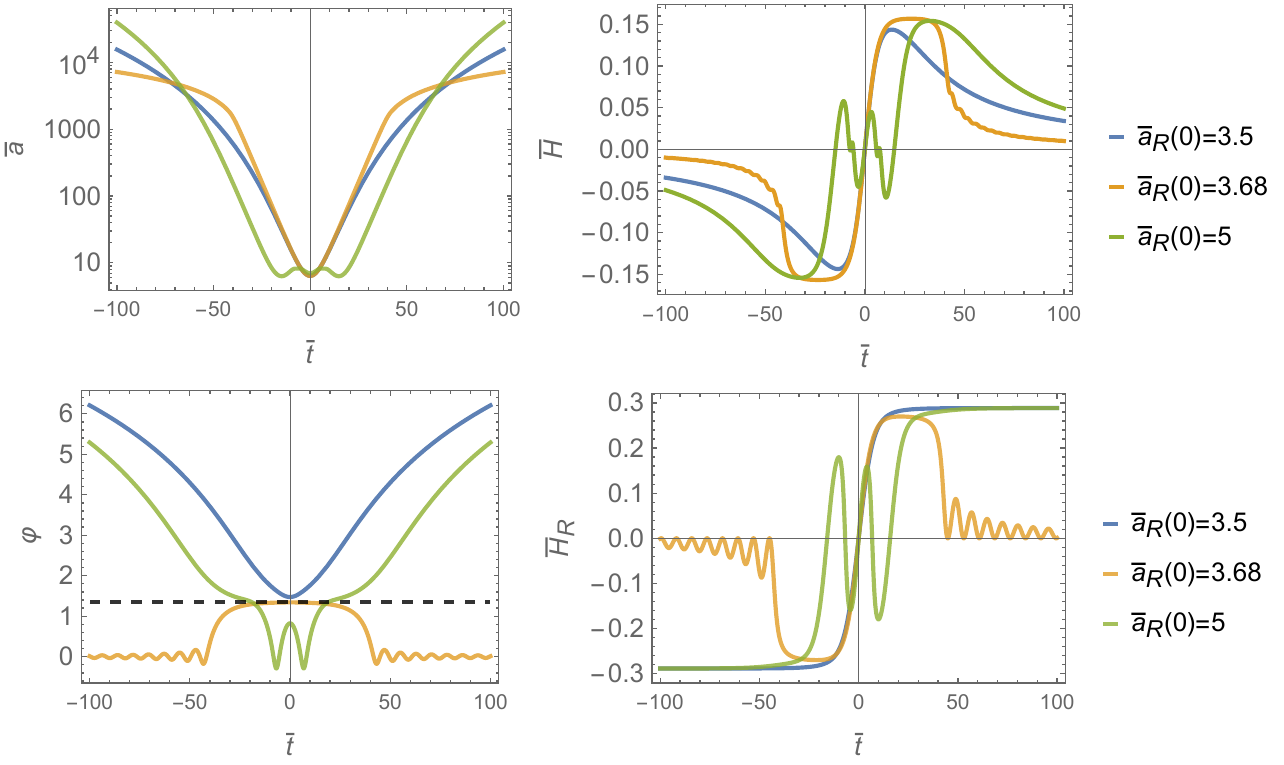}
\caption{Examples of different types of bouncing solutions (runaway, oscillatory, and runaway with triple-bounce) in the Einstein frame. Dashed black line represents the scalaron critical value $\varphi_{\rm cr}=\sqrt{\frac{3}{2}}\log{3}$ ($y_{\rm cr}=1/3$). Hubble function $\bar H_R$ of the $f(R)$ frame is shown in the bottom-right plot.}
\label{Fig_varphi_bounce}
\end{figure}

We can now reproduce these solutions in the Einstein frame by using $\bar a_R=\bar a\sqrt{y}$ and Eq. \eqref{a0_y0} to fix the initial conditions in this frame from a given $\bar a_R(0)$. In Fig. \ref{Fig_varphi_bounce}, we show examples of three types of solutions: runaway (with $\bar a_R(0)=3.5$), oscillatory (with $\bar a_R(0)=3.68$), and runaway with three bounces (with $\bar a_R(0)=5$). The plots show the evolution of the Einstein-frame scale factor (top-left), Hubble function (top-right), and the scalaron (bottom-left). Bottom-right plot shows the $f(R)$-frame Hubble function (Eq. \eqref{H_R_H}) for comparison. One can also change the time variable to $\bar t_R$ and confirm that $\bar H_R(\bar t_R)$ is in one-to-one correspondence with the results of the previous subsection -- see Figs. \ref{Fig_R_bounce} and \ref{Fig_R_odd_bounce}.

\begin{figure}
\centering
 \includegraphics[width=.5\textwidth]{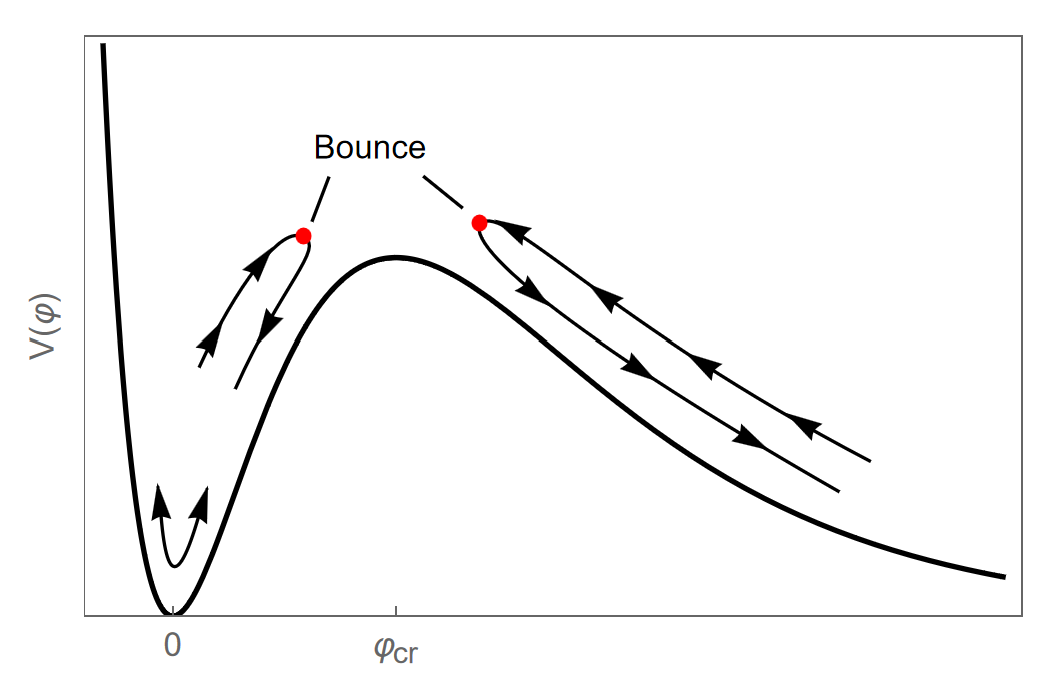}
\caption{Scalaron potential and a schematic representation of two types of single-bounce solutions: oscillatory solution (to the left from the local maximum) and runaway solution (to the right). The bounce occurs close to the local maximum $\varphi_{\rm cr}$.}
\label{Fig_varphi_scheme}
\end{figure}

The different types of bouncing solutions can be explained in terms of the scalaron dynamics and its potential as follows. Since the critical value $\varphi_{\rm cr}=\sqrt{\frac{3}{2}}\log{3}$ (or $y_{\rm cr}=1/3$) corresponds to the local maximum of the potential, this means that the runaway and oscillatory single-bounce solutions take place in two different regions of the scalaron potential, divided by the local maximum (see Fig. \ref{Fig_varphi_bounce}, bottom-left plot). Runaway evolution takes place at $\varphi>\varphi_{\rm cr}$, in the region with runaway Minkowski vacuum ($V\rightarrow 0$, as $\varphi\rightarrow\infty$). Oscillatory evolution takes place at $\varphi<\varphi_{\rm cr}$, in the region around the stable Minkowski minimum with $\varphi=0$. This explains the oscillatory behavior of such solutions: The scalaron rolls down from the hilltop (after the bounce) and starts oscillating around the minimum of the potential, similarly to the reheating phase after inflation. Eventually, the oscillatory stage (around the Minkowski vacuum) triggers the contraction phase and the collapse of the universe. This evolution is reflected in time for negative $\bar t$: the universe emerges from a singularity, then enters a contraction phase with growing oscillations of the scalaron around its minimum, and after a period of time it gains enough velocity to climb up the local maximum around which the bounce occurs. In the single-bounce runaway solutions, there are no oscillations because the whole evolution of the scalaron is confined to the runaway region $\varphi>\varphi_{\rm cr}$. A schematic representation of the two types of simplest single-bounce (symmetric) solutions is shown in Fig. \ref{Fig_varphi_scheme} in terms of the scalaron dynamics and its scalar potential. The Minkowski minimum can be seen at $\varphi=0$, where the scalaron oscillates before and after the bounce. Local maximum at $\varphi_{\rm cr}$ divides scalaron domains for the oscillatory and the runaway single-bounce solutions.

In multi-bounce and some asymmetric scenarios, the scalaron can overshoot the local maximum. For instance, in the multi-bounce runaway solutions (such as the $\bar a_R(0)=5$ solution in Fig. \ref{Fig_varphi_bounce}), the scalaron crosses the local maximum while triggering the first bounce, after which ``sub-bounces" can occur in the region $\varphi<\varphi_{\rm cr}$. After the sub-bounces, the scalaron crosses the local maximum for the second time (triggering the last bounce), and rolls back to the runaway minimum.

As mentioned earlier, asymmetric solutions include those which interpolate between the runaway and oscillatory solutions. In the simplest solution of this kind, the scalaron starts in the runaway region $\varphi>\varphi_{\rm cr}$ in the contraction phase, triggers a bounce while crossing the local maximum, and finally ends up oscillating around the Minkowski minimum after the bounce (this example is shown in Fig. \ref{Fig_R_bounce_asymmetric}). The reverse case is also possible, which amounts to flipping the time coordinate. The oscillatory part of such solutions, as always, takes place near the Minkowski minimum, so a positive cosmological constant is needed to prevent the eventual singularity.

\section{Effects of the Gauss--Bonnet term}\label{Sec_GB}

Having studied bouncing solutions in the BI-type $f(R)$ gravity, we can now generalize the results to the presence of the GB term, as given by the Lagrangian
\eqref{L_BIG}. This Lagrangian does not include any new parameters compared to the $f(R)$ case, but for generality, we introduce a new parameter $c$ for the GB term, and write the Lagrangian as
\begin{equation}
    \cl=\tfrac{1}{2}\sqrt{-g}f(L)~,\label{L_f(L)}
\end{equation}
where
\begin{equation}\label{f(L)_def}
    f(L)=\frac{2}{b^2}\Big(1-\sqrt{1-b^2L}\Big)~,~~~L=R+\frac{c\,b^2}{24}\cg~.
\end{equation}
For $c=1$, our BI-type gravity \eqref{L_BIG} is recovered. By considering $c\neq 1$, our model can be generalized if one gives up the idea of obtaining the BI electrodynamics from a 5D gravitational theory. A BI-type modified gravity of such kind (with general $c$) was considered in \cite{Comelli:2005tn,Garcia-Salcedo:2009lui,Kruglov:2012ja}. In this subsection, we will treat $c$ as a free parameter. As to be shown, the $c=1$ case leads to mild deformations of the bouncing solutions obtained earlier in the $f(R)$ model, while for negative $c$, there is a new class of bouncing solutions that are not smoothly connected to the corresponding $f(R)$ solutions (as $c\rightarrow 0$).

Einstein equations for the Lagrangian \eqref{L_f(L)} are
\begin{align}
\begin{aligned}
    \Big(g_{\mu\nu}-\frac{c\,b^2}{6}G_{\mu\nu}\Big)\Box f_L-\Big(1+\frac{c\,b^2}{12}R\Big)\nabla_{\mu\nu}f_L+\Big(R_{\mu\nu}+\frac{c\,b^2}{48}g_{\mu\nu}\cg\Big)f_L-\tfrac{1}{2}g_{\mu\nu}f\\
    +\frac{c\,b^2}{6}\Big({R_\mu}^\lambda\nabla_{\lambda\nu}f_L+{R_\nu}^\lambda\nabla_{\lambda\mu}f_L-g_{\mu\nu}R^{\rho\sigma}\nabla_{\rho\sigma}f_L-R_{\mu\rho\sigma\nu}\nabla^{\rho\sigma}f_L\Big)=0~,
\end{aligned}
\end{align}
where $f_L\equiv \partial f/\partial L$. The nondimensionalized ($\bar a\equiv a/b$ and $\bar t\equiv t/b$) $00$ equation is given by (in an FLRW background)
\begin{equation}\label{GB_background_EoM}
    \Big[1+\frac{c}{6}\Big(\bar H^2+\frac{K}{a^2}\Big)\Big]\Big[\bar H\mathring f_L-(\mathring{\bar H}+\bar H^2)f_L\Big]+\tfrac{1}{6}\bar f=0~,
\end{equation}
where $\bar f\equiv b^2f$, which in terms of $f_L$ can be written as $\bar f=2(1-f_L^{-1})$.

The initial condition $\mathring{\bar H}(0)$ for bouncing solutions (with $\bar H(0)=0$) is constrained by
\begin{equation}\label{H_root_0_GB}
    \mathring{\bar H}(0)=2\Big(-\frac{K}{\bar a^2(0)}\pm\sqrt{\frac{K}{6\bar a^2(0)}}\Big)\Big(1+\frac{c\,K}{6\bar a^2(0)}\Big)^{-1}~,
\end{equation}
where the ``$+$" branch is again restricted to $\bar a(0)>\sqrt{6}$. The GB modification in \eqref{H_root_0_GB} is not dramatic with respect to Eq. \eqref{H_root_0} of the $f(R)$ case, but the ``$-$" branch can now be viable for sufficiently large negative $c$. As in the $f(R)$ case, $K=+1$ is a necessary condition for $\rring{\bar a}(0)>0$, which will be assumed below. 

With a free parameter $c$, we can outline two possibilities to guarantee $\rring{\bar a}(0)>0$: (i) the ``$+$" branch of \eqref{H_root_0_GB} where $\bar a(0)>\sqrt{6}$ and $c>-6\bar a^2(0)$ (can be positive or negative), and (ii) the ``$-$" branch where $\bar a(0)$ is unconstrained and $c<-6\bar a^2(0)$. In the $f(R)$ case ($c=0$), only the ``$+$" branch was viable, but for sufficiently large negative $c$ ($c<-6\bar a^2(0)$), the ``$-$" branch can now produce new bouncing solutions supported by the GB term. 

\begin{figure}
 \includegraphics[width=1\textwidth]{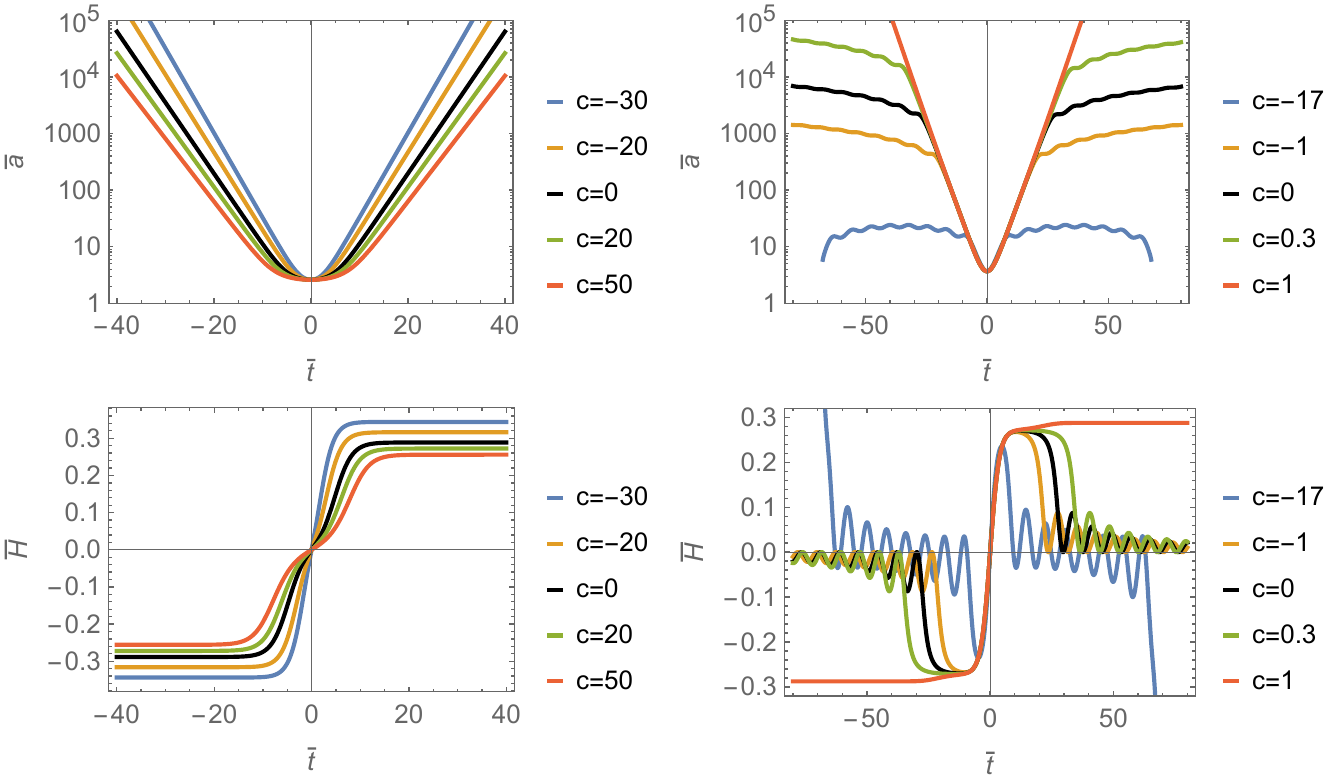}
\caption{Bouncing solutions in the presence of the GB term with $c$ as a free parameter. Runaway solutions with $\bar a(0)=2.6$ are shown in the left column, and oscillatory solutions with $\bar a(0)=3.68$ are in the right column.}
\label{Fig_GB_bounce}
\end{figure}

\begin{figure}
 \includegraphics[width=1\textwidth]{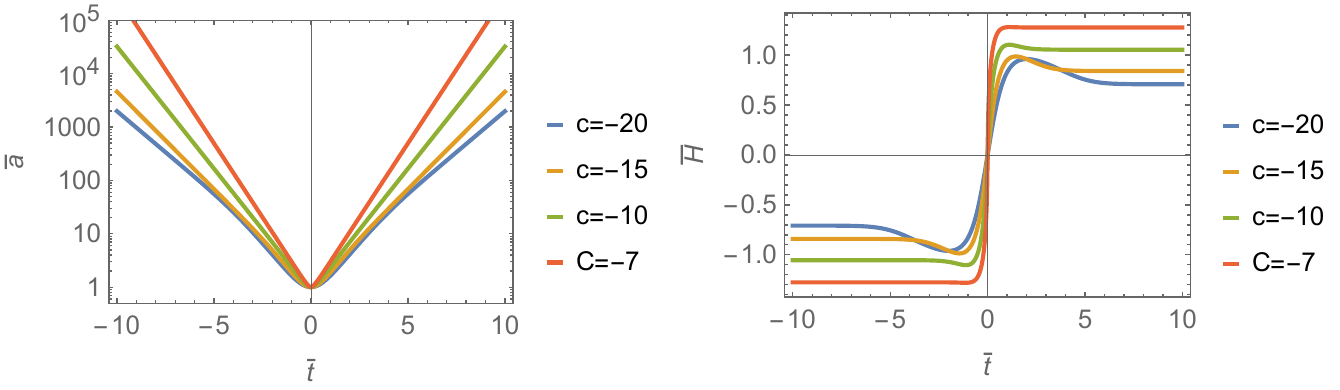}
\caption{New bouncing solutions supported by the GB term. Here $\bar a(0)=1$.}
\label{Fig_GB_bounce_new}
\end{figure}

In Fig. \ref{Fig_GB_bounce}, we plot the ``$+$" branch bouncing solutions which are GB-deformations of the previously discussed $f(R)$ solutions. The effect of non-zero $c$ is very limited for the runaway solutions. In particular, positive $c$ slightly flattens the scale factor at the bounce point. The oscillatory solutions are more sensitive to $c$. For example, as shown in Fig. \ref{Fig_GB_bounce} (right column), sufficiently negative $c$ introduces $\bar a=0$ singularities around $\bar t=\pm 70$ for $\bar a(0)=3.68$. It can also be seen that for positive $c$, such as $c=1$ in this case, the oscillatory solution becomes a runaway solution. This can be explained by the change in the effective potential for the scalaron in the presence of the GB term, as will be shown shortly. The solutions in Fig. \ref{Fig_GB_bounce} can be smoothly deformed into the $f(R)$ solutions by reducing $c$ to zero. As argued above, for $c<-6\bar a^2(0)$, we have new bouncing solutions that do not exist in the $f(R)$ case. In this case, $\bar a(0)$ does not have a lower bound, and as an example, we set $\bar a(0)=1$ and show the solutions for different choices of $c$ in Fig. \ref{Fig_GB_bounce_new}. For $\bar a(0)=1$, such solutions require $c<-6$.

\subsubsection*{In the Einstein frame}

The $f(L)$ Lagrangian \eqref{L_f(L)} with \eqref{f(L)_def} can be Weyl-transformed to the Einstein frame, where we find a Horndeski-type higher-derivative Lagrangian given by \eqref{L_BI_SVT} by setting $A_\mu=0$ (and including the parameter $c$ for the GB term):
\begin{equation}\label{L_BI_scalar}
    \sqrt{-g}^{\,-1}\cl=\tfrac{1}{2}R-\tfrac{1}{2}\partial\varphi\partial\varphi-V(\varphi)-\tfrac{1}{8}\xi(\varphi)\Big[\cg+\tfrac{8}{3}G^{\mu\nu}\partial_\mu\varphi\partial_\nu\varphi-\sqrt{\tfrac{8}{3}}\partial\varphi\partial\varphi\Box\varphi\Big]~,
\end{equation}
where
\begin{equation}\label{V_xi_defs}
    V(\varphi)=\frac{1}{2b^2}e^{-\sqrt{\frac{2}{3}}\varphi}\big(1-e^{-\sqrt{\frac{2}{3}}\varphi}\big)^2~,~~~\xi(\varphi)=-\frac{c\,b^2}{6}e^{\sqrt{\frac{2}{3}}\varphi}~.
\end{equation}
The Lagrangian \eqref{L_BI_scalar} describes the canonical scalaron in the Einstein frame, but unlike the $f(R)$ model, it contains higher derivatives (in the Lagrangian) and a non-minimal GB coupling $\xi(\varphi)$. The corresponding nondimensionalized FLRW equations are
\begin{align}
\begin{aligned}
    \rring{\varphi}+3\bar H\mathring{\varphi}+\bar V_{,\varphi}+\bar\xi_{,\phi}\Big[(\mathring\varphi^2+3\mathring{\bar H}+3\bar H^2)\Big(\bar H^2+\frac{K}{\bar a^2}\Big)-\frac{\mathring\varphi^2}{2\sqrt{6}}(\rring{\varphi}+3\bar H\mathring\varphi)\Big]\\
    -2\bar\xi\Big[\Big(\bar H^2+\frac{K}{\bar a^2}\Big)\rring\varphi+\Big(2\mathring{\bar H}+3\bar H^2+\frac{K}{\bar a^2}\Big)\bar H\mathring\varphi-\sqrt{\tfrac{3}{2}}\bar H\rring\varphi\mathring\varphi-\sqrt{\tfrac{3}{8}}(\mathring{\bar H}+3\bar H^2)\mathring\varphi^2\Big]\\
    -\mathring{\bar\xi}\Big[2\Big(\bar H^2+\frac{K}{\bar a^2}\Big)\mathring\varphi+\tfrac{1}{\sqrt{6}}\rring\varphi\mathring\varphi-\sqrt{\tfrac{3}{8}}\bar H\mathring\varphi^2\Big]-\tfrac{1}{2\sqrt{6}}\rring{\bar\xi}\mathring\varphi^2=0~,\label{GB_varphi_FLRW_KG}
\end{aligned}\\[10pt]
\begin{aligned}
    3(1-\mathring{\bar\xi}\bar H)\Big(\bar H^2+\frac{K}{\bar a^2}\Big)-\tfrac{1}{2}\Big[1-\tfrac{1}{\sqrt{6}}\mathring{\bar\xi}\mathring\varphi-2\bar\xi\Big(3\bar H^2+\frac{K}{\bar a^2}-\sqrt{\tfrac{3}{2}}\bar H\mathring\varphi\Big)\Big]\mathring{\varphi}^2-\bar V =0~,\label{GB_varphi_FLRW_Fried1}
\end{aligned}\\[10pt]
\begin{aligned}
    2(1-\mathring{\bar\xi}\bar H+\tfrac{1}{3}\bar\xi\mathring\varphi^2)\Big(\mathring{\bar H}-\frac{K}{\bar a^2}\Big)-(\rring{\bar\xi}-\mathring{\bar\xi}\bar H)\Big(\bar H^2+\frac{K}{\bar a^2}\Big)+\tfrac{1}{3}\bar\xi(4\bar H-\sqrt{\tfrac{3}{2}}\mathring\varphi)\rring\varphi\mathring\varphi\\
    +\Big[1+\tfrac{2}{3}\mathring{\bar\xi}\bar H-\tfrac{1}{\sqrt{6}}\mathring{\bar\xi}\mathring\varphi-2\bar\xi\Big(\bar H^2+\frac{K}{3\bar a^2}-\sqrt{\tfrac{3}{8}}\bar H\mathring\varphi\Big)\Big]\mathring{\varphi}^2 =0~,\label{GB_varphi_FLRW_Fried2}
\end{aligned}
\end{align}
where the dimensionless quantities are
\begin{equation}
    \bar a\equiv a/b~,~~~\bar t\equiv t/b~,~~~\bar H\equiv \mathring{\bar a}/\bar a~,~~~\bar V\equiv b^2V~,~~~\bar\xi\equiv\xi/b^2~.
\end{equation}

The relation between the $f(R)$ frame scale factor and its Einstein frame counterpart is given by Eq. \eqref{a_R_a}, as before, but the presence of the GB term modifies the effective potential of the scalaron. For example, by setting $\rring\varphi=\mathring\varphi=0$ in \eqref{GB_varphi_FLRW_KG}, we get the effective stationary/critical point equation,
\begin{equation}\label{GB_st_pt}
    \bar V_{{\rm eff},\varphi}\equiv\bar V_{,\varphi}+3\bar\xi_{,\varphi}(\mathring{\bar H}+\bar H^2)\Big(\bar H^2+\frac{K}{\bar a^2}\Big)=0~,
\end{equation}
where the second term comes from the GB contribution. The Friedmann equations at the critical point yield
\begin{equation}\label{Friedmann_GB_st}
    3\Big(\bar H^2+\frac{K}{\bar a^2}\Big)=\bar V~,~~~\mathring{\bar H}=\frac{K}{\bar a^2}~.
\end{equation}
Plugging \eqref{Friedmann_GB_st} in \eqref{GB_st_pt} leads to
\begin{equation}\label{y_GB_st}
    \bar V_{,\varphi}+\tfrac{1}{3}\bar\xi_{,\varphi}\bar V^2=0~~\Rightarrow~~1-3y+\frac{c}{36}(1-y)^3=0~,
\end{equation}
where $y\equiv e^{-\sqrt{2/3}\varphi}$, and $V$ and $\xi$ are given in \eqref{V_xi_defs}. From \eqref{y_GB_st} it can be seen that when $c=0$, the original critical point ($y_{\rm cr}=1/3$) of the BI scalar potential is recovered, which is a local maximum. When $c\neq 0$, however, the critical point gets effectively shifted. For example, for the BI gravity in Sec. \ref{sec_BIKK}, where $c=1$, we get $y_{\rm cr}\approx 0.336$. The corresponding critical value of $\bar a(0)$ (at the bounce) can also be found from the first equation of \eqref{Friedmann_GB_st} at $\bar H=0$, which yields
\begin{equation}
    \bar a_{\rm cr}(0)=\frac{\sqrt{6}}{\sqrt{y_{\rm cr}}(1-y_{\rm cr})}~.
\end{equation}
Then, the $f(L)$-frame (the Jordan frame) scale factor $\bar a_{L,{\rm cr}}(0)$ can be found from
\begin{equation}
    \bar a_{L,{\rm cr}}(0)=\bar a_{\rm cr}(0)\sqrt{y_{\rm cr}}=\frac{\sqrt{6}}{1-y_{\rm cr}}~,
\end{equation} 
as discussed in Sec. \ref{subsec_frames}. For $c=1$ we obtain $\bar a_{L,{\rm cr}}\approx 3.689$ (as opposed to the $c=0$ case where $\bar a_{L,{\rm cr}}\approx 3.674$). This explains why in Fig. \ref{Fig_GB_bounce} (right column plots with $\bar a_{L}(0)=3.68$) the oscillatory solution with $c=0$ becomes a runaway solution for $c=1$ because the effective critical scale factor is shifted and becomes larger than $3.68$.

Let us also show that the effective critical point $y_{\rm cr}$ is always a local maximum, regardless of $c$. For this, we look into the effective mass-squared $V_{{\rm eff},\varphi\varphi}$ at the critical point by using \eqref{y_GB_st},
\begin{equation}
    V_{{\rm eff},\varphi\varphi}(\varphi_{\rm cr})=-2b^{-2}y_{\rm cr}^3=-2b^{-2}e^{-\sqrt{6}\varphi_{\rm cr}}~.
\end{equation}
Here, the parameter $c$ only enters indirectly through the solution $\varphi_{\rm cr}$ to \eqref{y_GB_st}. So, as long as $\varphi_{\rm cr}$ exists (and is non-zero), it corresponds to a local maximum of the effective potential. The local maximum exists in the range $-36<c<+\infty$. For $c=-36$, the maximum is at $y=0$ or $\varphi=+\infty$, which implies that for $c\leq -36$, only oscillatory solutions are possible because no runaway minimum exists (it is effectively uplifted by the GB term).

The effective potential itself can be easily found by integrating \eqref{y_GB_st} (left side),
\begin{equation}
    V_{\rm eff}=\frac{y}{2b^2}(1-y)^2-\frac{c}{360b^2}(1-y)^5~,
\end{equation}
up to an irrelevant integration constant. Minkowski vacuum is located at $y=1$ ($\varphi=0$) as before.

\section{Stability of the solutions}

In this section, we will analyze the stability of the bouncing solutions in both the $f(R)$ and the $f(R,\cg)$ cases of BI-type gravity, following the approach of Ref. \cite{Nojiri:2010wj} where the Hubble function is perturbed assuming spatial homogeneity. Full cosmological perturbation theory (with all the relevant degrees of freedom taken into account) will be investigated in future works.

\subsection{Stability analysis for BI-type $f(R)$ gravity}
We start with the $f(R)$ case and Eq. \eqref{EFE_00_fR}. For this analysis, it is useful to introduce a new function $\bar G= \bar H^2$, and rewrite \eqref{EFE_00_fR} as
\begin{equation}\label{EFE_00_fR_G}
	6{{\bar f}_{\bar R\bar R}}\bar G\left( {2{{\bar G}'} + \frac{{{{\bar G}''}}}{2} - 2\frac{K}{{{{\bar a}^2}}}} \right) - \Big(\bar G + \frac{{{{\bar G}' }}}{2}\Big){{\bar f}_{\bar R}} + \frac{1}{6}\bar f = 0~.
\end{equation}
Here, $\bar f_{\bar R}$ denotes the derivative of $\bar f$ with respect to $\bar R\equiv b^2R= 6(\frac{1}{2}{\bar G}' + 2\bar G + K/\bar a^2) $, the prime denotes derivative with respect to the forward e-fold number $N$, defined as $\bar a =\bar a(0)e^{N}$, and $\bar f\equiv b^2f = 2\big( {1 - \sqrt {1 - \bar R} } \big)$ such that $\bar f_{\bar R}=(1-\bar R)^{-1/2}$.

The stability of the equation \eqref{EFE_00_fR_G} can be studied by introducing a linear perturbation of a given solution as $\bar G(N)=\bar G_0(N)+\delta \bar G(N)$, where $\delta \bar G(N)$ is a small perturbation around a given background solution $\bar G_0(N)$, which also satisfies Eq. \eqref{EFE_00_fR_G}. This yields a second-order ordinary differential equation for $\delta \bar G$:
\begin{equation}\label{EoM_delta_G}
	\mathcal{A}\delta\bar G^{\prime\prime}+\mathcal{B}\delta\bar G^{\prime}+\mathcal{C}\delta \bar G=0~,
\end{equation}
where 
\begin{align}
	\mathcal{A} &= 3 f_{\bar R\bar R}(\bar R_0)\,\bar G_0~,\label{A_coeff}\\
	\mathcal{B} &= 9 f_{\bar R\bar R}(\bar R_0)\,\bar G_0
	-\frac{3}{2}\,\bar G_0'\, f_{\bar R\bar R}(\bar R_0)
	+18 f_{\bar R\bar R\bar R}(\bar R_0)\,\bar G_0
	\left(2\bar G_0' + \frac{\bar G_0''}{2} - 2\frac{K}{\bar a^2}\right)~,\label{B_coeff}\\
	\mathcal{C} &= f_{\bar R}(\bar R_0)
	+72 f_{\bar R\bar R\bar R}(\bar R_0)\,\bar G_0
	\left(2\bar G_0' + \frac{\bar G_0''}{2} - 2\frac{K}{\bar a^2}\right)
	+6 f_{\bar R\bar R}(\bar R_0)
	\left(\bar G_0' - 2\bar G_0 + \frac{\bar G_0''}{2} - 2\frac{K}{\bar a^2}\right)~,\label{C_coeff}
\end{align}
and $\bar R_0$ denotes the rescaled Ricci scalar constructed from $\bar G_0$.

Solutions of a second-order differential equation can generally be decomposed into growing and decaying modes. Growing modes correspond to exponentially amplified perturbations and signal instability of the original solution $\bar G_0$, whereas decaying modes correspond to exponentially damped perturbations, ensuring linear stability of $\bar G_0$. A solution of the second-order differential equation is free of growing modes if and only if the associated characteristic equation admits no non-positive roots \cite{Nojiri:2010wj}. Therefore, stability of $\bar G_0$ means both $p\equiv\mathcal{B}/\mathcal{A}>0$ and $q\equiv\mathcal{C}/\mathcal{A}>0$, which gives
\begin{align}
	& 6{{\bar G}_0} - {{\bar G}_0}^\prime
	+ 6{{\bar G}_0}{{\bar f}_{\bar R\bar R\bar R}}\!\left( {{{\bar R}_0}} \right)
	{\left[ {{{\bar f}_{\bar R\bar R}}\!\left( {{{\bar R}_0}} \right)} \right]^{ - 1}}
	\left( {4{{\bar G}_0}^\prime  + {{\bar G}_0}^{\prime \prime }
		- 4\frac{K}{{{{\bar a}^2}}}} \right)
	\;>\; 0~,\label{vac_Stab_con_1}
	\\[1ex]
	& {{\bar f}_{\bar R}}\!\left( {{R_0}} \right)
	{\left[ {{{\bar f}_{\bar R\bar R}}\!\left( {{R_0}} \right)} \right]^{ - 1}}
	+ 36{{\bar G}_0}{{\bar f}_{\bar R\bar R\bar R}}\!\left( {{{\bar R}_0}} \right)
	{\left[ {{{\bar f}_{\bar R\bar R}}\!\left( {{{\bar R}_0}} \right)} \right]^{ - 1}}
	\left( {4{{\bar G}_0}^\prime  + {{\bar G}_0}^{\prime \prime }
		- 4\frac{K}{{{{\bar a}^2}}}} \right)
	\nonumber\\
	&\quad
	+ \left( {6{{\bar G}_0}^\prime  - 12{{\bar G}_0}
		+ 3{{\bar G}_0}^{\prime \prime }
		- 12\frac{K}{{{{\bar a}^2}}}} \right)
	\;>\; 0~.\label{vac_Stab_con_2}
\end{align}

Bouncing solutions require the initial conditions $\bar H(0)=0$ and $\mathring{\bar H}(0)>0$,~\footnote{We set the initial e-fold time at the bounce to $N=0$. Note that in terms of the e-fold time the analysis cannot be extended to the pre-bounce contraction phase, as $\bar G_0(N)$ becomes multi-valued. This is not a problem because one can apply the same analysis to the contraction phase by reflecting the solution in time, $\bar t\rightarrow -\bar t$, in which case the direction of growing $N$ corresponds to going back in physical time $\bar t$.} which correspond to $\bar G_0(0)=0$ and $\bar G^{\prime}_0(0)=2\mathring{\bar H}(0)>0$. Although this contradicts the condition \eqref{vac_Stab_con_1}, it does not imply that equation \eqref{EFE_00_fR_G} is initially unstable. This can be seen from the real part of the root of the characteristic equation, $ \chi\equiv\mathrm{Re}[(-p+\sqrt{p^2-4q})/2]$. At the initial time, when $\bar G_0(0)=0$, the coefficient of the second-derivative term in \eqref{EFE_00_fR_G} vanishes, such that \eqref{EFE_00_fR_G} degenerates into a first-order differential equation. Consequently, the stability of \eqref{EFE_00_fR_G} at $\bar t=0$ cannot be analyzed using the characteristic equation of \eqref{EoM_delta_G}.

\begin{figure}
	\centering
	\includegraphics[width=0.4\linewidth]{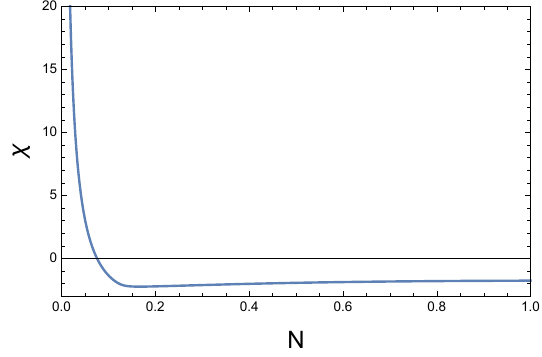}
	\captionsetup{width=1\linewidth}
	\caption{Evolution of $\chi$ for the $\bar a(0)=3$ solution.}\label{x-3_pic}
\end{figure}

As an example, consider the bouncing solution with $\bar{a}(0)=3$. Starting from $N=10^{-3}$, $\chi$ associated with the solution is shown in Fig. \ref{x-3_pic}. The transition of $\chi$ from positive to negative values occurs within only $O(10^{-2})$ e-folds, indicating that the growing mode is rapidly converted into a decaying mode before it can be significantly amplified. Therefore, this solution can be seen as initially stable.

\begin{figure}
	\centering
	\begin{subfigure}{0.48\linewidth}
		\includegraphics[width=\linewidth]{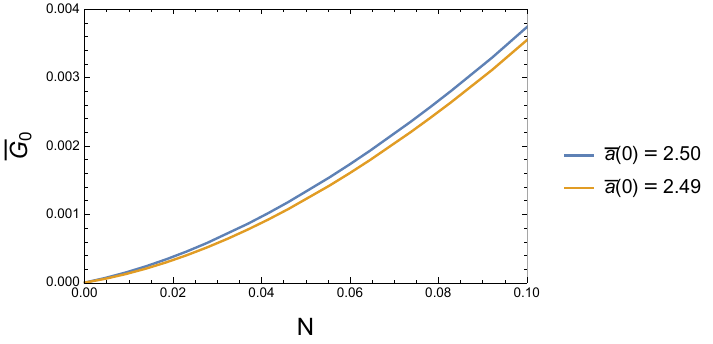}
		\caption{$\bar G_0 $ with $\bar a(0)\approx 2.50$.}
		
	\end{subfigure}\hspace{10pt}
	\begin{subfigure}{0.48\linewidth}
		\includegraphics[width=\linewidth]{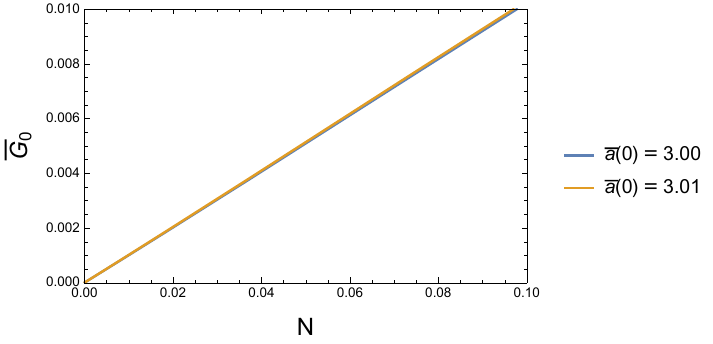}
		\caption{$\bar G_0 $ with $\bar a(0)\approx 3.00$.}
		
	\end{subfigure}
	\captionsetup{width=1\linewidth}
	\caption{$\bar G_0 $ with $\bar a(0)$ around different values.}\label{G-250-300-short}
\end{figure}

This can also be confirmed analytically. Near $N=0$, the solution of \eqref{EFE_00_fR_G} can be expanded as 
\begin{equation}\label{Expand_G_near_0}
	\bar{G}_0(N)=\frac{2}{3}\left(s-s^2\right) N+\frac{2 s^2+s-2}{3} N^2+O\left(N^3\right)~,
\end{equation}
with initial conditions $\bar G_0(0)=0$ and ${{G_0}^\prime }\left( 0 \right) = \frac{2}{3}\left( {\sqrt {\frac{6}{{\bar a{{\left( 0 \right)}^2}}}}  - \frac{6}{{\bar a{{\left( 0 \right)}^2}}}} \right)$, whereas $s\equiv\sqrt{6} / \bar{a}(0) \in(0,1)$. For $N\ll 1$, $\partial \bar G/\partial s$ for \eqref{Expand_G_near_0} takes the maximum absolute value at $s=1$, indicating that $\bar G_0$ will be sensitive to $\delta \bar a (0)$ if $\bar a(0)$ is close to $\sqrt{6}\approx 2.449$. This can be seen in Fig. \ref{G-250-300-short}.

From the analytical solution \eqref{Expand_G_near_0} one can calculate $p(N)=\mathcal{B}/\mathcal{A}\approx -1/2 N^{-1}+O(1)$ and $q(N)=\mathcal{B}/\mathcal{A}\approx -1/s N^{-1}+O(1)$ for $N\ll 1$. Therefore, Eq. \eqref{EoM_delta_G} near $N=0$ can be generally solved as 
\begin{equation}\label{Expand_deltaG_N_0}
	\delta \bar{G}(N)=c_1[1+O(N)]+c_2 N^{3 / 2}[1+O(N)]~,
\end{equation}
where $c_1$ and $c_2$ are some constants. This shows that $\delta \bar{G}(N)$ will not grow significantly for $N\ll 1$ even if the characteristic equation has positive roots near $N=0$, and the solutions are therefore stable near the initial time, provided that $\chi$ drops below zero sufficiently fast.

\begin{figure}
	\centering
	\begin{subfigure}{0.48\linewidth}
		\includegraphics[width=\linewidth]{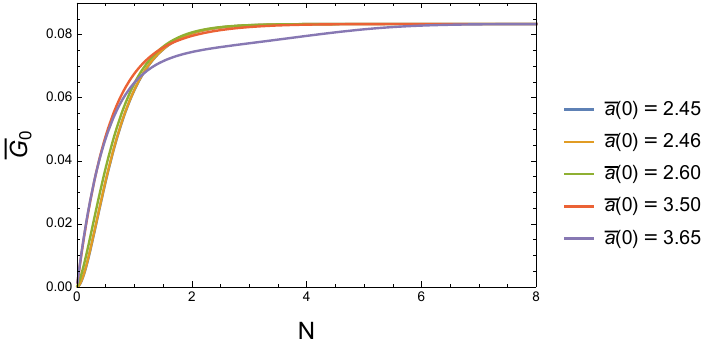}
		
	\end{subfigure}\hspace{10pt}
	\begin{subfigure}{0.48\linewidth}
		\includegraphics[width=\linewidth]{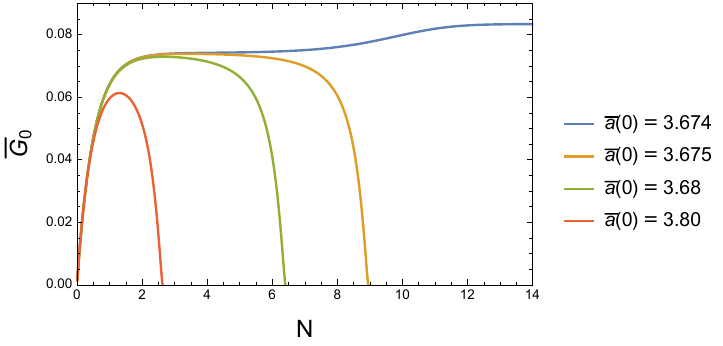}
		
	\end{subfigure}
	\captionsetup{width=1\linewidth}
	\caption{$\bar G_0$ corresponding to different $\bar a(0)$.}
	\label{GN-total}
\end{figure}

\begin{figure}
	\centering
	\begin{subfigure}{0.5\linewidth}
		\includegraphics[width=\linewidth]{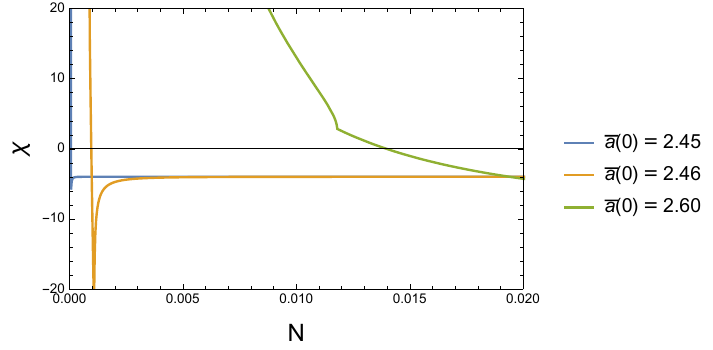}
		
	\end{subfigure}
	\begin{subfigure}{0.48\linewidth}
		\includegraphics[width=\linewidth]{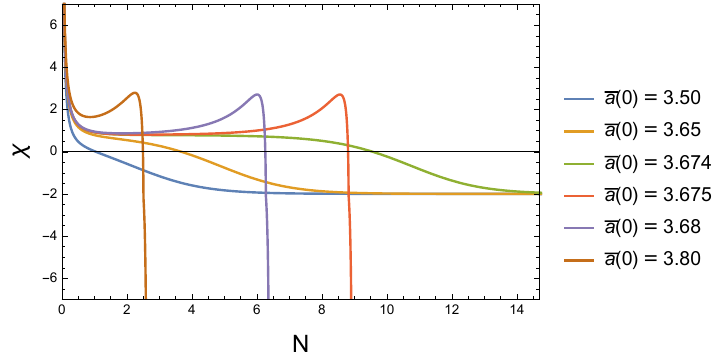}
		
	\end{subfigure}
	\captionsetup{width=1\linewidth}
	\caption{$\chi$ corresponding to different $\bar a(0)$. }\label{Total_chi}
\end{figure}

The late time behavior of $\chi$ for various choices of $\bar a(0)$ (from Fig. \ref{Fig_R_bounce}) are shown in Fig. \ref{Total_chi}. For $\sqrt{6}<\bar a(0)\lesssim 3.674$, which corresponds to the bouncing solutions, larger $\bar a(0)$ will lead to a longer duration of growing perturbation $\delta \bar G$. When $\bar a(0)>3.65$, $\delta \bar G$ may be exponentially amplified by a factor larger than $O(1)$. Eventually, $\chi$ asymptotically approaches $-2$, indicating late-time stability of the runaway solutions and that the solutions asymptotically converge to the attractor $\bar H\sim 1/\sqrt{12}$. In the case of $\bar a(0)\gtrsim 3.674$ (oscillatory bouncing solutions), $\chi$ is initially positive and reaches a local maximum before a sharp drop to $-\infty$ when $\bar G_0$ starts to roll down to zero. This indicates stability of the solution during the phase of decreasing $\bar G_0$, which is followed by the oscillatory phase to which our analysis cannot be extended.

\subsubsection*{Late-time approximation}
 The late-time asymptotic behavior of $\chi$ can be studied analytically in the case of the runaway solutions or pre-oscillation phase of the oscillatory solutions. With the help of \eqref{EFE_00_fR_G}, the $p$ and $q$ can be written as
\begin{align}
	p&=6 + \frac{1}{{{{\bar G}_0}}}\left( {\frac{1}{{{{\bar f}_{\bar R}}^2}} - \frac{1}{{{{\bar f}_{\bar R}}}} + {{\bar G}_0}^\prime } \right)~,\label{gen_p} \\
	q&= {8 - \frac{4}{{{{\bar f}_{\bar R}}{{\bar G}_0}}} + \frac{{4{{\bar G}_0}^\prime }}{{{{\bar G}_0}}}}  + \frac{2}{{{{\bar G}_0}}}\frac{1}{{3{{\bar f}_{\bar R}}^2}}\left( {8 + \frac{1}{{3{{\bar f}_{\bar R}}^2{{\bar G}_0}}} - \frac{1}{{3{{\bar f}_{\bar R}}{{\bar G}_0}}} + \frac{{{{\bar G}_0}^\prime }}{{2{{\bar G}_0}}}} \right)~.\label{gen_q}
\end{align}
At late time where $a\gg1$, the contribution of $K/\bar a^2$ in $\bar R$ can be neglected. Therefore, we can write ${{\bar G_0}^\prime } \approx  - 4\bar G_0 + \frac{{\bar R}}{3} = \frac{1}{3} - \frac{1}{{3{f_R}^2}} - 4\bar G_0$, which yields late-time approximations of $p$ and $q$ as
\begin{align}
	p&\approx 2 + \frac{1}{{{{\bar G}_0}}}\left( {\frac{2}{{3{{\bar f}_{\bar R}}^2}} - \frac{1}{{{{\bar f}_{\bar R}}}} + \frac{1}{3}} \right)~,\label{late_time_p} \\
	q&\approx  - 8 + \frac{{4\left( {{{\bar f}_{\bar R}} - 2} \right)\left( {{{\bar f}_{\bar R}} - 1} \right)}}{{3{{\bar G}_0}{{\bar f}_{\bar R}}^2}} + \frac{{{{\left( {{{\bar f}_{\bar R}} - 1} \right)}^2}}}{{9{{\bar f}_{\bar R}}^4{{\bar G}_0}^2}}~.\label{late_time_q}
\end{align}

For the runaway solutions, $\bar f_{\bar R}$ will roll down to infinity, leading to $\bar R\to 1$ and $\bar G_0\to 1/12$ (see Appendix \ref{App_B} or Figs.~\ref{Fig_R_bounce} and \ref{Fig_R_bounce_curvature}). From \eqref{late_time_p} and \eqref{late_time_q},  one can obtain $p\to 6$ and $q\to 8$, which leads to $\chi\to-2$. This is consistent with the numerical results of Fig. \ref{Total_chi}, showing that $\bar H\sim 1/\sqrt{12}$ is indeed a stable attractor. Late-time behavior of the oscillatory solutions requires a more careful treatment. While the true asymptotic limit corresponds to $\bar f_{\bar R} \to 1$ ($\bar R \to 0$), 
there is an intermediate stage when $\bar G_0$ approaches zero, which is characterized by $\bar G_0 \ll 1$ and $ 0 < \bar f_{\bar R} < 1 $. This is because the scalar curvature typically becomes negative before relaxing to the  Minkowski value. In this regime, Eqs.~\eqref{late_time_p} and \eqref{late_time_q} imply
\begin{equation}
	p \sim \frac{(\bar f_{\bar R}-1)(\bar f_{\bar R}-2)}{3 \bar f_{\bar R}^2} \frac{1}{\bar G_0}~, 
	\qquad 
	q \sim \frac{(\bar f_{\bar R}-1)^2}{9 \bar f_{\bar R}^4} \frac{1}{\bar G_0^2}~.
\end{equation}
For $0 < \bar f_{\bar R} < 1$, both coefficients are positive, while the discriminant behaves as
\begin{equation}
	p^2 - 4q 
	\sim 
	\frac{(\bar f_{\bar R}-1)^2 (\bar f_{\bar R}^2 - 4 \bar f_{\bar R})}{9 \bar f_{\bar R}^4} \frac{1}{\bar G_0^2} < 0~.
\end{equation}
Therefore, the characteristic roots form a complex conjugate pair, and their real part is
\begin{equation}
	\chi = -\frac{p}{2}
	\sim -\frac{(\bar f_{\bar R}-1)(\bar f_{\bar R}-2)}{6 \bar f_{\bar R}^2} \frac{1}{\bar G_0}~,
\end{equation}
leading to $\chi \to -\infty$ as $\bar G_0 \to 0^+$.
This explains the sharp negative divergence of $\chi$ observed numerically in Fig.~\ref{Total_chi}, when the oscillatory solution first approaches $\bar G_0 = 0$. 
Physically, this behavior originates from the fact that $\mathcal{A}$ ($\propto \bar G_0)$ in Eq.~\eqref{EoM_delta_G} vanishes in this limit, causing the second-order perturbation equation to effectively degenerate into a strongly damped first-order system. As a result, perturbations are rapidly suppressed when the solution passes through $\bar G_0 \approx 0$. Nevertheless, in scenarios where $\bar H$ changes from positive to negative, the e-folding number $N$ is not well defined, and the applicability of this method to multi-bounce solutions is limited to time intervals between the zeroes of $\bar G_0$.

\begin{figure}
	\centering
	\begin{subfigure}{0.48\linewidth}
		\includegraphics[width=\linewidth]{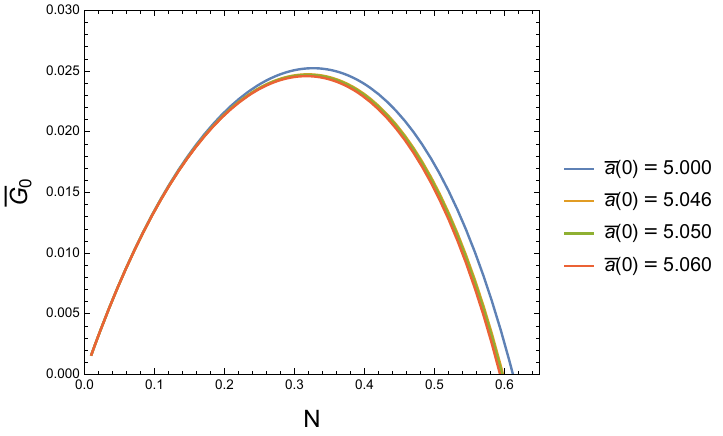}
	\end{subfigure}
	\begin{subfigure}{0.48\linewidth}
		\includegraphics[width=\linewidth]{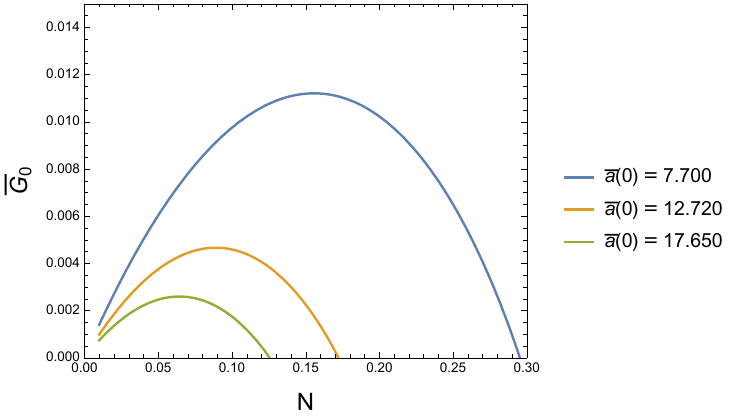}
		
	\end{subfigure}\\

	\begin{subfigure}{0.48\linewidth}
		\includegraphics[width=\linewidth]{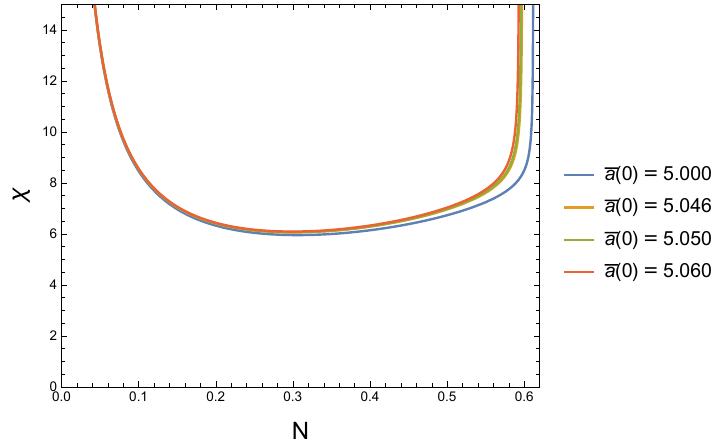}	
	\end{subfigure} 
	\begin{subfigure}{0.48\linewidth}
		\includegraphics[width=\linewidth]{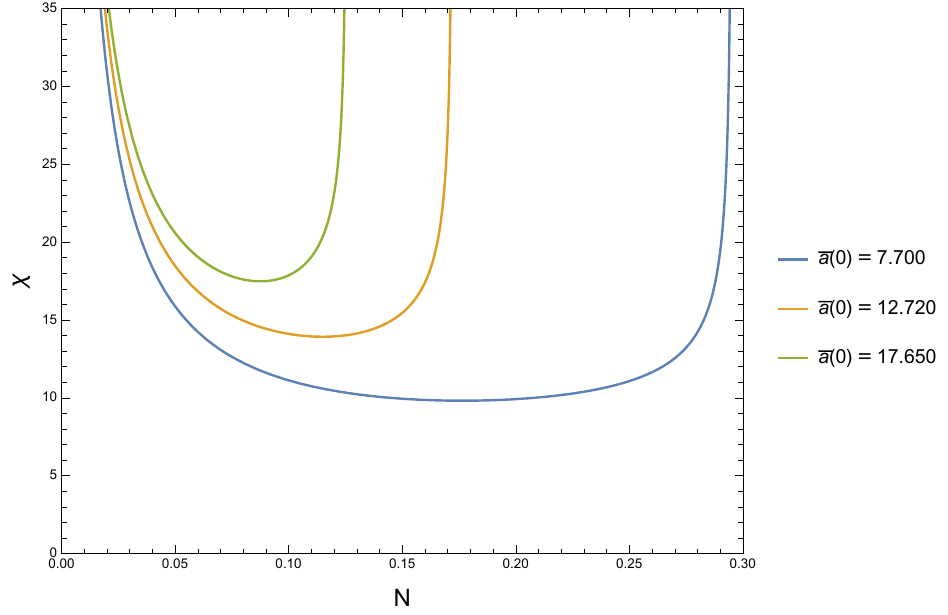}
		
	\end{subfigure}
	\captionsetup{width=1\linewidth}
	\caption{$\bar G_0(N)$ and $\chi(N)$ corresponding to odd-bounce solutions shown in Fig.~\ref{Fig_R_odd_bounce}.  Left column shows the results with $\bar a(0)$ close to $5$. Right column shows results corresponding to solutions with three, five, and seven bounces.}
	\label{G-chi-Odd-Bounc}
\end{figure}

\subsubsection*{Multi-bounce solutions}
 Here, we analyze the stability of multi-bounce solutions. Considering odd-bounce solutions in Fig.~\ref{Fig_R_odd_bounce} as an example, we plot the evolution of $\bar G_0$ and $\chi$ in Fig.~\ref{G-chi-Odd-Bounc}. For all the examples displayed there, $\chi(N)$ remains positive between the main ($N=0$) bounce and the first sub-bounce. This suggests the presence of a growing mode throughout this interval. Nevertheless, since the e-fold number in this interval is $N<1$, and according to the near-origin result in Eq.~\eqref{Expand_deltaG_N_0}, the perturbation $\delta \bar G(N)$ can still remain under control. That is, the time interval is too short for the perturbation to be amplified substantially.

Qualitatively, stability properties of even-bounce solutions are very similar to those of the odd-bounce solutions. In particular, in the interval between the anti-bounce point at $N=0$ and the next zero of $\bar G_0(N)$, the characteristic exponent $\chi(N)$ remains positive. This again indicates the presence of a growing mode during this stage, but since the relevant evolution takes place within one e-fold, the perturbation $\delta\bar G(N)$ does not have enough time to significantly grow before $\bar G_0(N)$ reaches zero. In this sense, the early-stage evolution of both odd-bounce and even-bounce solutions remains under perturbative control.

The asymptotic behavior (runaway or oscillatory) of multi-bounce solutions is qualitatively the same as in the single-bounce case discussed above, leading to the conclusion that the late-time stability of single-bounce solutions apply to the multi-bounce cases as well.

\subsubsection*{The case with a cosmological constant}
Let us extend the above stability analysis to the case with a cosmological constant ($\bar\Lambda\equiv b^2\Lambda$), where the background equation is Eq.~\eqref{EFE_00_fR_no_b_Lambda}. Introducing $\bar G=\bar H^2$ as before, Eq.~\eqref{EFE_00_fR_no_b_Lambda} can be rewritten as
\begin{equation}\label{EFE_00_fR_G_Lambda}
	6{{\bar f}_{\bar R\bar R}}\bar G\left( {2{{\bar G}'} + \frac{{{{\bar G}''}}}{2} - 2\frac{K}{{{{\bar a}^2}}}} \right) - \Big(\bar G + \frac{{{{\bar G}' }}}{2}\Big){{\bar f}_{\bar R}} + \frac{1}{6}\bar f_{\Lambda} = 0~,
\end{equation}
where
\begin{equation}
	\bar f_{\Lambda}=2\left(1-\bar\Lambda-\sqrt{1-\bar R}\right)~,\qquad 
	\bar f_{\bar R}=(1-\bar R)^{-1/2}~.
\end{equation}
Since the cosmological constant only shifts $\bar f$ by a constant, the derivatives $\bar f_{\bar R}$, $\bar f_{\bar R\bar R}$ and $\bar f_{\bar R\bar R\bar R}$ remain unchanged. Therefore, the perturbation equation still takes the form of Eq.~\eqref{EoM_delta_G}, and the coefficients $\mathcal{A}$, $\mathcal{B}$ and $\mathcal{C}$ are still given by Eqs.~\eqref{A_coeff}--\eqref{C_coeff}, with the only difference that the background solution $\bar G_0(N)$ must now satisfy Eq.~\eqref{EFE_00_fR_G_Lambda} instead of \eqref{EFE_00_fR_G}.

Using \eqref{EFE_00_fR_G_Lambda} to eliminate the combination $2\bar G_0'+\bar G_0''/2-2K/\bar a^2$, one finds the $\Lambda$-generalization of Eqs.~\eqref{gen_p} and \eqref{gen_q}:
\begin{align}
	p_{\Lambda}
	&=
	6+\frac{1}{{{{\bar G}_0}}}\left( {\frac{1}{{{{\bar f}_{\bar R}}^2}} - \frac{{1-\bar\Lambda}}{{{{\bar f}_{\bar R}}}} + {{\bar G}_0}^\prime } \right)~,
	\label{gen_p_lambda}
	\\
	q_{\Lambda}
	&=
	8+\frac{{4{{\bar G}_0}^\prime }}{{{{\bar G}_0}}}
	-\frac{{4(1-\bar\Lambda)}}{{{{\bar f}_{\bar R}}{{\bar G}_0}}}
	+\frac{{16}}{{3{{\bar f}_{\bar R}}^2{{\bar G}_0}}}
	+\frac{{{{\bar G}_0}^\prime }}{{3{{\bar f}_{\bar R}}^2{{\bar G}_0}^2}}
	-\frac{{2(1-\bar\Lambda)}}{{9{{\bar f}_{\bar R}}^3{{\bar G}_0}^2}}
	+\frac{{2}}{{9{{\bar f}_{\bar R}}^4{{\bar G}_0}^2}}~.
	\label{gen_q_lambda}
\end{align}
Near the initial point $N=0$, the leading behavior of the perturbation equation remains the same as in the $\bar\Lambda=0$ case: One still has $\mathcal{A}\propto N$, $p_{\Lambda}(N)\approx -\tfrac{1}{2}N^{-1}+O(1)$, and $q_{\Lambda}(N)=O(N^{-1})$. Consequently, the local solution for $\delta G(N)$ around $N=0$ still takes the form of Eq.~\eqref{Expand_deltaG_N_0}, which implies that the initial-stage regularity argument remains true in the presence of $\bar\Lambda$.

Late-time asymptotics can also be analyzed analytically. For the runaway branch, one still has $\bar f_{\bar R}\to\infty$, $\bar R\to 1$, $\bar G_0\to 1/12$, and $\bar G_0'\to 0$, such that \eqref{gen_p_lambda} and \eqref{gen_q_lambda} give
\begin{equation}\label{late_time_pq_runaway_lambda}
	p_{\Lambda}\to 6~,\qquad q_{\Lambda}\to 8~,
\end{equation}
and therefore
\begin{equation}\label{late_time_chi_runaway_lambda}
	\chi_{\Lambda}\equiv\mathrm{Re}\left[(-p_{\Lambda}+\sqrt{p_{\Lambda}^2-4q_{\Lambda}})/2\right]\to -2~.
\end{equation}
Hence, the runaway asymptotic solution $\bar H\sim 1/\sqrt{12}$ remains a stable attractor. Late-time behavior of the oscillatory branch is qualitatively modified by the presence of a positive cosmological constant. In contrast to the $\bar\Lambda=0$ case, where the solution asymptotically approaches $\bar G_0 \to 0$, the system can instead be driven to a de Sitter fixed point, as expected. We now analyze this possibility.

Assuming that the late-time evolution approaches a de Sitter configuration, we impose
\begin{equation}
	\bar G_0' \to 0~, \qquad \bar G_0 \to \bar G_{\ast}~, \qquad \bar f_{\bar R} \to F_{\ast}~
\end{equation}
on \eqref{EFE_00_fR_G_Lambda} and obtain an algebraic equation for $F_{\ast}$,
\begin{equation}\label{dS_fixed_point_lambda_clean}
	F_{\ast}^2 - 4(1-\bar\Lambda)F_{\ast} + 3 = 0~.
\end{equation}
A real de Sitter fixed point exists only if the discriminant of this equation is non-negative, which leads to the condition
\begin{equation}\label{dS_existence_condition_clean}
	\bar\Lambda \le 1 - \frac{\sqrt{3}}{2}~.
\end{equation}
When this condition is satisfied, Eq.~\eqref{dS_fixed_point_lambda_clean} admits two branches,
\begin{equation}\label{dS_two_branches_clean}
	F_{\pm} = 2(1-\bar\Lambda) \pm \sqrt{4(1-\bar\Lambda)^2 - 3}~.
\end{equation}
The branch that continuously connects to the $\bar\Lambda=0$ oscillatory solution is $F_{-}$ because $F_{-} \to 1$ as $\bar\Lambda \to 0$. The corresponding squared Hubble parameter $\bar G_*$ is determined by
\begin{equation}
	\bar G_{\ast} = \tfrac{1}{12}(1 - F_{\ast}^{-2})~.
\end{equation}
Evaluating \eqref{gen_p_lambda} and \eqref{gen_q_lambda} at the fixed point (where $\bar G_0' = 0$), one finds \footnote{
	It is important to distinguish this finite-$F_{\ast}$ de Sitter fixed point from the runaway asymptotics. 
	The latter corresponds to $\bar f_{\bar R} \to \infty$, $\bar R \to 1$, and $\bar G_0 \to 1/12$, which does not arise as a solution to \eqref{dS_fixed_point_lambda_clean}. 
	In particular, $\bar f_{\bar R} \to \infty$ is not a branch of the algebraic fixed-point equation, but rather represents a boundary-type asymptotic behavior in the field space. 
	Consequently, the runaway branch must be analyzed directly from \eqref{gen_p_lambda} and \eqref{gen_q_lambda}, yielding $p_{\Lambda} \to 6$ and $q_{\Lambda} \to 8$, instead of the finite-$F_{\ast}$ result $p_{\Lambda} \to 3$.
}
\begin{equation}\label{late_time_p_osc_lambda_clean}
	p_{\Lambda} \to 3~,~~~q_{\Lambda} \to -4 + \frac{8}{F_{-}^2 - 1}~,
\end{equation}
and therefore,
\begin{equation}\label{late_time_chi_osc_lambda_clean}
	\chi_{\Lambda} \to \tfrac{1}{2}\mathrm{Re}\Bigg[
	-3 + \sqrt{25 - \dfrac{32}{F_{-}^2 - 1}}\,
	\Bigg]~.
\end{equation}
On this branch, one has $1<F_{-} \le \sqrt{3}$, which implies $q_{\Lambda} \ge 0$. Consequently, $\chi_{\Lambda} \le 0$, and strictly $\chi_{\Lambda} < 0$ for $1 < F_{-} < \sqrt{3}$, indicating that the fixed point is stable.

It is important to clarify the nature of the $\bar\Lambda \to 0^+$ limit. At the level of the background solution, the de Sitter fixed point continuously degenerates into the Minkowski configuration, with $\bar f_{\bar R} \to 1$ and $\bar G_* \to 0$. Nevertheless, this limit is singular for the perturbation analysis. Indeed, the coefficients of the perturbation equation \eqref{EoM_delta_G} satisfy $\mathcal{A} \propto \bar G_0 \to 0$ making the second-order differential equation effectively first-order. As a result, the auxiliary quantities $p = \mathcal{B}/\mathcal{A}$ and $q = \mathcal{C}/\mathcal{A}$ do not admit a unique limit at $(\bar f_{\bar R}, \bar G_0) = (1,0)$ but instead depend on the path along which this point is approached, i.e. on the relative scaling of $(\bar f_{\bar R}-1)$ and $\bar G_0$. In particular, the $\bar\Lambda \to 0^+$ limit of the de Sitter branch corresponds to a specific trajectory in this plane, which does not coincide with the intrinsic late-time behavior of the $\bar\Lambda = 0$ oscillatory solutions. Therefore, continuity holds at the level of the background dynamics but not at the level of $(p,q,\chi)$.

\begin{figure}
	\centering
	\begin{subfigure}{0.48\linewidth}
		\includegraphics[width=\linewidth]{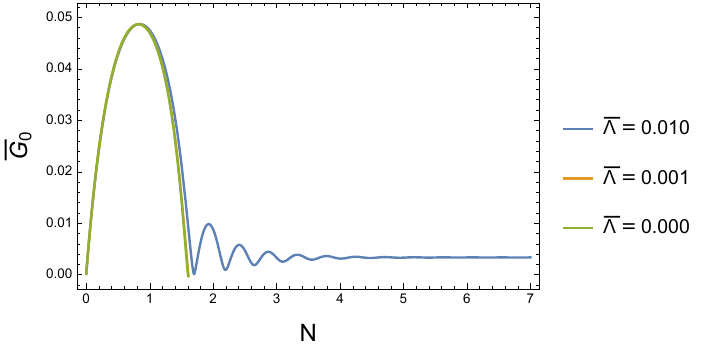}
		
	\end{subfigure}
	\begin{subfigure}{0.48\linewidth}
		\includegraphics[width=\linewidth]{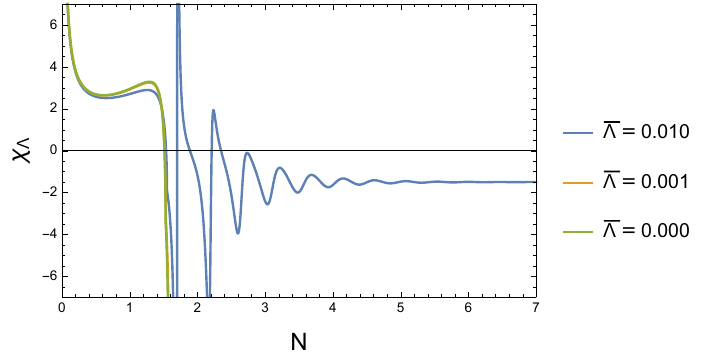}
		
	\end{subfigure}
	\captionsetup{width=1\linewidth}
	\caption{$\bar G_0(N)$ and $\chi_{\Lambda}(N)$ in the presence of non-vanishing cosmological constant with parameters given in Fig.~\ref{Fig_R_bounce_CC}. }\label{G-chi-Lambda}
\end{figure}

Figure~\ref{G-chi-Lambda} clarifies the domain of validity of the $G(N)$-based stability analysis in the presence of a non-vanishing cosmological constant. At early times, the three curves exhibit nearly identical behavior, which confirms that the near-origin expansion leading to Eq.~\eqref{Expand_deltaG_N_0} remains unaffected by $\bar\Lambda$. In particular, the rapid transition of $\chi_\Lambda$ from large positive values to negative ones is universal. The qualitative difference appears when the oscillatory branch approaches its next zero crossing of $\bar G_0$. For $\bar\Lambda=0$, this reproduces the behavior described by Eqs.~\eqref{late_time_p}--\eqref{late_time_q}: As $\bar G_0\to0^+$ with $0<\bar f_{\bar R}<1$, one has $\chi_\Lambda\to-\infty$. As emphasized above, this reflects the degeneration of the perturbation equation due to $\mathcal{A}\propto\bar G_0\to0$, and signals the breakdown of the $G(N)$ description once $\bar H$ changes sign, rather than a physical obstruction to continuing the solution.

The same limitation persists for small but non-vanishing $\bar\Lambda$, such as $\bar\Lambda=0.001$. Although the cosmological constant eventually increases the center of oscillation of $\bar{H}$ away from zero (in terms of $\bar t$), the dip in the oscillation in Fig.~\ref{G-chi-Lambda} can still reach $\bar G_0=0$ so long as $\bar\Lambda$ is sufficiently small. Consequently, once $\bar G_0=0$, the mapping between $\bar t$ and $N$ is no longer one-to-one, and the stability criterion based on Eq.~\eqref{EoM_delta_G} cannot be applied globally. In contrast, the $\bar\Lambda=0.01$ curve exhibits a qualitatively different behavior: $\bar G_0(N)$ turns around before reaching zero, implying that $\bar H$ never changes sign (after the bounce) and that $N$ remains a well-defined time variable for post-bounce evolution. Therefore, the stability analysis based on \eqref{dS_fixed_point_lambda_clean}--\eqref{late_time_chi_osc_lambda_clean} is valid throughout the entire evolution.

One can make use of this fact in late-time analysis. From Eqs.~\eqref{dS_fixed_point_lambda_clean}--\eqref{dS_two_branches_clean}, one finds that $\bar\Lambda=0.01$ gives $F_-\simeq1.0206$, which yields $\bar G_*\simeq3.33\times10^{-3}$. Plugging these into \eqref{late_time_p_osc_lambda_clean}, one obtains $p_\Lambda\to3$ and $q_\Lambda\gg1$, and hence \eqref{late_time_chi_osc_lambda_clean} yields $\chi_\Lambda\to-3/2$. This is in agreement with Fig.~\ref{G-chi-Lambda} where for $\bar\Lambda=0.01$, $\bar G_0$ asymptotes to a positive constant, and $\chi_\Lambda$ to a negative constant, both finite.

It is thus important to distinguish between small and large $\bar\Lambda$ regimes in the stability analysis. For small $\bar\Lambda$, the solution still reaches $\bar G_0=0$, such that the $G(N)$ formulation breaks down once $\bar H$ changes sign. For sufficiently large positive $\bar\Lambda$, however, the trajectory turns around before $\bar H$ vanishes, in agreement with the background behavior shown in Fig.~\ref{Fig_R_bounce_CC}. In this regime, the de Sitter fixed point is reached entirely within the domain where the present stability analysis is applicable.

The presence of matter (radiation and dust) does not significantly modify the bouncing solutions, as discussed in the previous section, and the stability analysis can be straightforwardly applied, producing expected results. We perform this analysis in Appendix \ref{App_matter}.

\subsection{Stability analysis with the Gauss--Bonnet term}
Now, we generalize the above analysis to the case with a non-vanishing Gauss--Bonnet contribution. Starting from Eq.~\eqref{GB_background_EoM}, and following the same logic as in the derivation of \eqref{EFE_00_fR_G}, we introduce
\begin{equation}
\bar L = \bar R + \frac{c}{24}\bar{\mathcal G}
= 3\bar G' + 12\bar G + 6\frac{K}{\bar a^2}
+ c\left(\bar G+\frac{K}{\bar a^2}\right)\left(\bar G+\frac{\bar G'}{2}\right)~.
\end{equation}
Then, the background equation \eqref{GB_background_EoM} can be re-expressed as
\begin{equation}\label{EFE_00_fL_G}
	\left[1+\frac{c}{6}\left(\bar G+\frac{K}{\bar a^2}\right)\right]
	\left[
	\bar G\,\bar f_{\bar L\bar L}\,\bar L'
	-\left(\bar G+\frac{\bar G'}{2}\right)\bar f_{\bar L}
	\right]
	+\frac{1}{6}\bar f=0~,
\end{equation}
which is the $f(L)$ counterpart of Eq.~\eqref{EFE_00_fR_G}. Here, $\bar f_{\bar L}\equiv\partial\bar f/\partial\bar L$.

As before, let $\bar G_0(N)$ be a background solution to Eq.~\eqref{EFE_00_fL_G}, and consider a small perturbation $\bar G(N)=\bar G_0(N)+\delta\bar G(N)$.
Expanding Eq.~\eqref{EFE_00_fL_G} to linear order in $\delta\bar G$, one again obtains a second-order ordinary differential equation of the form
\begin{equation}\label{EoM_delta_G_L}
	\mathcal{A}_L\,\delta\bar G''+\mathcal{B}_L\,\delta\bar G'+\mathcal{C}_L\,\delta\bar G=0~.
\end{equation}
For convenience, we introduce
\begin{align}
	P_0 &\equiv 1+\frac{c}{6}\left(\bar G_0+\frac{K}{\bar a^2}\right)~,\\
	U &\equiv 12+c\left(2\bar G_0+\frac{K}{\bar a^2}+\frac{\bar G_0'}{2}\right)~,\\
	W &\equiv 12+c\left(2\bar G_0+\bar G_0'\right)~,\\
	T &\equiv 2\bar G_0'+\frac{\bar G_0''}{2}-2\frac{K}{\bar a^2}~,\\
	X & \equiv \bar G_0 \, \bar f_{\bar L \bar L \bar L}(\bar L_0)\, \bar L_0'
	- \left( \bar G_0 + \frac{\bar G_0'}{2} \right)
	\bar f_{\bar L \bar L}(\bar L_0)~.
\end{align}
Then, the coefficients in Eq.~\eqref{EoM_delta_G_L} are given by
\begin{align}
	\mathcal{A}_L
	&= 3 P_0^2 \, \bar G_0 \, \bar f_{\bar L \bar L}(\bar L_0)~, \label{AL_GB}
	\\[0.3em]
	\mathcal{B}_L
	&= P_0 \, \bar G_0 \, \bar f_{\bar L \bar L}(\bar L_0)\, W + 3 P_0^2X~, \label{BL_GB}
	\\[0.3em]
	\mathcal{C}_L
	&= \bar f_{\bar L}(\bar L_0)
	+ P_0 U X
	+ P_0 c \, \bar G_0 \, \bar f_{\bar L \bar L}(\bar L_0)\, T
	+ \left( 1 + \frac{c}{6} \left( 2\bar G_0 + \frac{K}{\bar a^2} \right) \right)
	\bar f_{\bar L \bar L}(\bar L_0)\, \bar L_0'~. \label{CL_GB}
\end{align}
Here, $\bar L_0$ denotes the background value of $\bar L$ constructed from $\bar G_0$. As a consistency check, in the limit $c\to0$, Eq.~\eqref{EFE_00_fL_G} reduces to \eqref{EFE_00_fR_G}, while $\mathcal{A}_L$, $\mathcal{B}_L$, $\mathcal{C}_L$ reduce to $\mathcal{A}$, $\mathcal{B}$, $\mathcal{C}$ in Eq.~\eqref{EoM_delta_G}.

Following notations in $f(R)$ stability analysis, we define
\begin{align}
	p_{\rm GB} &\equiv \frac{\mathcal{B}_L}{\mathcal{A}_L}=\frac{W}{3P_0}
	+
	\frac{X}{\bar G_0\,\bar f_{\bar L\bar L}(\bar L_0)}\label{p_GB_def}~,\\ 
	q_{\rm GB}& \equiv \frac{\mathcal{C}_L}{\mathcal{A}_L}=\frac{\bar f_{\bar L}(\bar L_0)}
	{3P_0^2\,\bar G_0\,\bar f_{\bar L\bar L}(\bar L_0)}
	+\frac{U}{3P_0}
	\frac{X}{\bar G_0\,\bar f_{\bar L\bar L}(\bar L_0)}
	+\frac{c}{3P_0}T
	+\frac{D}{3P_0^2\,\bar G_0}\,\bar L_0'~,\label{q_GB_def}
\end{align}
where 
\begin{equation}
	D \equiv 1+\frac{c}{6}\left(2\bar G_0+\frac{K}{\bar a^2}\right)~.
\end{equation}
Then, $\chi_{\rm GB}$ is defined as
\begin{equation}
	\chi_{\rm GB}\equiv\mathrm{Re}\Big[(-p_{\rm GB}+\sqrt{p_{\rm GB}^2-4q_{\rm GB}})/2\Big]~.
\end{equation}
Employing the background Eq. \eqref{EFE_00_fL_G} to eliminate $\bar L_0'$, Eq.~\eqref{p_GB_def} becomes
\begin{equation}
	\label{pGB}
	p_{\rm GB}
	=
	6+\frac{\bar G_0'}{\bar G_0}
	-\frac{\bar f_{\bar L}(\bar L_0)-1}
	{P_0\,\bar G_0\,\bar f_{\bar L}^2(\bar L_0)}
	+\frac{c}{3P_0}
	\left(
	\bar G_0'-2\frac{K}{\bar a^2}
	\right).~
\end{equation}
By using the relation $\displaystyle \frac{X}{\bar G_0\,\bar f_{\bar L\bar L}(\bar L_0)}
=
p_{\rm GB}-\frac{W}{3P_0}$, implied by Eq.~\eqref{p_GB_def}, we can express $q_{\rm GB}$ as
\begin{equation}
	\label{qGB}
	q_{\rm GB}
	=
	\frac{U}{3P_0}\,p_{\rm GB}
	-\frac{U W}{9P_0^2}
	+\frac{c}{3P_0}T
	+\frac{2}{3P_0^2\,\bar G_0\,\bar f_{\bar L}^2(\bar L_0)}
	+\frac{D}{3P_0^2\,\bar G_0}\,\bar L_0'~.
\end{equation}
It is straightforward to verify that when $c \to 0$, one has $P_0 \to 1$ and $\bar L_0 \to \bar R_0$, and Eqs.~\eqref{pGB} and \eqref{qGB} reduce (smoothly) to \eqref{gen_p} and \eqref{gen_q}.

Near the bounce, $N=0$, general early-time behavior of $p_{\rm GB}$, $q_{\rm GB}$, and $\chi_{\rm GB}$ can be analyzed as follows: Introducing (for $K=+1$)
\begin{equation}
	\sigma=\pm1~,~~~s\equiv \frac{\sqrt6}{\bar a(0)}~,~~~P_{00}\equiv P_0(0)=1+\frac{c}{6\bar a(0)^2}
	=1+\frac{cs^2}{36}~,
	\label{eq:GB-sigma-def}
\end{equation}
one can write (from \eqref{H_root_0_GB})
\begin{equation}
	\bar G_0'(0)=2\dot{\bar H}(0)
	=\frac{2}{3}\,\frac{\sigma s-s^2}{P_{00}}~,
	\label{eq:GB-Gprime-sigma}
\end{equation}
leading to
\begin{equation}
	\bar G_0(N)=\alpha_\sigma N+O(N^2)~,~~~\alpha_\sigma\equiv \frac{2}{3}\frac{\sigma s-s^2}{P_{00}}~.
	\label{eq:GB-G-early}
\end{equation}
Therefore, near the bounce, $p_{\rm GB}$ and $q_{\rm GB}$ behave as
\begin{equation}
	p_{\rm GB}(N)=-\frac{1}{2N}+O(1)~,~~~q_{\rm GB}(N)=\frac{Q_\sigma(s,c)}{N}+O(1)~,
	\label{eq:GB-pq-early}
\end{equation}
where
\begin{equation}
	Q_\sigma(s,c)
	=
	-\,\frac{\sigma}{s\,P_{00}^2}
	\Big[
	1+\frac{cs^2}{36}\bigl(3\sigma-2s\bigr)
	\Big]~.
	\label{eq:GB-Qsigma}
\end{equation}
In the limit $c\to0$, $Q_+(s,c)$ reduces to $-1/s$, complying with the $f(R)$ case. Hence, the characteristic exponent, for both branches, behaves as
\begin{equation}
	\chi_{\rm GB}
	\sim \frac{1}{2N}+O(1)~,
	\label{eq:GB-chi-early}
\end{equation}
when $N\to 0^{+}$, which means that $\chi_{\rm GB}$ is generically positive sufficiently close to the
bounce. This behavior is identical to that in the pure $f(R)$ case. Although the characteristic exponent is positive, it does not signal a genuine instability because the coefficient of the second-derivative term in Eq.~\eqref{EoM_delta_G_L} satisfies $\mathcal{A}_L=3P_0^2G_0\bar f_{\bar L\bar L}(L_0)\propto N$ and vanishes at the bounce. The perturbation equation therefore degenerates to be first-order, and therefore the local solution remains in the form $\delta G(N)=c_1[1+O(N)]+c_2N^{3/2}[1+O(N)]$, implying that the perturbation does not have sufficient time to grow if $\chi_{\rm GB}$ changes sign within one e-fold.

Late-time behavior of the GB solutions can be analyzed analytically from \eqref{pGB} and \eqref{qGB}. For the runaway solutions, one has $\bar a\gg1$, $\bar G_0'\to0$, $\bar G_0''\to0$, and $K/\bar a^2\to0$. Equation~\eqref{EFE_00_fL_G} then implies that asymptotically $\bar G_0$ approaches a constant $\bar G_0\to \bar G_\ast$, with $\bar L_0\to 1$, and therefore, 
\begin{equation}
c\bar G_\ast^2+12\bar G_\ast-1=0~.
\end{equation}
The two roots of this equation can be written as
\begin{equation}\label{late-time-Runaway-GB-G}
\bar G_\ast^{(+)}=\frac{1}{6+\sqrt{36+c}}~,~~~\bar G_\ast^{(-)}=\frac{1}{6-\sqrt{36+c}}~.
\end{equation}
These two solutions are distinguished by their $c\to0$ limits. The first one satisfies $\bar G_\ast^{(+)}\to 1/12$, and is thus smoothly connected to the runaway solution of the pure $f(R)$ theory. In contrast, the second root diverges as $\bar G_\ast^{(-)}\sim -12/c$ when $c\to0$. From this point of view, the ``$+$'' branch of the initial condition \eqref{H_root_0_GB} must evolve towards $\bar G_\ast^{(+)}$, while the ``$-$'' branch, associated with $\bar G_\ast^{(-)}$, corresponds to the genuine GB-supported runaway solution.

In either case, we have $\bar f_{\bar L}(\bar L_0)\to\infty$, as well as $P_0\to 1+\frac{c}{6}\bar G_\ast$, $U\to 12+2c\bar G_\ast$ and $W\to 12+2c\bar G_\ast$. Substituting these into \eqref{pGB} and \eqref{qGB}, one finds $p_{\rm GB}\to 6$ and $q_{\rm GB}\to 8$, such that
\begin{equation}\label{late-time-chi_GB-behav}
	\chi_{\rm GB}
	={\rm Re}\!\Bigg[\frac{-p_{\rm GB}+\sqrt{p_{\rm GB}^2-4q_{\rm GB}}}{2}\Bigg]
	\to -2~.
\end{equation}
Therefore, while the GB term modifies the asymptotic value of $\bar G_0$ in the runaway case, it does not alter the late-time attractor values of $p_{\rm GB}$, $q_{\rm GB}$, or $\chi_{\rm GB}$.

For the oscillatory branch, the relevant late-time regime is when $\bar G_0\to0^+$, with $0<\bar f_{\bar L}(\bar L_0)<1$ (because the system is not well-defined beyond this point). In this limit, Eq.~\eqref{EFE_00_fL_G} implies $\bar L_0=3\bar G_0'+O(\bar G_0)$ and hence $\bar G_0'=(1-\bar f_{\bar L}^{-2})/3+O(\bar G_0)$. Also, from $P_0\to1$, Eq.~\eqref{pGB} yields
\begin{equation}
	p_{\rm GB}\sim
	\frac{(\bar f_{\bar L}-1)(\bar f_{\bar L}-2)}{3\bar f_{\bar L}^{\,2}}
	\frac{1}{\bar G_0}~.
\end{equation}
Using \eqref{EFE_00_fL_G}, one can find $\bar L_0'\sim\tfrac{1}{3}(\bar f_{\bar L}-1)^2\bar f_{\bar L}^{-4}\bar G_0^{-1}$,
which leads to
\begin{equation}
	q_{\rm GB}\sim
	\frac{(\bar f_{\bar L}-1)^2}{9\bar f_{\bar L}^{\,4}}
	\frac{1}{\bar G_0^2}~,
\end{equation}
and accordingly,
\begin{equation}
	p_{\rm GB}^2-4q_{\rm GB}
	\sim
	\frac{(\bar f_{\bar L}-1)^2(\bar f_{\bar L}^2-4\bar f_{\bar L})}
	{9\bar f_{\bar L}^{\,4}}
	\frac{1}{\bar G_0^2}
	<0~~~(0<\bar f_{\bar L}<1)~,
\end{equation}
such that the characteristic roots form a complex conjugate pair. Their real part behaves as
\begin{equation}\label{late-time-Osc-chi_GB-behav}
	\chi_{\rm GB}\sim
	-\frac{(\bar f_{\bar L}-1)(\bar f_{\bar L}-2)}
	{6\bar f_{\bar L}^{\,2}}
	\frac{1}{\bar G_0}
	\to -\infty
	\qquad {\rm as}\qquad \bar G_0\to0^+~.
\end{equation}
Thus, as in the pure $f(R)$ case, the oscillatory branch enters a strongly damped regime when $\bar G_0$ approaches zero. The GB term shifts the background evolution and the location of the branch, but does not modify the asymptotic structure of the damping in this regime.

\begin{figure}
	\centering
	\begin{subfigure}{0.48\linewidth}
		\includegraphics[width=\linewidth]{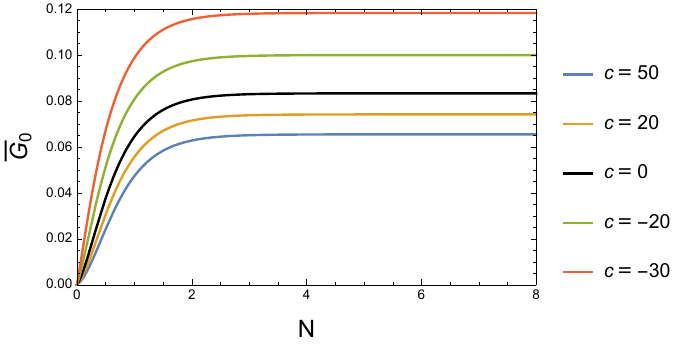}
	\end{subfigure}
	\begin{subfigure}{0.48\linewidth}
		\includegraphics[width=\linewidth]{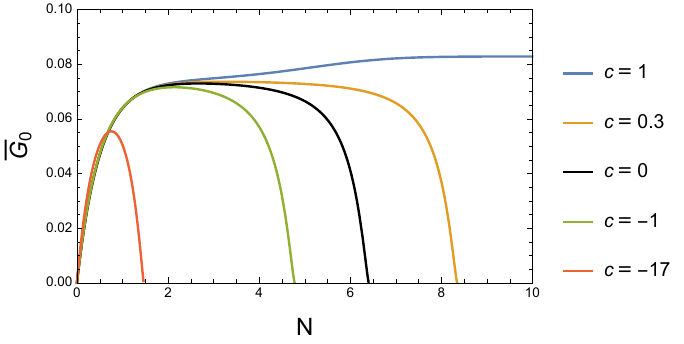}
		
	\end{subfigure}\\
	
	\begin{subfigure}{0.48\linewidth}
		\includegraphics[width=\linewidth]{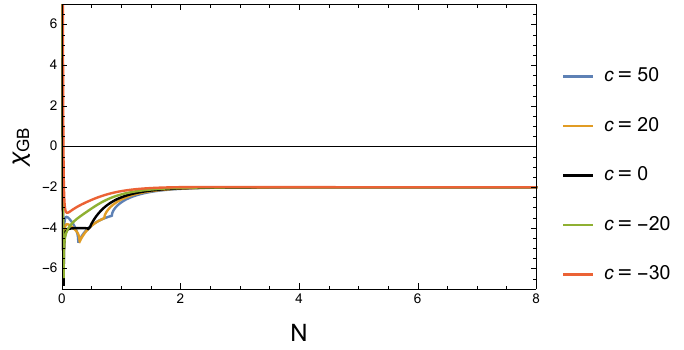}	
	\end{subfigure} 
	\begin{subfigure}{0.48\linewidth}
		\includegraphics[width=\linewidth]{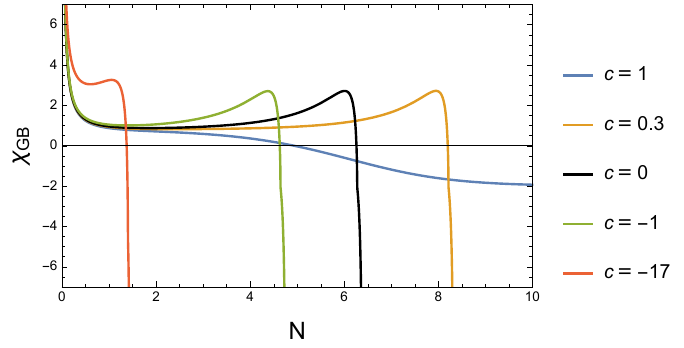}
	\end{subfigure}
	\captionsetup{width=1\linewidth}
	\caption{Evolution of $\bar G_0(N)$ and $\chi_{\rm GB}(N)$ for bouncing solutions in the presence of the GB term with the ``$+$" branch choice in \eqref{H_root_0_GB}, corresponding to  Fig.~\ref{Fig_GB_bounce}.  Left (right) column displays the evolution for runaway solutions with $\bar{a}(0)=2.6$ ($\bar{a}(0)=3.68$).}
	\label{Plus-branch-Runaway-G-chi-GB}
\end{figure}

Figure.~\ref{Plus-branch-Runaway-G-chi-GB} shows the numerical evolution of $\bar G_0(N)$ and $\chi_{\rm GB}(N)$ for representative values of $c$ with the ``$+$" branch of initial condition \eqref{H_root_0_GB}. The qualitative behavior closely parallels that of the pure $f(R)$ case but with characteristic shifts induced by the GB coupling.

\begin{figure}
	\centering
	\includegraphics[width=0.5\linewidth]{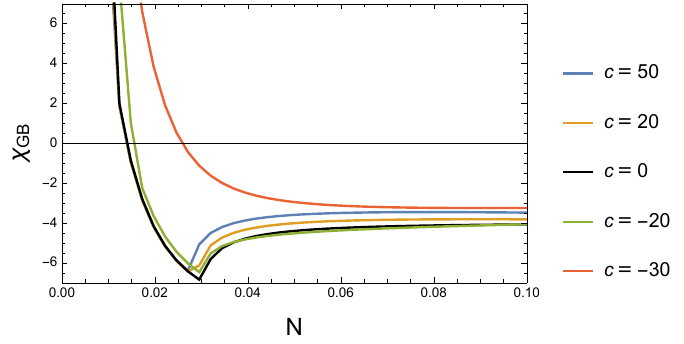}
	\captionsetup{width=1\linewidth}
	\caption{Early time evolution for $\chi_{\rm GB}$ with $\bar a(0)=2.6.$ and the ''$+$" branch choice of initial condition \eqref{H_root_0_GB}. }\label{ChiGBSN_26}
\end{figure}

For the runaway solutions (left column of Fig.~\ref{Plus-branch-Runaway-G-chi-GB}), the qualitative behavior of $\chi_{\rm GB}$ remains the same as in the pure $f(R)$ case: It is initially positive, then crosses zero at a finite e-fold, and finally asymptotes to $\chi_{\rm GB}\to -2$, in agreement with the late-time result \eqref{late-time-chi_GB-behav}. Nevertheless, Fig.~\ref{ChiGBSN_26} shows that the e-fold at which $\chi_{\rm GB}$ first crosses zero is affected by the GB parameter $c$. In particular, for $c>0$, this shift is rather mild, whereas for $c<0$, the zero-crossing e-fold is much more sensitive to $c$. This asymmetry can be understood from the subleading $O(1)$ term in the early-time expansion \eqref{eq:GB-chi-early}. Although the leading-order behavior is universal, the $O(1)$ part explicitly contains contributions proportional to $Q_{\sigma}$ given in \eqref{eq:GB-Qsigma}, and hence involves factors of $P_{00}^{-2}$. Since $P_{00}=1+\frac{cs^2}{36}$ from Eq.~\eqref{eq:GB-sigma-def}, its inverse remains of order unity for positive $c$, so varying $c>0$ does not significantly change $1/P_{00}$ or $1/P_{00}^2$. In contrast, for negative $c$, one has $P_{00}<1$, and as a result $1/P_{00}$ and especially $1/P_{00}^2$ are rapidly enhanced, which in turn leads to a much stronger shift of the first zero-crossing e-fold. Meanwhile, the numerical evolution of $\bar G_0(N)$ in the left column of Fig.~\ref{Plus-branch-Runaway-G-chi-GB} shows that the runaway $\bar G_0$ approaches a constant at late times, agreeing with the analytic prediction that $\bar G_0\to \bar G_\ast$, where $\bar G_\ast$ is given by the first root $ \bar G_\ast^{(+)}$ in Eq.~\eqref{late-time-Runaway-GB-G}. The first zero-crossing e-fold in Fig.~\ref{ChiGBSN_26} is at the order of $O(10^{-2})$, indicating that the growing modes of $\delta\bar{G}$ are under perturbative control and the solutions are effectively stable.

For the oscillatory solutions (right column of Fig.~\ref{Plus-branch-Runaway-G-chi-GB}), the behavior of $\chi_{\rm GB}$ is qualitatively similar to that in the $f(R)$ case. In particular, $\chi_{\rm GB}$ starts from a positive value near the bounce and remains positive until $\bar G_0$ approaches zero, at which point $\chi_{\rm GB}$ crosses zero and rapidly diverges to negative values. This is consistent with the analytic result in Eq.~\eqref{late-time-Osc-chi_GB-behav}. As shown in Fig.~\ref{Plus-branch-Runaway-G-chi-GB}, the corresponding zero-crossing e-fold is relatively large, indicating that the oscillatory solutions remain unstable throughout the evolution until $\bar G_0$ approaches zero. The GB parameter $c$ affects the zero-crossing e-fold number of $\bar G_0$: For a given oscillatory solution, positive $c$ shifts this e-fold to larger values, whereas negative $c$ reduces it. Moreover, when $c$ becomes sufficiently large, the oscillatory branch ceases to exist and is replaced by a runaway solution.

\begin{figure}
	\centering
	\begin{subfigure}{0.48\linewidth}
		\includegraphics[width=\linewidth]{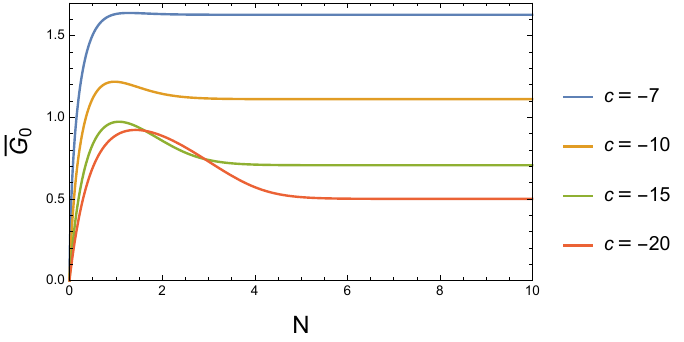}
	\end{subfigure}
	\begin{subfigure}{0.48\linewidth}
		\includegraphics[width=\linewidth]{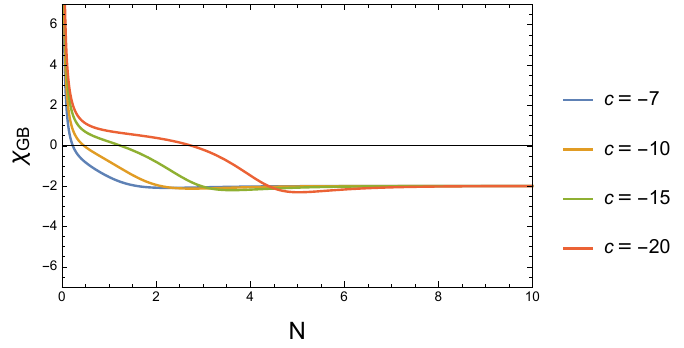}
		
	\end{subfigure}
	\captionsetup{width=1\linewidth}
	\caption{Evolution of $\bar G_0(N)$ and $\chi_{\rm GB}(N)$ for bouncing solutions in the presence of the GB term for the ``$-$" branch choice, corresponding to  Fig.~\ref{Fig_GB_bounce_new}.}
	\label{Minus-IC-G-chi-GB-Runaway}
\end{figure}

For the ``$+$'' branch of \eqref{H_root_0_GB}, the resulting runaway solutions exhibit qualitatively the same behavior of $\bar G_0(N)$, and $\chi_{\rm GB}(N)$ as in the pure $f(R)$ case, and for the (GB-supported) ``$-$'' branch, $\bar G_0$ approaches a constant and $\chi_{\rm GB}$ eventually becomes negative. The only difference with the ``$+$" branch runaway case is that the late-time asymptotic value is now given by the second root $\bar G^{(-)}_\ast$ in \eqref{late-time-Runaway-GB-G}. See in Fig.~\ref{Minus-IC-G-chi-GB-Runaway}, the magnitude of $c$ plays an important role: For sufficiently large $|c|$, $\chi_{\rm GB}$ does not cross zero at small $N$, implying that the solution remains effectively unstable at early times and is stabilized only at later stages when $\bar G_0$ approaches its asymptotic regime.

\section{Conclusion and discussion}\label{Sec_Conclusion}

In this work, we used the Kaluza--Klein approach to construct Born--Infeld-type modifications of gravity. The idea is to embed 4D Born--Infeld electrodynamics (in flat spacetime) in a 5D modified gravity by using the KK metric (for simplicity, the radial modulus scalar, or radion, is assumed to be constant). This yields the relations between the spacetime curvature and the electromagnetic field strength
\begin{equation}
    R\leftrightarrow F_{\mu\nu}F^{\mu\nu}~,~~~\cg\leftrightarrow (\epsilon_{\mu\nu\rho\sigma}F^{\mu\nu}F^{\rho\sigma})^2~,
\end{equation}
up to field strength derivatives. The first relation is what gives rise to the original idea of unifying GR and Maxwell's theory within an underlying gravitational theory in 5D. The second relation allows one to go further and embed the BI theory in a 5D modified gravity. The latter can then be dimensionally reduced to yield a vector-tensor theory in four dimensions, whose vector part coincides with the BI theory non-minimally coupled to the BI-type $f(R,\cg)$ gravity. The theory is a ghost-free one-parameter extension of GR. It has second-order equations of motion and propagates a massless graviton, a massless BI gauge field, and a massive scalar of gravitational origin (scalaron). In the Einstein frame, the theory becomes a Horndeski-type scalar-vector-tensor theory.

There are two ways to view this theory. One is to consider it as a genuine higher-dimensional gravity reduced to four dimensions, in which case the stabilization of the radial modulus should eventually be addressed. The other is to simply treat it as a formalism to construct the BI-type gravity. In this paper, we adopt the latter viewpoint and focus on the gravitational sector only. 

We studied bouncing cosmological solutions in the FLRW background in both the Jordan and Einstein frames, and found that positive spatial curvature is needed to satisfy the initial conditions for successful bounce. Most of the important features of the bouncing solutions can be described within a simplified $f(R)$ gravity model of the BI type, $\cl\sim \sqrt{1-b^2R}$, which was considered first. The BI parameter $b$ can be absorbed by appropriate rescalings, such that the resulting (third-order) equation of motion for the scale factor is parameter-free. Furthermore, the simplest (symmetric under time reflection) bouncing solutions are fully characterized by a single number -- the scale factor value at the bounce point. By scanning the allowed (by bounce conditions) values of the initial scale factor, we found solutions with one or multiple bounces centered around the main bounce at $t=0$. All of the bouncing solutions can be categorized, depending on their asymptotic behavior, as either runaway (Hubble function approaches a constant) or oscillatory (Hubble function oscillates around zero) solutions. More complicated, asymmetric, bouncing solutions can be obtained from the symmetric ones by introducing a small, non-zero first derivative of the scale factor at $t=0$. Such solutions can interpolate between oscillatory and runaway behavior at $t\rightarrow\pm\infty$.

For a better understanding of the bouncing solutions, we switched to the Einstein frame and studied the solutions in terms of the scalaron dynamics and its scalar potential which is shown in Fig. \ref{Fig_varphi_scheme}. The potential features a local maximum dividing a stable Minkowski minimum and a runaway minimum at $\varphi\rightarrow\infty$. The oscillatory solutions can be understood as those whose scalaron oscillates around the stable minimum, while for the runaway solutions the scalaron follows the runaway minimum of the potential. One can also notice the difference in the evolution of the Hubble function between the Jordan and Einstein frames, as seen in Fig. \ref{Fig_varphi_bounce}. For example, the runaway solutions feature a plateauing Hubble function in the Jordan frame, and an exponentially decaying one in the Einstein frame.

When adding the GB term, in order to complete the BI-like structure of the Lagrangian as
\begin{equation}\label{L_BIg_conc}
    \cl=\frac{1}{b^2}\Big(1-\sqrt{1-b^2R-\frac{c\,b^4}{24}\cg}\Big)~,
\end{equation} 
we also introduced a new parameter $c$ for generality. Our BI-type gravity is obtained for $c=1$, which leads to mild deformations of the bouncing solutions that we obtained in the $f(R)$ model. If, on the other hand, $c$ is treated as a free parameter, it can lead to new bouncing solutions that are supported by the GB term with sufficiently negative $c$. The addition of the GB term does not lead to new degrees of freedom, because it enters via a linear combination $R+\cg$ (which is a 4D Lovelock invariant), but it modifies the effective potential for the scalaron, as we discussed in the context of bouncing solutions.

We performed a linear stability analysis of the bouncing solutions in both the $f(R)$ and $f(R,\cg)$ cases by perturbing the Hubble function using e-fold number as the time variable. Although we found an initial growing perturbation mode (at the bounce point), it eventually decays, in certain cases within an e-fold. The runaway solutions were shown to be future attractors, while the oscillatory solutions become stable close to the first post-bounce zero crossing of the Hubble function. Nevertheless, full linear perturbation theory should still be analyzed in order to show the absence of gradient instabilities of the perturbations which were shown to plague Horndeski-type theories with bouncing solutions \cite{Libanov:2016kfc,Kobayashi:2016xpl}. Although the BI-type gravity considered in this work is equivalent to a Horndeski theory, our bouncing solutions are supported by positive spatial curvature, rather than higher-derivative corrections in the Lagrangian (with the exception of the Gauss--Bonnet-supported solutions discussed in Sec. \ref{Sec_GB}, which rely on both positive spatial curvature and the Gauss--Bonnet term).

The generalized BI-type Lagrangian \eqref{L_BIg_conc} has previously appeared in the literature. For example, Comelli \cite{Comelli:2005tn} arrived at this Lagrangian by considering generalizations of the determinant-based BI-type gravity of Deser and Gibbons \cite{Deser:1998rj}, and showed the existence of regular black holes~\footnote{Some observational signatures of regular black holes are discussed in \cite{Eiroa:2010wm,Guo:2024svn,Errehymy:2025zuj,Yuan:2026qxe,Hsiao:2026oti}.}. A more detailed study of such regular black holes can be found in \cite{Feigenbaum:1998wy}, where a similar BI-type gravity was considered. Taking into account the results of our work, one can therefore conclude that both cosmological and black hole singularities can be removed in the BI-type gravity \eqref{L_BIg_conc}. In another related study \cite{Garcia-Salcedo:2009lui}, cosmological solutions of this theory were considered in the case of flat FLRW metric. The authors of \cite{Garcia-Salcedo:2009lui} used dynamical system analysis, and concluded that the theory generically leads to a Big Bang singularity. This is compatible with our results suggesting the necessity of positive spatial curvature to satisfy initial conditions for the bounce. The simplified model of the type $\cl\sim \sqrt{1-b^2R}$ has also been considered before -- for instance in \cite{Kruglov:2012ja}, cosmological solutions in flat FLRW spacetime and some black hole solutions were studied.

There are several directions for the future studies of the BI-type gravity \eqref{L_BIg_conc}. First, the model should be able to generate cosmological perturbations with correct power spectra to match observations. In general bouncing cosmologies, the observed power spectra can be generated during the contraction phase, or during post-bounce inflation. Both scenarios can in principle be considered. A more detailed study of regular black hole solutions is also warranted. One can study, for example, regular black holes with the inclusion of the non-minimally coupled BI electrodynamics given by Eq. \eqref{L_BI_SVT}.

Finally, multi-bounce scenarios in general also deserve further study. This is due to both their genericness and their potential in distinguishing different beyond-$\Lambda$CDM models. Our BI-type gravity offers a concrete case of oscillating bounces in the early universe. Nevertheless, multiple bounces can also arise from quantum corrections induced by discreetness of spacetime \cite{Alesci:2016xqa}. An initial investigation of the evolution of linear scalar perturbations in multi-bounce scenarios was carried out using simplified toy models in \cite{Brandenberger:2017pjz}. One key feature uncovered was the reddening of the tilt for small-scale (scalar) modes, which offers some reassurance for the viability of multi-bounce scenarios. In future work, it would be worthwhile to extend the analysis beyond linear curvature perturbations.
In terms of non-Gaussianities, one might expect multiple bounces to further relieve tensions in the CMB via modulation effects \cite{Agullo:2020cvg}. Multiple bounces in the BI-type gravity might also evade the non-Gaussianity excess problem \cite{Gao:2014eaa} commonly found in bounce scenarios with small kinetic energy.
For gravitational waves, multi-bounce scenarios effectively provide multiple barriers for tensor perturbations, which might induce detectable resonant tunneling-like effects \cite{Zhu:2026rbl}.

\section*{Acknowledgements}

DD is supported by China Postdoctoral Science Foundation (Certificate No.2023M730704). YW is supported by NSFC Grant No. 12475001, the Shanghai Municipal Science and Technology Major Project (Grant No. 2019SHZDZX01), Science and Technology Commission of Shanghai Municipality (Grant No. 24LZ1400100), and the Innovation Program for Quantum Science and Technology (No. 2024ZD0300101). YW is grateful for the hospitality of the Perimeter Institute during his visit, where this work was partially done. This research was supported in part by the Perimeter Institute for Theoretical Physics. Research at Perimeter Institute is supported by the Government of Canada through the Department of Innovation, Science and Economic Development and by the Province of Ontario through the Ministry of Research, Innovation and Science.

\appendices

\section{Dimensional reduction results}\label{App_dimred}

We consider a $(4+1)$-dimensional metric $\hat{g}_{MN}$ with KK decomposition (with constant radial scalar):
\begin{equation}
\hat{g}_{MN} = 
\begin{pmatrix}
g_{\mu\nu} + 2A_\mu A_\nu & \sqrt{2} A_\mu \\
\sqrt{2} A_\nu & 1
\end{pmatrix}~,
\end{equation}
where $g_{\mu\nu}$ is the $(3+1)$-dimensional metric, and $A_\mu$ is the KK vector. Upon dimensional reduction, the relation between the higher-dimensional and reduced metric determinants is given by $\sqrt{-\hat{g}} = \sqrt{-g}$. The relevant curvature invariants are then reduced as follows:
\begin{align}
\hat{R} &= R - \tfrac{1}{2}F^2~, \\
\hat{R}_{MN}\hat{R}^{MN} &= R_{\mu\nu}R^{\mu\nu} + 2R_{\mu\nu} F^{\mu\lambda} {F_\lambda}^\nu + \tfrac{1}{4}F^4 + F^\mu_{\phantom{\mu}\nu} F^\nu_{\phantom{\nu}\lambda} F^\lambda_{\phantom{\lambda}\rho} F^\rho_{\phantom{\sigma}\mu} + \nabla^\lambda F_{\lambda\mu}\nabla_\rho F^{\rho\mu}~, \label{eq:RR} \\
\hat{R}_{MNPQ}\hat{R}^{MNPQ} &= R_{\mu\nu\rho\sigma}R^{\mu\nu\rho\sigma} - 3R_{\mu\nu\rho\sigma}F^{\mu\nu}F^{\rho\sigma} + \tfrac{3}{2}F^4 + \tfrac{5}{2}F^\mu_{\phantom{\mu}\nu} F^\nu_{\phantom{\nu}\lambda} F^\lambda_{\phantom{\lambda}\rho} F^\rho_{\phantom{\sigma}\mu} + 2\nabla_\lambda F_{\mu\nu}\nabla^\lambda F^{\mu\nu}~, \label{eq:R2}
\end{align}
where $F^4\equiv(F_{\mu\nu}F^{\mu\nu})^2$. Defining the dual field strength tensor as $\tilde{F}^{\mu\nu} = \frac{1}{2}\epsilon^{\mu\nu\rho\sigma} F_{\rho\sigma}$, and using the identities
\begin{align}
    F^4-2F^\mu_{\phantom{\mu}\nu} F^\nu_{\phantom{\nu}\lambda} F^\lambda_{\phantom{\lambda}\rho} F^\rho_{\phantom{\sigma}\mu} &=2(F_{\mu\nu}\tilde F^{\mu\nu})^2~,\\
    2R^{\mu\nu\rho\sigma}F_{\mu\rho}F_{\nu\sigma} &=R^{\mu\nu\rho\sigma}F_{\mu\nu}F_{\rho\sigma}~,\\
    \nabla_\lambda F_{\mu\nu}\nabla^\lambda F^{\mu\nu}-2\nabla^\lambda F_{\lambda\mu}\nabla_\rho F^{\rho\mu} &=2\nabla_\lambda\nabla^\mu(F_{\mu\nu}F^{\nu\lambda})+\Box F^2+2R_{\mu\nu}F^{\nu}_{\phantom\nu\lambda}F^{\lambda\mu}+R^{\mu\nu\rho\sigma}F_{\mu\nu}F_{\rho\sigma}~,
\end{align}
we can write the reduction of the 5D GB invariant as
\begin{equation}
\hat{\mathcal{G}} = \mathcal{G} - RF^2 - 4R_{\mu\nu}F^{\nu}_{\phantom\nu\lambda}F^{\lambda\mu} - R^{\mu\nu\rho\sigma}F_{\mu\nu}F_{\rho\sigma} + \tfrac{3}{2}(F\tilde{F})^2 + 2\Box F^2 + 4\nabla_\lambda \nabla^\mu (F_{\mu\nu}F^{\nu\lambda})~. \label{eq:G}
\end{equation}
Consequently, $\hat{L} \equiv \hat{R} + \frac{b^2}{24}\hat{\mathcal{G}}$ takes the explicit form:
\begin{align}
\begin{aligned}
\hat{L} = R - \tfrac{1}{2}F^2 + \tfrac{b^2}{24} \Bigl[ & \mathcal{G} - RF^2 - 4R_{\mu\nu}F^{\nu}_{\phantom\nu\lambda}F^{\lambda\mu} - R^{\mu\nu\rho\sigma}F_{\mu\nu}F_{\rho\sigma} \\
& + \tfrac{3}{2}(F\tilde{F})^2 + 2\Box F^2 + 4\nabla_\lambda \nabla^\mu (F_{\mu\nu}F^{\nu\lambda}) \Bigr]~.
\end{aligned}
\end{align}

\section{Runaway and oscillatory bouncing solutions in $f(R)$ frame}\label{App_B}

In this Appendix, we discuss an intuitive picture for runaway and oscillatory bouncing solutions of Sec. \ref{FR_type} (without cosmological constant and matter) directly in the $f(R)$ frame. The trace of the Einstein equations \eqref{fR_EFE} of an $f(R)$ gravity yields (in terms of the rescaled variables)
\begin{equation}\label{Tr_bar_fR_EFE}
	\bar\Box{\bar{f}_{\bar{R}}} - \frac{2}{3}\bar{f} + \frac{1}{3}{\bar{f}_{\bar{R}}}\bar{R}=0~,
\end{equation}
where $\bar\Box \equiv b^2 \Box$ .
Equation \eqref{Tr_bar_fR_EFE} can be interpreted as an equation for the scalar field $\bar{f}_{\bar{R}}$ with the effective potential slope
\begin{equation}\label{f_R_V_eff}
	{V^{{\rm{eff}}}}_{,{{\bar f}_{\bar R}}}\left( {{{\bar f}_{\bar R}}} \right) = \frac{2}{3}\bar f - \frac{1}{3}{{\bar f}_{\bar R}}\bar R = \frac{4}{3} - \frac{1}{{{{\bar f}_{\bar R}}}} - \frac{{{{\bar f}_{\bar R}}}}{3}~.
\end{equation}
The scalar $\bar f_{\bar R}$ is non-canonical, as can be seen from how it enters the Einstein equations (the canonical scalar $\varphi$ is discussed in Sec. \ref{subsec_frames}). 

\begin{figure}
	\centering
		\includegraphics[width=0.5\linewidth]{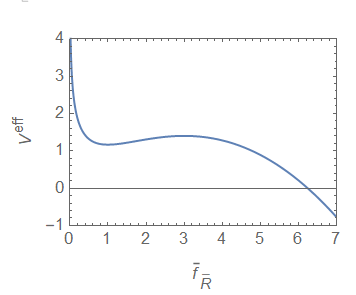}
		\caption{Effective potential ${V^{{\rm{eff}}}}\left( {{{\bar f}_{\bar R}}} \right)$.}\label{f_R_Veff_pic}
\end{figure}

The potential ${V^{{\rm{eff}}}}\left( {{{\bar f}_{\bar R}}} \right)$ has one local minimum at $\bar{f}_{\bar R}=1$ and one local maximum at $\bar{f}_{\bar R}=3$ (see Fig. \ref{f_R_Veff_pic} for an illustration). Solving Eq. \eqref{Tr_bar_fR_EFE} requires two initial conditions -- $\bar f_{\bar R}(0)$ and $\mathring{\bar f}_{\bar R}(0)$. During our calculation in Sec. \ref{FR_type}, Eq. \eqref{H_root_0} and the bounce condition $\bar H(0)=0$ provide the initial value of $\bar f_{\bar R}(0)$. Smaller $\bar a(0)$ results in a larger  $\bar f_{\bar R}(0)$. We also used $\bar a^{(3)}(0)=0$, which gives $\mathring{\bar f}_{\bar R}(0)=0$ (and therefore $\mathring{\varphi}=0$ in the Einstein frame). Thus, the process can be understood as putting $\bar f_{\bar R}(0)$ somewhere along its potential $V^{\rm{eff}}\left( {{{\bar f}_{\bar R}}} \right)$ with zero initial velocity (non-zero initial velocity would correspond to non-zero $\bar a^{(3)}$). If $\bar f_{\bar R}(0)>3$, $\bar f_{\bar R}$ will roll down to $\bar f_{\bar R}\to \infty$, which corresponds to $\bar R\to 1$. The corresponding solutions are the ``runaway" bouncing solutions given in Figs. \ref{Fig_R_bounce} and \ref{Fig_R_bounce_curvature}. In contrast, in the case $\bar f_{\bar R}(0)<3$,  $\bar f_{\bar R}$ will oscillate around the local minimum $\bar f_{\bar R}=1$, which corresponds to $\bar R=0$. Moreover, if $\bar H$ is positive and sufficiently large, the damping term $\propto \bar H \bar f_{\bar R}$ will suppress the amplitude of the oscillations. The corresponding solutions are the oscillatory bouncing solutions shown in Figs.  \ref{Fig_R_bounce} and \ref{Fig_R_bounce_curvature}. Nevertheless, if $\bar H$ become negative, the damping term turns into an anti-damping (driving) term, such that the amplitude of oscillations grows with time. The critical initial value of $\bar f_{\bar R}(0)=3$ corresponds to $\bar a(0)=9/\sqrt{6}\approx 3.67423$, which is indeed the  critical value of $\bar a(0)\approx 3.674$ shown in Figs.  \ref{Fig_R_bounce} and \ref{Fig_R_bounce_curvature}.

\section{Stability analysis with non-vanishing matter}\label{App_matter}

In the presence of matter source, the background equation \eqref{EFE_00_fR_G} is be modified as
\begin{equation}\label{EFE_00_fR_G_rho}
	6{{\bar f}_{\bar R\bar R}}\bar G\left( {2{{\bar G}'} + \frac{{{{\bar G}''}}}{2} - 2\frac{K}{{{{\bar a}^2}}}} \right) - \Big(\bar G + \frac{{{{\bar G}' }}}{2}\Big){{\bar f}_{\bar R}} + \frac{1}{6}\bar f = \frac{1}{3}\bar\rho~,
\end{equation}
where
\begin{equation}
	\bar\rho(N)=\bar\rho(0)e^{-3(1+\omega)N}~.
\end{equation}
Matter only modifies the background equation by the term on the right-hand side. Therefore, the perturbation equation still takes the form of Eq.~\eqref{EoM_delta_G}, and the coefficients $\mathcal{A}$, $\mathcal{B}$ and $\mathcal{C}$ are still given by Eqs.~\eqref{A_coeff}--\eqref{C_coeff}, with the only difference that the background solution $\bar G_0(N)$ must now satisfy Eq.~\eqref{EFE_00_fR_G_rho} instead of \eqref{EFE_00_fR_G}.

Using Eq.~\eqref{EFE_00_fR_G_rho} to eliminate the combination $2\bar G_0'+\bar G_0''/2-2K/\bar a^2$, one finds the matter generalizations of Eqs.~\eqref{gen_p} and \eqref{gen_q}:
\begin{align}
	p_{\rho}
	&=
	6+\frac{1}{{{{\bar G}_0}}}\left( {\frac{1}{{{{\bar f}_{\bar R}}^2}} - \frac{{1-\bar\rho}}{{{{\bar f}_{\bar R}}}} + {{\bar G}_0}^\prime } \right)~,
	\label{gen_p_rho}
	\\
	q_{\rho}
	&=
	8+\frac{{4{{\bar G}_0}^\prime }}{{{{\bar G}_0}}}
	-\frac{{4(1-\bar\rho)}}{{{{\bar f}_{\bar R}}{{\bar G}_0}}}
	+\frac{{16}}{{3{{\bar f}_{\bar R}}^2{{\bar G}_0}}}
	+\frac{{{{\bar G}_0}^\prime }}{{3{{\bar f}_{\bar R}}^2{{\bar G}_0}^2}}
	-\frac{{2(1-\bar\rho)}}{{9{{\bar f}_{\bar R}}^3{{\bar G}_0}^2}}
	+\frac{{2}}{{9{{\bar f}_{\bar R}}^4{{\bar G}_0}^2}}~.
	\label{gen_q_rho}
\end{align}
Near the initial point $N=0$, the leading behavior of the perturbation equation remains unchanged. Indeed, one still has $\mathcal{A}\propto N$, while $\bar\rho(N)$ is regular at the bounce, such that $p_{\rho}(N)\approx -\tfrac{1}{2}N^{-1}+O(1)$ and $q_{\rho}(N)=O(N^{-1})$. Consequently, the local solution for $\delta G(N)$ around $N=0$ keeps the form \eqref{Expand_deltaG_N_0},
which implies that the initial-stage regularity argument remains unchanged in the presence of matter.

The late-time asymptotics can also be analyzed analytically. For $\bar a\gg1$, we have
\begin{align}
	p_{\rho}
	&\approx
	2+\frac{1}{{{{\bar G}_0}}}\left(
	\frac{2}{3{{\bar f}_{\bar R}}^2}
	-\frac{1-\bar\rho}{{{{\bar f}_{\bar R}}}}
	+\frac{1}{3}
	\right)
	=
	2+
	\frac{\left( {{{\bar f}_{\bar R}} - 1} \right)\left( {{{\bar f}_{\bar R}} - 2} \right)+3\bar\rho{{\bar f}_{\bar R}}}{3{{\bar f}_{\bar R}}^2{{\bar G}_0}}~,
	\label{late_time_p_rho}
	\\
	q_{\rho}
	&\approx
	-8+
	\frac{4\left[\left( {{{\bar f}_{\bar R}} - 1} \right)\left( {{{\bar f}_{\bar R}} - 2} \right)+3\bar\rho{{\bar f}_{\bar R}}\right]}{3{{\bar f}_{\bar R}}^2{{\bar G}_0}}
	+
	\frac{\left( {{{\bar f}_{\bar R}} - 1} \right)^2+2\bar\rho{{\bar f}_{\bar R}}}{9{{\bar f}_{\bar R}}^4{{\bar G}_0}^2}~.
	\label{late_time_q_rho}
\end{align}

For ordinary matter with $\omega>-1$, the density redshifts exponentially, $\bar\rho\propto e^{-3(1+\omega)N}$, during the expanding late-time evolution. Therefore, for the runaway branch, one still has $\bar f_{\bar R}\to\infty$, $\bar R\to1$, $\bar G_0\to1/12$, $\bar G_0'\to0$ and $\bar\rho\to0$, such that Eqs.~\eqref{gen_p_rho} and \eqref{gen_q_rho} reduce asymptotically to the vacuum result,
\begin{equation}\label{late_time_pq_runaway_rho}
	p_{\rho}\to 6~,\qquad q_{\rho}\to 8~,
\end{equation}
and therefore
\begin{equation}\label{late_time_chi_runaway_rho}
\chi_{\rho}\equiv\mathrm{Re}\left[(-p_{\rho}+\sqrt{p_{\rho}^2-4q_{\rho}})/2\right]\to -2~.
\end{equation}
Hence, the runaway asymptotic solution $\bar H\sim 1/\sqrt{12}$ remains a stable attractor in the presence of a redshifting matter source.

\begin{figure}
	\centering
	\begin{subfigure}{0.48\linewidth}
		\includegraphics[width=\linewidth]{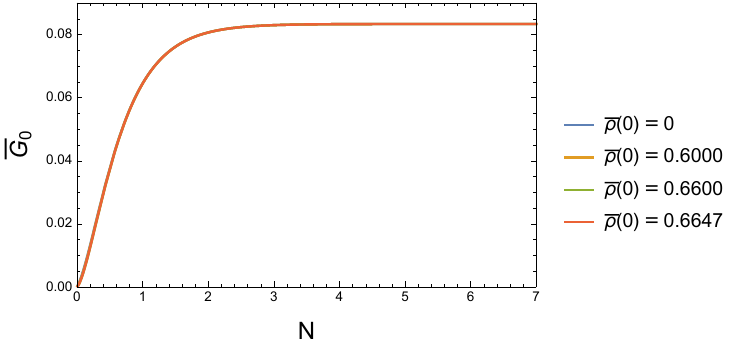}
	\end{subfigure}
	\begin{subfigure}{0.48\linewidth}
		\includegraphics[width=\linewidth]{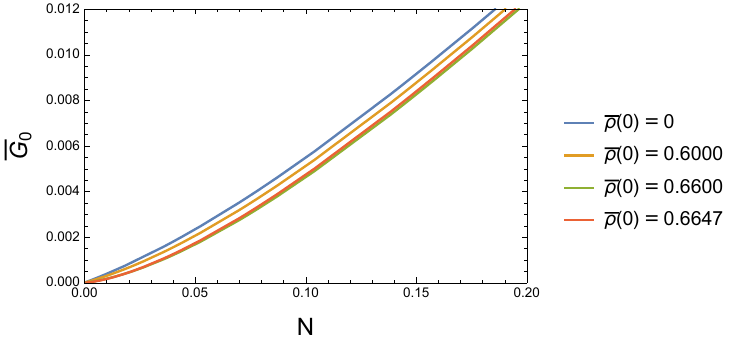}
		
	\end{subfigure}\\
	
	\begin{subfigure}{0.48\linewidth}
		\includegraphics[width=\linewidth]{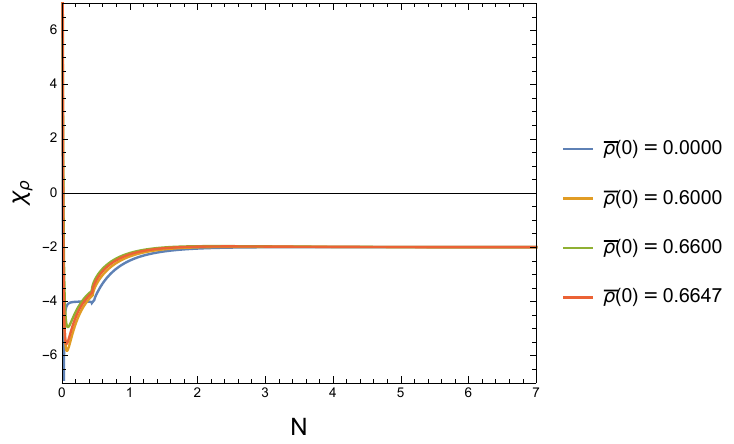}	
	\end{subfigure} 
	\begin{subfigure}{0.48\linewidth}
		\includegraphics[width=\linewidth]{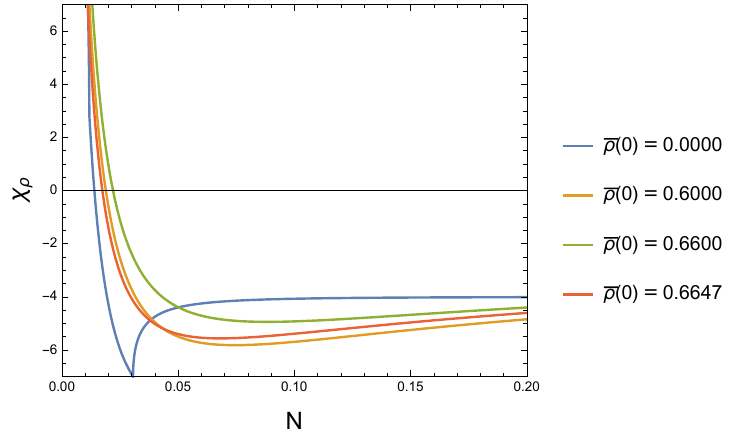}
	\end{subfigure}
	\captionsetup{width=1\linewidth}
	\caption{Evolution of $\bar G_0(N)$ and $\chi_{\rho}(N)$ for the simplest runaway bouncing solutions in the presence of matter, corresponding to the left column Fig.~\ref{Fig_R_bounce_matter}. Left (Right) column displays the evolution over $N=0\sim7$ ($N=0\sim 0.2$) .}
	\label{Runaway-G-chi-rho}
\end{figure}

Figure~\ref{Runaway-G-chi-rho} shows that for the runaway solutions, the inclusion of matter does not qualitatively modify the behavior of $\chi_{\rho}(N)$. In all cases, $\chi_{\rho}$ starts from positive values near $N=0$, indicating the presence of a growing mode, but rapidly decreases and crosses zero within a very small $N$, after which it becomes negative and enters a decaying regime. This implies that the growing mode does not have sufficient time to develop significantly, which means that perturbations remain under control and the solutions are effectively stable even in the presence of matter. At the same time, matter has a noticeable quantitative effect on the timing of the first zero crossing of $\chi$: As $\bar\rho(0)$ varies, the corresponding e-fold $N_{\chi=0}$ shifts in a non-monotonic way, which can be traced back to the non-trivial dependence of $p_\rho$ and $q_\rho$ on both the background evolution $\bar G_0(N)$ and the matter density $\bar\rho(N)$, as shown in Eqs.~\eqref{gen_p_rho} and \eqref{gen_q_rho}.

For the oscillatory solutions, the true asymptotic limit still corresponds to $\bar f_{\bar R}\to1$ and $\bar G_0\to0$. As in the vacuum case, during the first stage, when $\bar G_0$ approaches zero, one can obtain
\begin{equation}\label{osc_p_q_with_rho_G_to_zero}
	p_{\rho}\sim
	\frac{\left( {{{\bar f}_{\bar R}} - 1} \right)\left( {{{\bar f}_{\bar R}} - 2} \right)+3\bar\rho{{\bar f}_{\bar R}}}{3{{\bar f}_{\bar R}}^2}\frac{1}{\bar G_0}~,
	\qquad
	q_{\rho}\sim
	\frac{\left( {{{\bar f}_{\bar R}} - 1} \right)^2+2\bar\rho{{\bar f}_{\bar R}}}{9{{\bar f}_{\bar R}}^4}\frac{1}{\bar G_0^2}~.
\end{equation}
For $\omega>-1$, the matter density is exponentially diluted along the expanding solution, such that the leading-order behavior reduces to
\begin{equation}
	p_{\rho}\sim
	\frac{\left( {{{\bar f}_{\bar R}} - 1} \right)\left( {{{\bar f}_{\bar R}} - 2} \right)}{3{{\bar f}_{\bar R}}^2}\frac{1}{\bar G_0}~,
	\qquad
	q_{\rho}\sim
	\frac{\left( {{{\bar f}_{\bar R}} - 1} \right)^2}{9{{\bar f}_{\bar R}}^4}\frac{1}{\bar G_0^2}~.
\end{equation}
The discriminant then behaves as
\begin{equation}
	p_{\rho}^2-4q_{\rho}
	\sim
	\frac{\left( {{{\bar f}_{\bar R}} - 1} \right)^2\left( {{{\bar f}_{\bar R}}^2 - 4{{\bar f}_{\bar R}}} \right)}{9{{\bar f}_{\bar R}}^4}\frac{1}{\bar G_0^2}<0
	~
\end{equation}
  because $0<\bar{f}_{\bar{R}}<1$. Therefore, the characteristic roots form a complex conjugate pair, and their real part satisfies
\begin{equation}\label{osc_chi_with_rho_G_to_zero}
	\chi_{\rho}
	=
	-\frac{p_{\rho}}{2}
	\sim
	-\frac{\left( {{{\bar f}_{\bar R}} - 1} \right)\left( {{{\bar f}_{\bar R}} - 2} \right)}{6{{\bar f}_{\bar R}}^2}\frac{1}{\bar G_0}~.
\end{equation}
As $\bar G_0\to0^+$, this again leads to $\chi_{\rho}\to-\infty$. Hence, near the first zero crossing of $\bar G_0$, the oscillatory solutions remain strongly damped and linearly stable, in complete analogy with the vacuum case. As before, if $\bar H$ changes sign, the e-fold time $N$ is no longer globally well defined, and the applicability of this method to multi-bounce solutions becomes limited.

\begin{figure}
	\centering
	\begin{subfigure}{0.48\linewidth}
		\includegraphics[width=\linewidth]{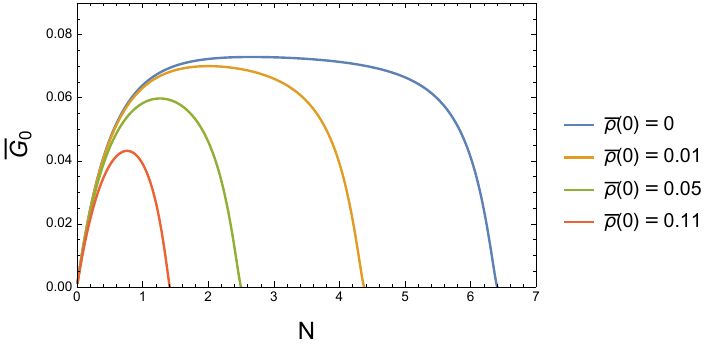}
	\end{subfigure}
	\begin{subfigure}{0.48\linewidth}
		\includegraphics[width=\linewidth]{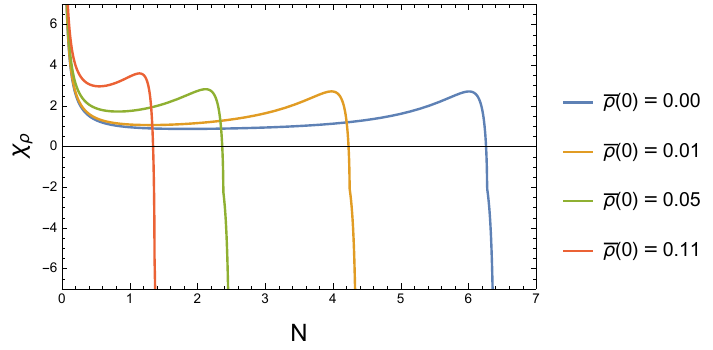}
		
	\end{subfigure}
	\captionsetup{width=1\linewidth}
	\caption{Evolution of $\bar G(N)$ and $\chi_{\rho}(N)$ for the simplest oscillatory bouncing solutions in the presence of matter, corresponding to the right column Fig.~\ref{Fig_R_bounce_matter}.}
	\label{Osc-G-chi-rho}
\end{figure}

Figure~\ref{Osc-G-chi-rho} illustrates the evolution of $\bar G_0(N)$ and $\chi_{\rho}(N)$ for the simplest oscillatory bouncing solutions in the presence of matter. Qualitatively, the presence of matter does not modify the behavior of $\chi_{\rho}(N)$: In all cases $\chi_{\rho}$ starts from positive values near $N=0$, indicating the presence of a growing mode, and then undergoes a rapid transition to negative values when $\bar G_0(N)$ approaches zero for the first time. This sharp drop is consistent with the behavior discussed around Eqs.~\eqref{osc_p_q_with_rho_G_to_zero}--\eqref{osc_chi_with_rho_G_to_zero}, where the vanishing of $\bar G_0$ leads to a strongly damped regime. On the other hand, matter has a significant impact on the background evolution $\bar G_0(N)$ itself. As $\bar\rho(0)$ increases, the e-fold at which $\bar G_0$ returns close to zero is substantially shifted, which in turn directly changes the e-fold at which $\chi_{\rho}$ drops to negative values. Therefore, while the qualitative structure of the oscillatory branch remains unchanged, the presence of matter can strongly affect the timing of the transition from the growing-mode regime to the decaying regime.

\clearpage

\providecommand{\href}[2]{#2}\begingroup\raggedright\endgroup

\end{document}